\documentclass[12pt]{article}
\pdfoutput=1
\usepackage{amsmath,amssymb,amsthm,bm,graphicx,float,array,multirow,multicol,rotfloat,caption,subcaption,hyperref,cleveref,enumerate,geometry,mathdots,adjustbox,booktabs,parskip,mathtools,tikz,tikz-cd,pdflscape,csquotes,lscape,rotating,empheq,mathrsfs,afterpage,bookmark}
\usepackage[para]{threeparttable}
\usepackage[all]{xy}
\usepackage[normalem]{ulem}
\usepackage[numbers,sort&compress]{natbib}
\usetikzlibrary{positioning}
\numberwithin{equation}{section}
\numberwithin{figure}{section}
\allowdisplaybreaks
\hypersetup{bookmarksdepth = 4}
\makeatletter
\usetikzlibrary{arrows,shapes.misc,positioning,decorations.pathmorphing,decorations.markings,matrix,patterns,backgrounds}
\tikzset{gauge/.style={rounded rectangle, draw=black!100,dashed, thick, minimum size=5mm},d2/.style={rounded rectangle, draw=white!100, thick, minimum size=5mm},flavor/.style={rectangle, draw=black!100, thick, minimum size=5mm},gaugeN1/.style={rounded rectangle, draw=black!100, thick, minimum size=5mm},scale cd/.style={every label/.append style={scale=#1}, cells={nodes={scale=#1}}}}

\newcommand*\redboxed[1]{\tikz[baseline=(char.base)]{\node[shape=rectangle,draw=red,thick,font=\LARGE,inner sep=5pt] (char) {#1};}}

\theoremstyle{plain}
\newtheorem*{thm*}{Theorem}

\theoremstyle{definition}

\newtheorem*{defn*}{Definition}

\makeatother

\newcommand{\trace}{\operatorname{Tr} }
\newcommand{\rank }{\operatorname{rank} }

\newcommand{\CN}{\mathcal{N}}
\newcommand{\CD}{\mathcal{D}}
\newcommand{\CO}{\mathcal{O}}

\begin{document}

\begin{titlepage}
\vspace*{-3cm} 
\begin{flushright}
{\tt DESY-24-171}\\
\end{flushright}
\begin{center}
\vspace{2cm}
{\LARGE\bfseries 
A landscape of 4d $\mathcal{N}=1$ SCFTs with $a = c$\\}
\vspace{1.2cm}

{\large
Monica Jinwoo Kang,$^1$ Craig Lawrie,$^2$ Ki-Hong Lee,$^{3,4}$ and Jaewon Song$^3$\\}
\vspace{.7cm}
{$^1$ Department of Physics and Astronomy, University of Pennsylvania\\
Philadelphia, PA 19104, U.S.A.}\par
\vspace{.2cm}
{$^2$ Deutsches Elektronen-Synchrotron DESY,}\par
{Notkestr.~85, 22607 Hamburg, Germany}\par
\vspace{.2cm}
{$^3$ Department of Physics, Korea Advanced Institute of Science and Technology}\par
{Daejeon 34141, Republic of Korea}\par
\vspace{.2cm}
{$^4$ Department of Physics and Haifa Center for Physics and Astrophysics}\par
{ University of Haifa, Haifa 3498838, Israel}\par
\vspace{.2cm}

\vspace{.3cm}

\scalebox{.78}{\tt monica6@sas.upenn.edu, craig.lawrie1729@gmail.com, khlee11812@gmail.com, jaewon.song@kaist.ac.kr}\par
\vspace{1.2cm}
\textbf{Abstract}
\end{center}

\noindent We study a landscape of four-dimensional $\mathcal{N}=1$ superconformal field theories (SCFTs) with identical central charges. These theories are obtained by renormalization group flows triggered by supersymmetry-preserving superpotential deformations of the $\mathcal{N}=1$ gauging of the flavor symmetry of a collection of $\mathcal{N}=2$ $\mathcal{D}_p(G)$ Argyres--Douglas SCFTs. In this work, we focus on the fixed points in the landscape of the $SU(3)$ gauging of three copies of the $\mathcal{D}_2(SU(3)) = H_2$ theory together with an adjoint-valued chiral multiplet. We catalogue the network of $a = c$ fixed points, and, along the way, we find a variety of dualities and instances of supersymmetry enhancement.

\vfill 
\end{titlepage}

\tableofcontents

\newpage

\section{Introduction}

Recently, it has been discovered \cite{Kang:2021lic,Kang:2021ccs} that there exists a vast collection of 4d $\mathcal{N}=1$ and $\mathcal{N}=2$ SCFTs where the two central charges are exactly equal, $a=c$, without taking any large $N$ limit. These theories are obtained via the diagonal gauging of the common flavor symmetry of a collection of Argyres--Douglas type theories. The resulting $a = c$ SCFTs have many novel properties, such as emergent symmetry enhancement from minimal to maximal supersymmetry \cite{Kang:2023dsa}, new vacuum moduli structure such as a mixed Higgs--Coulomb branch \cite{Kang:2022zsl}, quasi-modular structure \cite{Kang:2021lic}, and holographic supergravity duals exhibiting $R^2$ cancellations. We expect that this is sourced by $a=c$ being a characteristic that is typically a feature of theories that possess a larger supersymmetry; some of these properties have been explored in \cite{Kang:2021lic,Kang:2021ccs,Kang:2022zsl,Kang:2023dsa,Kang:2022vab,Buican:2020moo,Closset:2021lwy,Pan:2021mrw,Honda:2022hvy,Huang:2022bry,Hatsuda:2022xdv,Carta:2023bqn,Guo:2023mkn,Arakawa:2023cki,Du:2023kfu,Jiang:2024baj,Beccaria:2024szi,Kimura:2022yua,Kang:2024elv}.

In this paper, we initiate the study of an even more expansive landscape of $a = c$ theories, obtained via superpotential deformations by relevant operators of the SCFTs previously constructed via the gauging of Argyres--Douglas theories. 
Inspired from the analysis of SQCD with both fundamental and adjoint chiral multiplets \cite{Intriligator:2003mi}, where the different families of superpotential deformations are classified and their structure elucidated, we analyze the renormalization group (RG) flows triggered by superpotential deformations of SQCD with adjoint chiral multiplets and strongly-coupled Argyres--Douglas matter. Specifically, we begin with the $a=c$ SCFTs previously constructed, and consider relevant deformations that lead to interacting infrared SCFTs, in particular those that preserve the $a=c$ property.

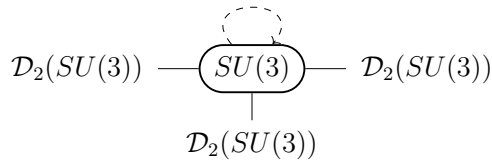
\begin{figure}[t]
    \centering
    \scalebox{0.90}{\begin{tikzpicture}
      \node[gaugeN1] (s0) {$SU(3)$};
      \node[d2] (c1) [left=0.6cm of s0] {$\mathcal{D}_{2}(SU(3))$};
      \node[d2] (c2) [right=0.6cm of s0] {$\mathcal{D}_{2}(SU(3))$};
      \node[d2] (c3) [below=0.4cm of s0] {$\mathcal{D}_{2}(SU(3))$};
      \draw (s0.east) -- (c2.west);
      \draw (s0.west) -- (c1.east);
      \draw (s0.south) -- (c3.north);
      \draw[dashed, ->] (s0) to[out=130, in=410, looseness=4] (s0);
    \end{tikzpicture}}
    \caption{An asymptotically-free $\mathcal{N}=1$ gauge theory via the diagonal gauging of the $SU(3)$ flavor symmetry of three copies of the $\mathcal{D}_2(SU(3))$ Argyres--Douglas theory, via $\mathcal{N}=1$ vector multiplets, with an adjoint-valued chiral multiplet, which is depicted as a dashed arrow line.}
    \label{fig:bigboy}
\end{figure}

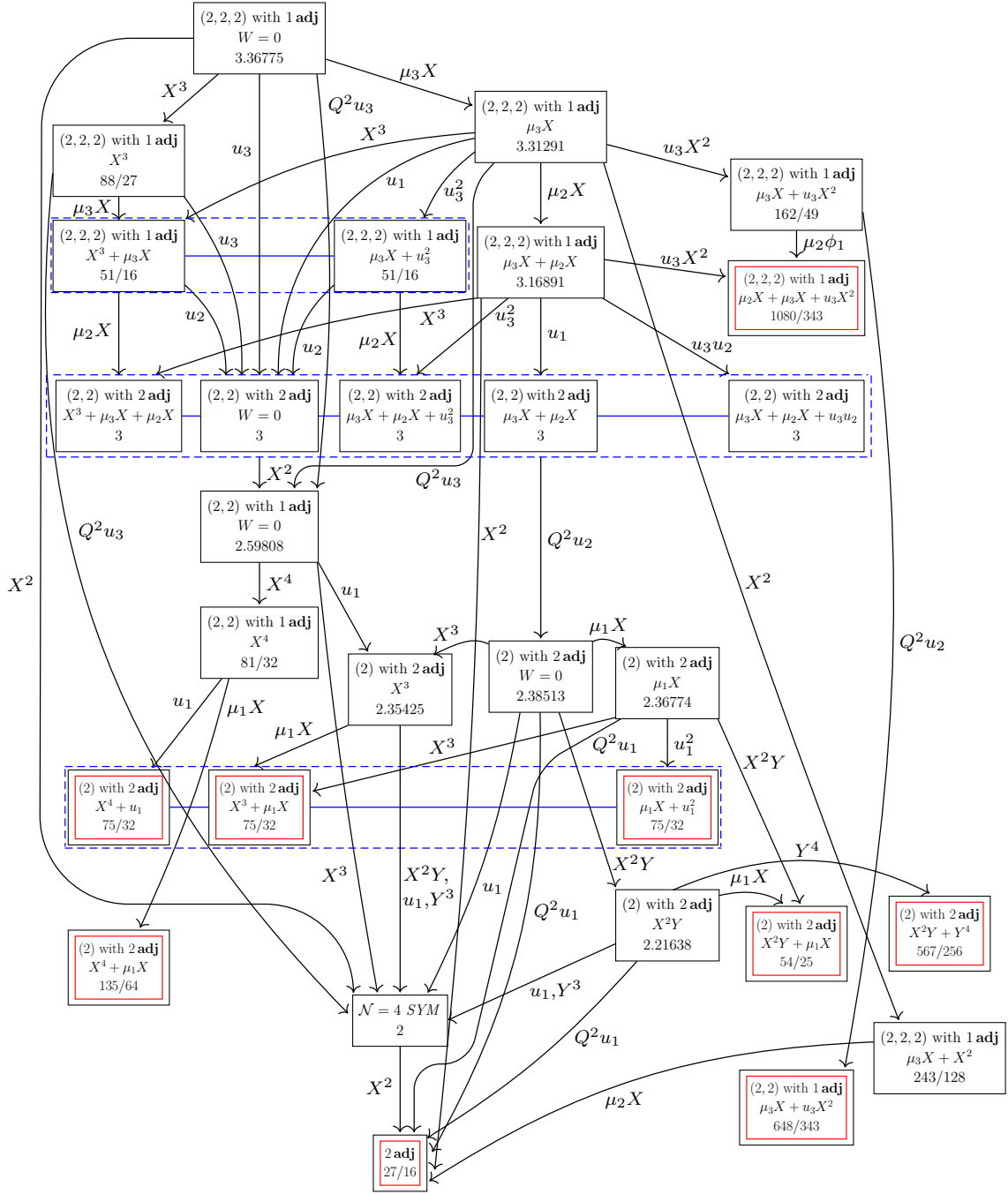
\begin{figure}[p]
\hspace{-37pt}
\begin{tikzcd}[cells={nodes={draw=black}},column sep=0.3em,row sep=1.6em,scale cd=0.54]
     & \begin{tabular}{@{}c@{}}
        $(2,2,2)$ with 1$\,\textbf{adj}$ \\
         $W=0$ \\
        $3.36775$
    \end{tabular}\arrow[phantom,bend right=68,"{X^2}"{font=\scriptsize}]{rddddddddddddddddddd}[]{}\arrow[to path={ -- ([xshift=-65pt]\tikztostart.west) -- ([xshift=-134pt,yshift=50pt]\tikztotarget.west) -| ([xshift=-20pt]\tikztotarget.north)},
    rounded corners=38pt,shorten >=1pt]{rddddddddddddddddddd}[swap]{X^2} \arrow[shorten >=2pt]{ldd}[yshift=-4pt,pos=.4,swap]{X^3}\arrow[shorten >=1pt]{rrd}[yshift=-3pt,pos=.45]{\mu_3 X}\arrow[shorten >=2pt]{ddddddd}[pos=.25,swap]{u_3}\arrow[shorten >=1pt,bend left=6,start anchor={[xshift=23pt]},end anchor={[xshift=23pt]}]{ddddddddd}[pos=.1]{Q^2u_3} & & & & \\[-1em]
     & & & \begin{tabular}{@{}c@{}}
        $(2,2,2)$ with 1$\,\textbf{adj}$ \\
        $\mu_3 X$ \\
        $3.31291$
    \end{tabular}\arrow[shorten >=1pt,bend right=2,start anchor={[xshift=21pt]},end anchor={[xshift=-10pt]}]{rrrddddddddddddddddddd}[xshift=-2pt,pos=.5]{X^2} \arrow[shorten >=1pt,bend right=12,start anchor={[yshift=0pt]}]{lllddd}[pos=.25,swap,yshift=-2pt]{X^3} \arrow[shorten >=2pt,bend right=40,start anchor={[yshift=-2pt]},end anchor={[xshift=7pt]}]{lldddddd}[yshift=3pt,pos=.3]{u_1} \arrow[shorten >=1pt,bend right=15,start anchor={[yshift=4pt]}]{lddd}[pos=.65,yshift=8pt]{u_3^2} \arrow[phantom,bend left=10,"{Q^2u_3}"{font=\scriptsize}]{llddddddddddddd}[]{} \arrow[to path={([xshift=-20pt]\tikztostart.south) -- ([xshift=-29pt,yshift=-10pt]\tikztostart.south) -- ([xshift=-29pt,yshift=-130pt]\tikztostart.south) -| ([xshift=15pt]\tikztotarget.north)}, rounded corners=8pt,shorten >=1pt]{lldddddddd}[swap]{Q^2u_3} \arrow[shorten >=1pt]{dddd}[pos=.45]{\mu_2X}\arrow[shorten >=1pt]{rrdd}[yshift=-2pt,pos=.4]{u_3X^2} & & & \\[-3em]
    \begin{tabular}{@{}c@{}}
        $(2,2,2)$ with 1$\,\textbf{adj}$ \\
        $X^3$ \\
        $88/27$
    \end{tabular}\arrow[shorten >=0pt,bend right=25,start anchor={[xshift=-26.5pt,yshift=10pt]},end anchor={[xshift=-12pt,yshift=-8pt]}]{rrddddddddddddddddd}[pos=.4]{Q^2u_3} \arrow[shorten >=0pt]{dd}[pos=.5,swap]{\mu_3 X}\arrow[shorten >=2pt,bend left=19,start anchor={[xshift=11pt]},end anchor={[xshift=-5pt]}]{rddddd}[xshift=-2pt,yshift=-2pt,pos=.3]{u_3} & & & & & & \\[-3em]
    |[draw=none]|\arrow[dash,dashed,blue,start anchor={[xshift=-32pt,yshift=-10pt]},end anchor={[xshift=93pt,yshift=-10pt]}]{r}\arrow[dash,dashed,blue,start anchor={[xshift=-32pt,yshift=-42pt]},end anchor={[xshift=93pt,yshift=-42pt]}]{r}\arrow[dash,dashed,blue,start anchor={[xshift=-29pt,yshift=-8pt]},end anchor={[xshift=-29pt,yshift=-30pt]}]{d}\arrow[dash,dashed,blue,start anchor={[xshift=150pt,yshift=-8pt]},end anchor={[xshift=150pt,yshift=-30pt]}]{d} & |[draw=none]| & & & & \begin{tabular}{@{}c@{}}
        $(2,2,2)$ with 1$\,\textbf{adj}$  \\
        $\mu_3X+u_3X^2$ \\
        $162/49$
    \end{tabular}\arrow[shorten >=1pt]{ddd}[pos=.4]{\mu_2\phi_1}\arrow[shorten >=1pt,bend left=9,start anchor={[xshift=26pt,yshift=8pt]},end anchor={[xshift=18pt]}]{dddddddddddddddddd}[pos=.5]{Q^2u_2} \\[-2em]
    \begin{tabular}{@{}c@{}}
        $(2,2,2)$ with 1$\,\textbf{adj}$ \\
        $X^3+\mu_3X$ \\
        $51/16$
    \end{tabular}\arrow[dash,blue]{rr}\arrow[shorten >=2pt]{ddd}[pos=.45,swap]{\mu_2X} \arrow[shorten >=2pt,bend left=25,end anchor={[xshift=-10pt]}]{rddd}[yshift=6pt,pos=.4,swap]{u_2}
    & & \begin{tabular}{@{}c@{}}
        $(2,2,2)$ with 1$\,\textbf{adj}$ \\
        $\mu_3X+u_3^2$ \\
        $51/16$
    \end{tabular}\arrow[shorten >=2pt,bend right=25,end anchor={[xshift=10pt]}]{lddd}[yshift=4pt,pos=.7]{u_2} \arrow[shorten >=2pt]{ddd}[pos=.55,swap,xshift=2pt]{\mu_2X} & & & & \\[-4em]
    & & & \begin{tabular}{@{}c@{}}
        $(2,2,2)\,$with$\,$1$\,\textbf{adj}$ \\
        $\mu_3X+\mu_2X$ \\
        $3.16891$
    \end{tabular}\arrow[shorten >=3pt,bend left=3,start anchor={[xshift=-24pt]},end anchor={[xshift=12pt,yshift=-18pt]}]{lddddddddddddddddd}[xshift=-2pt,pos=.25]{X^2} \arrow[yshift=5pt,shorten >=1pt]{rrd}[yshift=-3pt,pos=.4]{u_3X^2} \arrow[shorten >=4pt,end anchor={[xshift=-10pt]}]{ldd}[yshift=7pt,pos=.2]{u_3^2} \arrow[shorten >=2pt]{dd}[pos=.45]{u_1} \arrow[shorten >=5pt,xshift=1pt]{rrdd}[yshift=-5pt,pos=.6]{u_3u_2} \arrow[yshift=-12.5pt,shorten >=1pt,bend right=6,start anchor={[yshift=4pt]},end anchor={[xshift=-12pt,yshift=17pt]}]{llldd}[yshift=2pt,pos=.18]{X^3} & &  \\[-3em]
    |[draw=none]|\arrow[dash,dashed,blue,start anchor={[xshift=-31pt,yshift=-2 9pt]},end anchor={[xshift=-31pt]}]{dd}\arrow[dash,dashed,blue,start anchor={[xshift=-31pt,yshift=-33pt]},end anchor={[xshift=60pt,yshift=-33pt]}]{rrrrr} & & &  & & \redboxed{\begin{tabular}{@{}c@{}}
        $(2,2,2)$ with 1$\,\textbf{adj}$ \\
        $\mu_2X+\mu_3X+u_3X^2$ \\
        $1080/343$
    \end{tabular}}\arrow[dash,dashed,blue,start anchor={[xshift=32pt,yshift=-17pt]},end anchor={[xshift=32pt]}]{dd} \\
    \begin{tabular}{@{}c@{}}
        $(2,2)$ with 2$\,\textbf{adj}$ \\
        $X^3+\mu_3X+\mu_2X$ \\
        $3$
    \end{tabular}\arrow[dash,blue]{r}
    & \begin{tabular}{@{}c@{}}
        $(2,2)$ with 2$\,\textbf{adj}$ \\
        $W=0$ \\
        $3$
    \end{tabular}\arrow[dash,blue]{r}
    & \begin{tabular}{@{}c@{}}
        $(2,2)$ with 2$\,\textbf{adj}$ \\
        $\mu_3X+\mu_2X+u_3^2$ \\
        $3$
    \end{tabular}\arrow[dash,blue]{r}
    & \begin{tabular}{@{}c@{}}
        $(2,2)\,$with$\,$2$\,\textbf{adj}$ \\
        $\mu_3X+\mu_2X$ \\
        $3$
    \end{tabular}\arrow[dash,blue]{rr}
    & & \begin{tabular}{@{}c@{}}
        $(2,2)$ with 2$\,\textbf{adj}$ \\
        $\mu_3X+\mu_2X+u_3u_2$ \\
        $3$
    \end{tabular}
    \\[-1.4em]
    |[draw=none]|\arrow[dash,dashed,blue,start anchor={[xshift=-34pt,yshift=1pt]},end anchor={[xshift=35pt,yshift=1pt]}]{rrrrr} & |[draw=none]|\arrow[shorten >=1pt,start anchor={[yshift=5pt]}]{d}[pos=.45]{X^2} & |[draw=none]| & |[draw=none]|\arrow[shorten >=1pt,start anchor={[yshift=5pt]}]{ddd}[pos=.45]{Q^2u_2} & & |[draw=none]|\\[-.8em]
     & \begin{tabular}{@{}c@{}}
        $(2,2)$ with 1$\,\textbf{adj}$ \\
        $W=0$ \\
        $2.59808$
    \end{tabular}\arrow[shorten >=1pt]{d}[pos=.45]{X^4} \arrow[shorten >=1pt,start anchor={[xshift=12pt]}]{rdddd}[xshift=-3pt,pos=.45]{u_1} \arrow[shorten >=1pt,bend right=2,start anchor={[xshift=21pt]},end anchor={[xshift=-6pt]}]{rdddddddddd}[pos=.7,swap]{X^3} & & & & \\
     & \begin{tabular}{@{}c@{}}
        $(2,2)$ with 1$\,\textbf{adj}$ \\
        $X^4$ \\
        $81/32$
    \end{tabular}\arrow[shorten >=1pt,start anchor={[xshift=-3pt]}]{ldddd}[yshift=-4pt,pos=.3,swap]{u_1}\arrow[shorten >=1pt,bend left=5,start anchor={[xshift=-8pt]}]{ldddddddd}[xshift=-2pt,pos=.05]{\mu_1X} & & & & \\[-3em]
     & & & \begin{tabular}{@{}c@{}}
        $(2)$ with 2$\,\textbf{adj}$ \\
        $W=0$ \\
        $2.38513$
        \end{tabular} \arrow[shorten >=1pt,bend left=8,start anchor={[xshift=-5pt]},end anchor={[xshift=5pt]}]{ldddddddd}[yshift=3pt,pos=.6]{u_1} \arrow[shorten >=3pt,bend left=15,end anchor={[xshift=5pt,yshift=-10pt]}]{lddddddddddd}[yshift=3pt,pos=.4]{Q^2u_1} \arrow[yshift=6pt,shorten >=-9pt,bend right=25]{ldd}[yshift=-3pt,pos=.6,swap]{X^3}\arrow[yshift=6pt,shorten >=-5pt,bend left=25]{rd}[xshift=2pt,yshift=-1pt,pos=.5]{\mu_1X}\arrow[shorten >=1pt,end anchor={[xshift=-14pt]}]{rdddd}[xshift=-3pt,pos=.93]{X^2Y} & & \\[-4em]
     & & & & \begin{tabular}{@{}c@{}}
        $(2)$ with 2$\,\textbf{adj}$ \\
        $\mu_1X$ \\
        $2.36774$
    \end{tabular}\arrow[shorten >=1pt,start anchor={[yshift=-8pt]}]{llldd}[pos=.5,yshift=-2.5pt,swap]{X^3} \arrow[phantom,bend right=48,"{Q^2u_1}"{font=\scriptsize}]{dd}[]{} \arrow[to path={([xshift=-20pt]\tikztostart.south) -- ([xshift=-55pt,yshift=-20pt]\tikztostart.south) -- ([xshift=-85pt,yshift=-148pt]\tikztostart.south) -| ([xshift=6pt]\tikztotarget.north)}, rounded corners=11pt,shorten >=1pt]{lldddddddddd}[swap]{Q^2u_1} \arrow[shorten >=1pt]{dd}[pos=.5]{u_1^2}\arrow[shorten >=1pt,start anchor={[xshift=14pt]},end anchor={[xshift=10pt]}]{rddddd}[xshift=-3pt,pos=.3]{X^2Y} & \\[-4em]
    |[draw=none]|\arrow[dash,dashed,blue,start anchor={[xshift=-26pt,yshift=-33pt]},end anchor={[xshift=160pt,yshift=-33pt]}]{rr}\arrow[dash,dashed,blue,start anchor={[xshift=-26pt,yshift=-68pt]},end anchor={[xshift=160pt,yshift=-68pt]}]{rr}\arrow[dash,dashed,blue,start anchor={[xshift=-23pt,yshift=-30pt]},end anchor={[xshift=-23pt,yshift=-33pt]}]{d}\arrow[dash,dashed,blue,start anchor={[xshift=258pt,yshift=-30pt]},end anchor={[xshift=258pt,yshift=-33pt]}]{d} & & \begin{tabular}{@{}c@{}}
        $(2)$ with 2$\,\textbf{adj}$ \\
        $X^3$ \\
        $2.35425$
    \end{tabular}\arrow[shorten >=4pt,start anchor={[xshift=-4pt]},end anchor={[xshift=-25pt]}]{ld}[pos=.45,swap,xshift=8pt]{\mu_1X}\arrow[shorten >=1pt]{dddddd}[xshift=-4pt,pos=.6]{\substack{X^2Y,\\[1pt] \;u_1,Y^3}} & & & \\
    \redboxed{\begin{tabular}{@{}c@{}}
        $(2)$ with 2$\,\textbf{adj}$ \\
        $X^4+u_1$ \\
        $75/32$
    \end{tabular}}\arrow[dash,blue]{r} & \redboxed{\begin{tabular}{@{}c@{}}
        $(2)$ with 2$\,\textbf{adj}$ \\
        $X^3+\mu_1X$ \\
        $75/32$
    \end{tabular}}\arrow[dash,blue]{rrr} & & & \redboxed{\begin{tabular}{@{}c@{}}
        $(2)$ with 2$\,\textbf{adj}$ \\
        $\mu_1X+u_1^2$ \\
        $75/32$
    \end{tabular}} & \\
     & & & & \begin{tabular}{@{}c@{}}
        $(2)$ with 2$\,\textbf{adj}$ \\
        $X^2Y$ \\
        $2.21638$
    \end{tabular} \arrow[shorten >=5pt,,bend left=25,start anchor={[yshift=6pt]},end anchor={[xshift=20pt,yshift=0pt]}]{rdd}[xshift=-10pt,pos=.35]{\mu_1X} \arrow[shorten >=5pt,,bend left=25,start anchor={[xshift=-20pt,yshift=5.5pt]},end anchor={[xshift=22pt,yshift=4pt]}]{rrd}[yshift=-1pt,pos=.5]{Y^4} \arrow[shorten >=0pt,bend left=8,end anchor={[yshift=3pt]}]{llddddddd}[yshift=5pt,pos=.35]{Q^2u_1} \arrow[shorten >=0pt,end anchor={[yshift=-7pt]}]{lldddd}[yshift=5pt,pos=.55]{\substack{u_1,Y^3}} & \\[-4em]
     & & & & & & \redboxed{\begin{tabular}{@{}c@{}}
        $(2)$ with 2$\,\textbf{adj}$ \\
        $X^2Y+Y^4$ \\
        $567/256$
    \end{tabular}} \\[-4em]
     & & & & & \redboxed{\begin{tabular}{@{}c@{}}
        $(2)$ with 2$\,\textbf{adj}$ \\
        $X^2Y+\mu_1X$ \\
        $54/25$
    \end{tabular}} \\[-3.5em]
    \redboxed{\begin{tabular}{@{}c@{}}
        $(2)$ with 2$\,\textbf{adj}$ \\
        $X^4+\mu_1X$ \\
        $135/64$
    \end{tabular}} & & & & & \\[-2em]
     & & \begin{tabular}{@{}c@{}}
        $\mathcal{N}=4$ \textit{SYM} \\
        $2$
     \end{tabular}\arrow[shorten >=1pt]{ddd}[pos=.45,swap]{X^2} & & & \\[-2.5em]
     & & & & & & \begin{tabular}{@{}c@{}}
        $(2,2,2)$ with 1$\,\textbf{adj}$ \\
        $\mu_3X+X^2$ \\
        $243/128$
    \end{tabular}\arrow[shorten >=1pt,bend right=15,start anchor={[yshift=5pt]},end anchor={[yshift=-14pt]}]{lllldd}[yshift=3pt,pos=.6]{\mu_2X} \\[-2.5em]
     & & & & & \redboxed{\begin{tabular}{@{}c@{}}
        $(2,2)$ with 1$\,\textbf{adj}$ \\
        $
        \mu_3X+u_3X^2$ \\
        $648/343$
    \end{tabular}} \\[-2em]
     & & \redboxed{\begin{tabular}{@{}c@{}}
        2$\,\textbf{adj}$ \\
        $27/16$
    \end{tabular}} & & &
\end{tikzcd} 
\caption{The network of $a=c$ SCFT preserving superpotential deformations of the gauged Argyres--Douglas theory depicted in Figure \ref{fig:deformations}. In each box, we describe the fixed point (possibly in a dual frame) as well as giving the value of the coefficient of the central charges. We write $(2, \cdots, 2)$ to denote the number of gauged $\mathcal{D}_2(G)$s. See Section \ref{sec:222landscape} for the derivation of this landscape.}
\label{fig:deformations}
\end{figure}

In order to demonstrate the nature of such a landscape, we consider as a detailed example the theory obtained via the diagonal gauging of the $SU(3)$ flavor symmetry of three copies of the $\mathcal{D}_2(SU(3))$ Argyres--Douglas theory, together with an additional adjoint-valued chiral multiplet under the gauge group. This theory, which we depict in Figure \ref{fig:bigboy}, will be the main ``starting point'' or ``seed'' $a=c$ SCFT. We analyze the full network of relevant superpotential deformations of this theory, such that one ends up with an interacting SCFT in the infrared. One of the principle results is Figure \ref{fig:deformations}, which depicts the landscape of such superpotential deformations. Along the way, we find multiple different sequences of deformations that land on the (conformal manifold of the) same theory evincing interesting dualities.

An additional surprising result is that many of the $a = c$ theories can be deformed to flow to a point on the ($\mathcal{N}=1$)-preserving conformal manifold of $\mathcal{N}=4$ super-Yang--Mills (SYM). 
This provides many more examples of minimal to maximal supersymmetry enhancement, such as was discussed in our previous work \cite{Kang:2023dsa}. We also find other examples of supersymmetry enhancement, e.g., where the deformed $\mathcal{N}=1$ SCFT is a collection of free $\mathcal{N}=2$ vector multiplets. It is striking that these deformations of the $a = c$ SCFTs realize supersymmetry enhancements in intriguing ways; we will explore this general pattern in future work.


This paper provides a focused analysis of a particular theory, serving as a preview to the broader comprehensive study of the landscape of superpotential deformations for $\mathcal{N}=2$ and $\mathcal{N}=1$ $a=c$ SCFTs in the forthcoming works \cite{LANDSCAPEII} and \cite{LANDSCAPEIII}, building on the results of \cite{Kang:2021lic,Kang:2021ccs}.

The structure of this paper is as follows. First, in Section \ref{sec:acdefs}, we review the construction of $a=c$ SCFTs via the gauging of Argyres--Douglas theories, and explain when $c-a$ is preserved under a relevant deformation. In Section \ref{sec:ADdeformations}, we explain interesting classes of relevant deformations of $\mathcal{D}_p(G)$ theories, which can become relevant deformations also in the gauged theory. In Section \ref{sec:method}, we explain how to exhaustively explore the landscape of interacting SCFTs that can be obtained via superpotential deformations of an arbitrary SCFT obtained in the infrared after gauging $\mathcal{D}_p(G)$s together with adjoint-valued chiral multiplets. We then apply this algorithm to the theory in Figure \ref{fig:bigboy} and extract the landscape in Section \ref{sec:222landscape}. 
We have added \emph{unicode bookmarks for all 97 superpotential deformations} to aid the reader in the navigation of Section \ref{sec:222landscape}.
We summarize some of our findings/future directions in Section \ref{sec:discussion}.

\section{\texorpdfstring{$a=c$}{a=c} SCFTs and \texorpdfstring{$(a=c)$}{(a=c)}-preserving deformations}\label{sec:acdefs}

We first describe the setup that we consider in more detail. We begin with the Argyres--Douglas theories $\mathcal{D}_p(G)$, where $G$ is a simple and simply-laced Lie algebra and $p$ is an integer greater than one.\footnote{More accurately, $p=1$ is allowed but that corresponds to a trivial theory.} The $\mathcal{D}_p(G)$ theories admit a class $\mathcal{S}$ construction \cite{Gaiotto:2009we,Gaiotto:2009hg} with a regular puncture and an irregular puncture \cite{Cecotti:2012jx,Xie:2012hs,Cecotti:2013lda,Wang:2015mra}.
Some of the physical properties of such theories have recently been reviewed in \cite{Couzens:2023kyf}, and we refer there for notation and conventions. 

From a collection of $\mathcal{N}=2$ SCFTs $\mathcal{D}_{p_\alpha}(G)$, where $\alpha$ is some indexing integer, we construct an $\mathcal{N}=1$ theory by introducing an $\mathcal{N}=1$ vector multiplet gauging the $G$ flavor symmetry, possibly with additional adjoint-valued chiral multiplets under $G$, and trivial superpotential.\footnote{We sometimes refer to the $\mathcal{D}_p(G)$ factors as ``Argyres--Douglas matter''.} All possible solutions leading to $\mathcal{N}=1$ SCFTs must have their one-loop $\beta$-function to either vanish, which are listed in Table \ref{tbl:betais0} or be negative, as the latter type flow to an interacting SCFT, if they satisfy the certain conditions, and they are listed in Table \ref{tbl:negativebeta}.

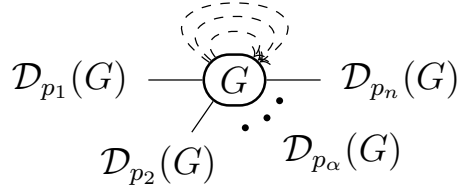
\begin{figure}[t]
    \centering
    \scalebox{1.2}{\begin{tikzpicture}
      \node[gaugeN1] (s0) {$G$};
      \node[d2] (c1) [left=0.6cm of s0] {$\mathcal{D}_{p_1}(G)$};
      \node[d2] (c2) [right=0.6cm of s0] {$\mathcal{D}_{p_n}(G)$};
      \node[d2] (c4) [below left=0.4cm of s0] {$\mathcal{D}_{p_2}(G)$};
      \node (c5) at (1.2,-.8) {$\mathcal{D}_{p_\alpha}(G)$};
      \node (cd1) at (.12,-.55) {\LARGE $\cdot$};
      \node (cd2) at (.36,-.44) {\LARGE $\cdot$};
      \node (cd3) at (.5,-.25) {\LARGE $\cdot$};
      \draw (s0.east) -- (c2.west);
      \draw (s0.west) -- (c1.east);
      \draw (s0.225) -- (c4.45);
      \draw[dashed, ->] (s0) to[out=130, in=410, looseness=3] (s0);
      \draw[dashed, ->] (s0) to[out=140, in=400, looseness=5] (s0);
      \draw[dashed, ->] (s0) to[out=145, in=395, looseness=7] (s0);
    \end{tikzpicture}}
    \caption{We schematically depict our construction of 4d $\mathcal{N}=1$ gauge theories via a collection of $n$ Argyres--Douglas theories $\mathcal{D}_{p_\alpha}$ that are diagonally gauged via an $\mathcal{N}=1$ vector multiplet. We can also include up to three adjoint-valued chiral multiplets.}
    \label{fig:gaugingDpG}
\end{figure}
\begin{table}[ht]
    \centering
    \begin{threeparttable}
\centering
\renewcommand{\arraystretch}{1.2}
    $\begin{array}{cccccc}
    \toprule
      p_1 & p_2 & p_3 & p_4 & p_5 & p_6 \\\midrule
      1 & 1 & 2 & 3 & 7 & 42 \\
      1 & 1 & 2 & 3 & 8 & 24 \\
      1 & 1 & 2 & 3 & 9 & 18 \\
      1 & 1 & 2 & 3 & 10 & 15 \\
      1 & 1 & 2 & 3 & 12 & 12 \\
      1 & 1 & 2 & 4 & 5 & 20
      \\\bottomrule
    \end{array}$
\quad
    $\begin{array}{cccccc}
    \toprule
      p_1 & p_2 & p_3 & p_4 & p_5 & p_6 \\\midrule
      1 & 1 & 2 & 4 & 6 & 12 \\
      1 & 1 & 2 & 4 & 8 & 8 \\
      1 & 1 & 2 & 5 & 5 & 10 \\
      1 & 1 & 2 & 6 & 6 & 6 \\
      1 & 1 & 3 & 3 & 4 & 12 \\
      1 & 1 & 3 & 3 & 6 & 6 
      \\\bottomrule
    \end{array}$
\quad
    $\begin{array}{cccccc}
    \toprule
      p_1 & p_2 & p_3 & p_4 & p_5 & p_6 \\\midrule
      1 & 1 & 3 & 4 & 4 & 6 \\
      1 & 1 & 4 & 4 & 4 & 4 \\
      1 & 2 & 2 & 3 & 3 & 3 \\
      1 & 2 & 2 & 2 & 4 & 4 \\
      1 & 2 & 2 & 2 & 3 & 6 \\
      2 & 2 & 2 & 2 & 2 & 2
    \\\bottomrule
    \end{array}$
\end{threeparttable}
    \caption{The collections of $p_\alpha$ such that diagonal gauging yields an $\mathcal{N}=1$ gauge theory where the one-loop $\beta$-function of the gauge coupling vanishes.}
    \label{tbl:betais0}
    \vspace{0.5cm}
\begin{threeparttable}
\centering
\renewcommand{\arraystretch}{1.2}
    $\begin{array}{ccccc}
    \toprule
      p_1 & p_2 & p_3 & p_4 & p_5 \\\midrule
      1 & 1 & 1 & 1 & p_5 \\
      1 & 1 & 1 & p_4 & p_5 \\
      1 & 1 & p_3 & p_4 & p_5 \\
      1 & 2 & 2 & p_4 & p_5 \\
      1 & 2 & 3 & \leq 6 & p_5 \\
      1 & 2 & 3 & 7 & \leq 41 \\
      1 & 2 & 3 & 8 & \leq 23 \\
      1 & 2 & 3 & 9 & \leq 17 
      \\\bottomrule
    \end{array}$
\quad
    $\begin{array}{ccccc}
    \toprule
      p_1 & p_2 & p_3 & p_4 & p_5 \\\midrule
      1 & 2 & 3 & 10 & \leq 14 \\
      1 & 2 & 3 & 11 & \leq 13\\
      1 & 2 & 4 & 4 & p_5 \\
      1 & 2 & 4 & 5 & \leq 19 \\
      1 & 2 & 4 & 6 & \leq 11 \\
      1 & 2 & 4 & 7 & \leq 9 \\
      1 & 2 & 5 & 5 & \leq 9 \\
      1 & 2 & 5 & 6 & \leq 7
      \\\bottomrule
    \end{array}$
\quad
    $\begin{array}{ccccc}
    \toprule
      p_1 & p_2 & p_3 & p_4 & p_5 \\\midrule
      1 & 3 & 3 & 3 & p_4 \\
      1 & 3 & 3 & 4 & \leq 11 \\
      1 & 3 & 3 & 5 & \leq 7 \\
      1 & 3 & 4 & 4 & \leq 5 \\
      2 & 2 & 2 & 2 & p_5 \\
      2 & 2 & 2 & 3 & 3 \\
      2 & 2 & 2 & 3 & 4 \\
      2 & 2 & 2 & 3 & 5
    \\\bottomrule
    \end{array}$
\end{threeparttable}
    \caption{All possible tuples of $p_\alpha$ such that $\mathcal{N}=1$ gauging of the common flavor symmetry of the associated $\mathcal{D}_{p_\alpha}(G)$ leads to a gauge theory with an asymptotically-free gauge coupling. An entry that is left as $p_\alpha$ indicates that the theory will be asymptotically free for any positive integer $p_\alpha$. We assume throughout that $p_\alpha \leq p_{\alpha+1}$.}
    \label{tbl:negativebeta}
\end{table}
\clearpage

The conditions under which such an $\mathcal{N}=1$ gauge theory flows in the infrared to an interacting SCFT with identical central charges are enumerated in \cite{Kang:2021lic,Kang:2021ccs}; a necessary, but not sufficient condition, is as follows:
\begin{equation}\label{eqn:GCDCOND}
    a = c \qquad \Leftrightarrow \qquad \prod_\alpha \gcd(p_\alpha, h_G^\vee) = 1 \,.
\end{equation}
The identical condition holds when the one-loop $\beta$-function vanishes and the gauge theory is directly conformal.
The condition in equation \eqref{eqn:GCDCOND} is equivalent to requiring that none of the Argyres--Douglas matter has any Coulomb branch operators with integer scaling dimension. 

We can also construct $\mathcal{N}=2$ theories by gauging the $G$ flavor symmetry as well, that is, if a collection of $\mathcal{D}_{p_\alpha}$ theories are gauged via an $\mathcal{N}=2$ vector multiplet. All possible solutions leading to $\mathcal{N}=2$ SCFTs, that have vanishing one-loop $\beta$-functions, are simply
\begin{align}
    (p_1, p_2, p_3, p_4) = (2, 2, 2, 2), (1, 3, 3, 3), (1, 2, 4, 4), (1, 2, 3, 6) \,.
\end{align}
The same condition as in equation \eqref{eqn:GCDCOND} is necessary for the theory thus constructed to be an SCFT with $a=c$. These $\mathcal{N}=2$ SCFTs with $a=c$ have surprising and interesting properties, demonstrating a deep connection to $\mathcal{N}=4$ super-Yang--Mills; some of these features were highlighted in \cite{Kang:2021lic}.

When the gauge coupling is asymptotically-free, equation \eqref{eqn:GCDCOND} is not sufficient and we have to make sure that the theory is actually interacting in the infrared. A necessary condition is to have an anomaly-free superconformal $U(1)$ R-symmetry. Typically there are multiple $U(1)$ symmetries, and the superconformal R-symmetry is determined via the $a$-maximization procedure \cite{Intriligator:2003jj}. We first need to find a real solution to $a$-maximization, and also check all the gauge-invariant operators, $\mathcal{O}$, satisfy the unitarity constraint $R[\mathcal{O}]>2/3$. If not, operators decouple along the RG flow and become free \cite{Kutasov:2003iy} and this renders the (interacting part of) IR fixed point to have $a \neq c$. Even if all the operators in the chiral ring satisfy the unitarity constraint, sometimes a more refined check of unitarity is necessary. This can be done by computing the superconformal index \cite{Evtikhiev:2017heo, Maruyoshi:2018nod, Cho:2024civ}. For low-rank $G$, this refined test of the unitarity of $a=c$ theories has been carried out in \cite{Kang:2022vab} by explicitly computing the superconformal index.

Coupling an $\CN=1$ vector multiplet with $\CD_{p_\alpha}(G)$ theories and adjoint-valued chiral multiplets by diagonal gauging explicitly breaks the $\CN=2$ supersymmetry. After gauging, each $\CD_{p_\alpha}(G)$ theory contains one classical $U(1)$ flavor symmetry with generator
\begin{align}
    F_\alpha=2I_{3,\alpha}-r_{\alpha} \,,
\end{align}
where $I_{3,\alpha}$ and $r_{\alpha}$ are the Cartan generator of the $SU(2)_R$ symmetry and the generator of the $U(1)_R$ symmetry of each $\CD_{p_\alpha}(G)$ theory, respectively. Also, there exist $U(1)$ symmetries $f_\nu$ rotating each $\nu$th chiral multiplet. In total, there are $m+n$ classical $U(1)$ flavor symmetries when $m$ $\CD_{p_\alpha}(G)$ theories and $n$ adjoint-valued chiral multiplets are glued by diagonal gauging; one of them is broken due to an ABJ anomaly. The anomaly-free $m+n-1$ $U(1)$ symmetries are captured by the following generators:
\begin{align}
  \mathcal{F}_l=\sum_{\alpha=1}^{m}c_{l,\alpha} F_\alpha+\sum_{\nu=1}^n c'_{l,\nu} f_\nu \,.   
\end{align}
These remaining $U(1)$s have no ABJ anomaly, and thus they must satisfy the following condition:
\begin{align}
\begin{split}
    \trace \mathcal{F}_lGG =& 
    \sum_{\alpha=1}^mc_{l,\alpha}\left( 2 \trace I_{3,\alpha}GG - \trace r_\alpha GG \right) + \sum_{\nu=1}^nc_{l,\nu}h^{\vee}\\
    =& \left(\sum_{\alpha=1}^m\frac{(p_\alpha-1)}{p_\alpha}c_{l,\alpha}+\sum_{\nu=1}^nc'_{l,\nu}f_l\right)h^\vee_G =0 \,,
\end{split}
\end{align}
for each $l$. Here, we used the known anomaly coefficients of the $\CD_{p}(G)$ building blocks:
\begin{align}
    \trace r GG=-\frac{1}{2}k_G=-\frac{p-1}{p}h^\vee_G\,,\quad \trace I_{3 }GG=0 \,.
\end{align}
The generator of the $\CN=1$ $U(1)_R$ R-symmetry in the UV is 
\begin{align}
    R_0=\sum_{\alpha=1}^m\left(\frac{1}{3}r_\alpha+\frac{4}{3}I_{3,\alpha}\right)+\sum_{\nu=1}^n\frac{2}{3}f_\nu \,,
\end{align}
where the coefficient $2/3$ of $f_\nu$ is the R-charge of the free chiral multiplet. When the gauging is conformal, $R_0$ is the $\CN=1$ superconformal R-charge. On the other hand, when the gauge coupling runs, the flavor symmetries (with generators $\mathcal{F}_l$) mix with $R_0$ to form the superconformal R-symmetry at the IR fixed point
\begin{align}
\begin{split}
    R
    =&\,R_0+\sum_{\alpha=1}^m\varepsilon_\alpha F_\alpha+\sum_{\nu=1}^n\left(r_\nu-\frac{2}{3}\right)f_\nu \\ 
    =& \sum_{\alpha=1}^m \left[ \left( \frac{1}{3}-\varepsilon_\alpha \right) r_\alpha + \left( \frac{4}{3} + 2 \varepsilon_\alpha \right) I_{3, \alpha} \right] + \sum_{\nu=1}^n r_\nu f_\nu  \,.
\end{split}
\label{eq:mix}
\end{align}
The values of mixing coefficients $\varepsilon_\alpha$ and the infrared R-charge of the chiral multiplets $r_\nu$ are determined by anomaly cancellation condition $\trace R GG = 0$ and $a$-maximization \cite{Intriligator:2003jj} in such a way that they maximize the trial $a$ central charge
\begin{align}
    a(\varepsilon_\alpha,r_\nu)=\frac{3}{32}\left( 3\trace R(\varepsilon_\alpha,r_\nu)^3-\trace R(\varepsilon_\alpha,r_\nu)\right) \,.
\end{align}

An important observation in \cite{Kang:2021ccs} is that, when the condition in equation \eqref{eqn:GCDCOND} for the gauging is satisfied, the trace of $R$ is proportional to its ABJ anomaly $\trace RGG$,
\begin{align}
    \trace R(\varepsilon_\alpha,r_\nu)=\frac{\dim (G)}{h^\vee_G}\trace R(\varepsilon_\alpha,r_\nu)GG\,,
\label{eq:RtoRGG}
\end{align}
regardless of the actual values of the mixing coefficients $\varepsilon_\alpha$ and the R-charges of the chiral multiplets $r_\nu$. Now, consider that the theory is deformed in the infrared by some relevant superpotential $W$. Using the holomorphic renormalization scheme, $W$ is preserved along the RG-flow and should satisfy
\begin{align}
    R_{\text{IR}}[W]=2\,,\qquad\mathcal{F}_{\text{IR},l}[W]=0\,,    
    \label{eq:IRcond}
\end{align}
at the IR fixed point with respect to the IR R-charge and flavor charges. The second condition typically reduces the number of $U(1)$ flavor symmetries by one for each relevant operator appearing in $W$. Effectively, the superpotential puts extra constraints on the $a$-maximization procedure, thereby changing the maximizing values of $\varepsilon_l$ and $r_\nu$ at the fixed point. Nevertheless, the structure in equation \eqref{eq:RtoRGG} is untouched unless some symmetries emerge along the RG flow and is satisfied at every possible IR fixed point of the RG-flow triggered by any relevant superpotential deformation with its specific mixing coefficients $\varepsilon_l$. This implies that any IR SCFT obtained by such a superpotential deformation also has equal central charges, since any four dimensional SCFT satisfies the condition
\begin{align}
    a-c=16\,\trace R\,,\quad \trace RGG=0 \,.
\end{align}

Here, we have sketched the argument that asymptotically-free $\mathcal{N}=1$ gauging of Argyres--Douglas matter leads to infrared SCFTs with identical central charges, regardless of any subsequent infrared superpotential deformation. This is subject to the condition that there are no emergent symmetries in the flow into the infrared after the deformation, or else that there are no operators that decouple along the flow. In this paper, we focus on the specific SCFT realized as the infrared of the gauge theory in Figure \ref{fig:bigboy}. This example exemplifies the analysis of relevant deformations, and highlights interesting features like supersymmetry enhancement. Also, the choice of a low-rank gauge algebra enables the use of the superconformal index. This allows for a more detailed investigation of the possible deformations from the Argyres--Douglas matter, going beyond the chiral ring. And more importantly, we observe in Figure \ref{fig:deformations} that most of the flows from the theory in Figure \ref{fig:bigboy} eventually flow down to the maximally supersymmetric theory: $\mathcal{N}=4$ SYM. While this is quite surprising, this illustrates how $a=c$ theories have an exciting commonality with $\mathcal{N}=4$ SYM. 

\section{Deformations of Argyres--Douglas theories}\label{sec:ADdeformations}

In this section, we consider two classes of deformations for an individual $\mathcal{D}_p(G)$ theory with $\gcd(p,h^{\vee}_G)=1$. These theories constitute the basic building blocks for the landscape of $a=c$ theories we construct. 
In each $\mathcal{D}_p(G)$ theory, there are three types of scalar operators preserving $\CN=1$ supersymmetry that we can use to trigger relevant deformations:
\begin{itemize}
  \item Coulomb branch operators $u_i$, with $(r,I_3)=(2\Delta_i,0)$, where $i=1,\,\cdots,\,\rank (\mathcal{D}_p(G))$,
  \item scalar superpartners of the $u_i$, $Q^2 u_i$, with $(r,I_3)=(2\Delta_i-2,1)$,
  \item a moment map operator, $\mu$, with $(r,I_3)=(0,1)$, 
\end{itemize}
where $r$ and $I_3$ denote the $\CN=2$ $U(1)$ and $SU(2)$ R-symmetries, respectively, and $\Delta_i$ denotes the scaling dimension of $u_i$.\footnote{We use the normalization for the $\CN=2$ $U(1)$ R-symmetry such that a Coulomb branch operator satisfies the condition $r=2\Delta$.}
Note that the first two sets of operators are singlets under the flavor symmetry $G$, while the last one is in the adjoint representation of $G$, since the moment map operator is the superconformal primary operator of the current multiplet.
The spectrum, i.e., the scaling dimensions, of the Coulomb branch operators is given as
\begin{align}\label{eqn:CBdims}
\mathcal{C}(p,G)=\left\{j-\left.\frac{h^{\vee}_G}{p}s\,\right|\,j-\frac{h^{\vee}_G}{p}s>1,\,j\in\text{Cas}(G),\,s=1,\cdots,p-1\right\}\,,
\end{align}
where the $\text{Cas}(G)$ denotes the degrees of the Casimir invariants of $G$. 

\subsection{Deformation via Coulomb branch operators}

Now, let us consider deforming the $\mathcal{D}_p(G)$ theory by turning on the $\mathcal{N}=1$ superpotential 
\begin{align}
    W=u_i \,, 
\end{align} 
whose scaling dimension is less than three. 
Such deformations of Argyres--Douglas theories are considered in \cite{Bolognesi:2015wta, Xie:2019aft,Xie:2021omd,Kang:2023dsa, Maruyoshi:2023mnv}. 
It breaks the $\mathcal{N}=2$ supersymmetry to $\mathcal{N}=1$ and triggers a RG-flow to the IR fixed point. At the IR fixed point, the $\mathcal{N}=1$ $U(1)$ R-charge is obtained by a linear combination of $r$ and $I_3$,
\begin{align}
R_{\text{IR}}=R_{\text{UV}}+\varepsilon \mathcal{F}=\left(\frac{1}{3}-\varepsilon\right)r+\left(\frac{4}{3}+2\varepsilon\right)I_3\,,
\end{align}
subject to the condition that the superpotential at the IR fixed point has R-charge two: 
\begin{align}
R_{\text{IR}}[u_i]=\left(\frac{1}{3}-\varepsilon\right)2\Delta_i=2\,.
\label{eq:uconst}
\end{align}
Here, $R_{\text{UV}}$ and $\mathcal{F}$ are $\mathcal{N}=1$ R-symmetry and flavor symmetry embedded in the $SU(2)\times U(1)$ $\mathcal{N}=2$ R-symmetry such that
\begin{align}
    R_{\text{UV}}=\frac{1}{3}r+\frac{4}{3}I_3\,,\qquad
    \mathcal{F}=2I_3-r \,.
\end{align}
Equation \eqref{eq:uconst} fixes the mixing parameter $\varepsilon$ and the infrared R-symmetry as
\begin{align} \label{eq:DpGdeformR}
\varepsilon=\frac{1}{3}-\frac{1}{\Delta_i}\,,\qquad R_{\text{IR}}=\frac{1}{\Delta_i}r+\frac{2(\Delta_i-1)}{\Delta_i}I_3\,.
\end{align}
Assuming the theory deformed by $u_i$ flows to a SCFT, we can plug-in the R-charge as above to the following formula 
\cite{Anselmi:1997am}
\begin{align}
    a=\frac{3}{32}(3\trace R_{\text{IR}}^3-\trace R_{\text{IR}})\,,\qquad c=\frac{1}{32}(9\trace R_{\text{IR}}^3-5\trace R_{\text{IR}})\, ,
    \label{eq:ac}
\end{align}
to obtain the central charge of the infrared SCFT. 

However, the situation is not so simple in general, since this deformation quite often results in unitarity violating operators. When this happens, it usually means that these operators get decoupled along the RG flow so that we have accidental $U(1)$ symmetries that can alter the superconformal $U(1)$ R-symmetry. In this case, in order to correctly identify the IR theory, we have to remove the would-be-decoupled operators and redo the $a$-maximization, until we do not find unitarity violating operators, as discussed in \cite{Kutasov:2003iy}. 

In addition to the linear deformations by Coulomb branch operators, we can also consider quadratic deformations of the form 
\begin{align}
    W = u_i u_j \qquad \text{ such that } \qquad  \Delta[u_i] + \Delta[u_j] < 3 \,.
\end{align}
Such deformations can be analyzed straightforwardly by simply replacing $\Delta_i$ with $\Delta_i + \Delta_j$ in equation \eqref{eq:DpGdeformR}. 
For example, $u_1^2$ is always relevant for $p>2$. Such quadratic deformations typically give rise to ``good'' $\CN=1$ SCFTs with smaller central charges and without any decoupled operators \cite{Buican:2016hnq, Xie:2016hny, Maruyoshi:2018nod}. For the $W=u_1^2$ deformation, we simply need to plug in $2(p+1)/p$ to $\Delta_i$ in equation \eqref{eq:DpGdeformR}.

\paragraph{\texorpdfstring{$W=u_1$ deformation}{W = u₁}}
We can always deform the theory by the Coulomb branch operator $u_1$ with the smallest scaling dimension $\Delta_1= (p+1)/p$.\footnote{We emphasize once more that we are assuming that $\gcd(p, h_G^\vee) = 1$.} It turns out this triggers a rather radical deformation of the theory since $u_1$ has a very small conformal dimension compared to the other operators of the SCFT. For this specific case, assuming we have a SCFT at the end of the flow, the infrared R-charge is fixed by $p$ as
\begin{align}
  R_{\text{IR}} =\frac{p}{p+1}r+\frac{2}{p+1}I_3\,.
\label{eq:Ru}
\end{align}
Upon flow, the three kinds of operators we considered above have scaling dimensions as
\begin{itemize}
  \item Coulomb branch operators $u_i$: $\Delta_{\text{IR}}[u_i]=\dfrac{3}{2}R_{\text{IR}}[u_i]=\dfrac{3p}{p+1}\Delta_i$,
  \item the superpartner of $u_i$: $\Delta_{\text{IR}}[Q^2u_i]=\dfrac{3}{2}R_{\text{IR}}[Q^2u_i]=\dfrac{3(p\Delta_i+1-p)}{p+1}$,
  \item the moment map $\mu$: $\Delta_{\text{IR}}[\mu]=\dfrac{3}{2}R_{\text{IR}}[\mu]=\dfrac{3}{p+1}$.
\end{itemize}
The Coulomb branch operators after the flow are always irrelevant as 
\begin{align}
    \Delta_{i\neq 1}>\frac{p+1}{p} .
\end{align}
However, some of the $Q^2u_i$ decouple as they hit the unitarity bound $\Delta_{\text{IR}}[Q^2u_i]\geq 1$ when $p\geq 5$, because the smallest value of the scaling dimension is $\frac{6}{p+1}$. Moreover, the moment map operators always decouple for the same reason. Since the moment map operators have the smallest R-charges, we first remove the moment map operators from the spectrum by introducing the flip field $M_{\mu}$ coupled via superpotential $W=\mu M_{\mu}$.\footnote{If there are multiple operators hitting the unitarity bound, one might ask which operators to decouple. Our prescription is to decouple the operator with the smallest dimension first, and then re-analyze other operators. This procedure is motivated by considering the $a$-theorem. See \cite{Agarwal:2020pol} for a detailed discussion.} After flipping the moment map $\mu$, the central charge for the remainder of the theory (after removal of the decoupled operator) can be obtained from the triangle anomalies using equation \eqref{eq:ac}. 
Substituting equation \eqref{eq:Ru} into equation \eqref{eq:ac}, including the contribution from the flip operator $M_{\mu}$, the central charge for the non-decoupled infrared sector of the theory is 
\begin{align} \label{eq:aDpG}
\begin{split}
a=\frac{3}{32}\Bigg(&\frac{3}{(p+1)^3}(p^3\text{Tr}\,r^3+6p^2\text{Tr}\,r^2I_3+12p\text{Tr}\,rI_3^2+8\text{Tr}\,I_3^3)-\frac{1}{p+1}(p\text{Tr}\,r+2\text{Tr}\,I_3)\\
&+\dim (G)\left(3\left(\frac{2p}{p+1}-1\right)^3-\left(\frac{2p}{p+1}-1\right)\right)\Bigg)\,.
\end{split}
\end{align}
The first line is the contribution from the $\mathcal{D}_p(G)$ theory and the second line is the contribution from $M_{\mu}$ that effectively subtracts the contribution of decoupled $\mu$. Let us remind the reader of the relation between the central charges of $\mathcal{D}_p(G)$ theories satisfying $\gcd(p,h^{\vee}_G)=1$, and the non-vanishing triangle anomaly coefficients \cite{Cecotti:2013lda,Kuzenko:1999pi}:
\begin{equation}
\begin{gathered}
a(p, G)=\frac{1}{48}\frac{(4p-1)(p-1)}{p}\dim (G),\,\quad c(p, G)=\frac{1}{12}(p-1)\dim (G) \,,\\
\text{Tr}\,r^3=\text{Tr}\,r=48(a(p,G)-c(p,G)) \,,\quad \text{Tr}\,rI_3^2=4a(p,G)-2c(p,G) \,. 
\label{eq:anom_coeff}
\end{gathered}
\end{equation} 
Entering the relations in equation \eqref{eq:anom_coeff} into equation \eqref{eq:aDpG}, we obtain the central charge for the non-decoupled sector in the infrared as
\begin{align}
\begin{split}
a=&\,\frac{3}{32}\Bigg(\frac{3}{(p+1)^3}\left(p^3\left(-\frac{p-1}{p}\right)+12p\left(\frac{(2p-1)(p-1)}{12p}\right)\right)-\frac{p}{p+1}\left(-\frac{p-1}{p}\right)\\
&\quad\quad+3\frac{(p-1)^3}{(p+1)^3}-\frac{p-1}{p+1}\Bigg)\dim (G)
\,\,=\,\,\,0\,.
\end{split}
\label{eq:aurest}
\end{align}
A similar computation yields that $c=0$ for the non-decoupled sector. Hence, the IR theory upon decoupling the moment map operator is empty. Therefore, the $\mathcal{D}_p(G)$ theory flows to $\dim (G)$ free chiral multiplets after it is deformed by $W=u_1$.

The situation is slightly modified when $G$ is gauged. The moment map operators are no longer gauge-invariant, so they are not decoupled. Rather, the $\mathcal{D}_p(G)$ theory behaves as a chiral multiplet $\phi$ coupled to the gauge field in the adjoint representation of $G$ as studied in \cite{Maruyoshi:2023mnv}, whose R-charge $R_\phi=\frac{2}{p+1}$ is fixed by the superpotential $W=\trace \phi^{p+1}$. 
One way to see this is by computing the central charges for the infrared theory by plugging in equation \eqref{eq:Ru} to equation \eqref{eq:ac}. We find that it agrees with that of the free chiral multiplet with R-charge $\frac{2}{p+1}$.
Alternatively, one can see that the Schur index after the deformation by $u_1$ agrees with that of the adjoint chiral $\phi$ with $W = \trace \phi^{p+1}$.\footnote{The Schur index is only well-defined in $\CN=2$ SCFT \cite{Gadde:2011ik}. However, for a certain $\CN=1$ deformation of $\CN=2$ SCFT, one can take a limit of $\CN=1$ superconformal index to agree with that of the Schur index of the \emph{undeformed} $\CN=2$ theory \cite{Buican:2016hnq}.} Such deformations have also been associated with monopole deformations in three-dimensional quiver gauge theories which exhibit deconfinement \cite{Bajeot:2023gyl,Benvenuti:2024glr,Hwang:2024hhy,CLM}.

\paragraph{\texorpdfstring{$W = Q^2 u_i$}{Q²uᵢ} deformation}
We can also deform $\mathcal{D}_p(G)$ by turning on a superpotential proportional to the $\mathcal{N}=2$ descendants of the Coulomb branch operators: 
\begin{equation}
    W=Q^2 u_i \,.
\end{equation}
Once again, assuming the theory flows to an SCFT, the mixing parameter $\varepsilon$ and the IR R-charge are fixed by the superpotential condition $R[Q^2u_i]=2$ as
\begin{align}
\varepsilon=\frac{1}{3}\,,\qquad R=2I_3\,.
\end{align}
This renders all the Coulomb branch operators $u_i$ to decouple since they are all have $R[u_i]=0$. 
Upon removing (or flipping) these operators, thee central charges for the remaining part are easily obtained from equation \eqref{eq:ac}. By including the contribution from the flip fields $M_{u_i}$ which has $R[M_{u_i}]=2$, and using the fact that $\trace I_3=\trace I_3^3=0$, we get 
\begin{align}
\begin{split}
    a=\frac{3}{32}\left( 24\trace I_3^3-2\trace I_3+\rank (\mathcal{D}_p(G)) (3(2-1)^3-(2-1))\right)
    =\frac{3}{16}\rank (\mathcal{D}_p(G))\,,\\
    c=\frac{1}{32}\left( 72\trace I_3^3-5\trace I_3+\rank (\mathcal{D}_p(G))(9(2-1)^3-5(2-1))\right)
    =\frac{1}{8}\rank (\mathcal{D}_p(G))\,.
\end{split}
\end{align}
This is exactly identical to the central charges of $\rank (\mathcal{D}_p(G))$ copies of the $\mathcal{N}=1$ vector multiplet. 
Together with the decoupled $u_i$, the $\mathcal{D}_p(G)$ theory deformed by $W=Q^2u_i$, flows to $\rank (\mathcal{D}_p(G))$ copies of the $\mathcal{N}=2$ free vector multiplet. When $G$ is gauged, the $\mathcal{D}_p(G)$ theory decouples from the rest of the system since the $\mathcal{N}=2$ vectors are all neutral under $G$.

\subsection{Summary of deformations in \texorpdfstring{$\mathcal{D}_p(G)$}{DpG}}

We considered two kinds of deformations that are available in any $\mathcal{D}_p(G)$ theories (with $\gcd(p, h_G^\vee) = 1$) and we identify the endpoints of the flows as:
\begin{enumerate}
  \item $W=u_1$: a $\mathcal{D}_p(G)$ theory flows to $\dim (G)$ free chiral multiplets. When $G$ is gauged, $\mathcal{D}_p(G)$ flows to a chiral multiplet $\phi$ in the adjoint representation of $G$, with an additional superpotential $W=\trace \phi^{p+1}$ turned on.
  \item $W=Q^2u_1$: a $\mathcal{D}_p(G)$ theory flows to $\rank (\mathcal{D}_p(G))$ free $\mathcal{N}=2$ vector multiplets.
\end{enumerate}
Depending on the value of $p$, additional Coulomb branch operators, and thus relevant deformations, may exist. However, these are not relevant in this paper, as we wish to study the theory given in Figure \ref{fig:bigboy} and its deformations. This theory in Figure \ref{fig:bigboy} involves the gauging of the rank one building blocks $\mathcal{D}_2(SU(3))$, and thus, the only linear Coulomb branch deformations available are $u_1$ and $Q^2u_1$.

We can also deform the $\mathcal{D}_p(G)$ theory via moment map $W=\mu$. This in general does not give rise to a SCFT in the infrared, since it sets $\varepsilon=1/3$ as in the case of $Q^2 u_i$ deformation. Instead, one can also consider nilpotent Higgsing of the moment map, which realizes partial closure of the puncture in the class $\mathcal{S}$ language \cite{Gaiotto:2009we, Xie:2012hs}. 
But since this breaks the flavor symmetry $G$ (which is to be gauged), we do not consider such deformations here. 

In general we can furthermore consider flipping a relevant operator $\CO$ with R-charge less than $4/3$ via adding an additional chiral multiplet ($M$) and coupling via $W = M \CO $ \cite{Barnes:2004jj} which gives rise to much larger landscape of fixed points \cite{Maruyoshi:2018nod, Cho:2024civ}.  
Especially, we can consider a flip field for the moment map and then nilpotent Higgs the flip field \cite{Gadde:2013fma, Agarwal:2014rua, Maruyoshi:2016aim, Agarwal:2016pjo} which realizes the color-flipped puncture in the language of class $\mathcal{S}$. Since these deformations break the $a=c$ property, we do not consider such deformations here. 

\section{How to generate the landscape}\label{sec:method}

In this section, we describe how to generate a landscape of 4d SCFTs with $a=c$. We start from a ``seed'' $a=c$ theory and consider all possible relevant deformations. A large sample of such ``seed SCFTs'' can be constructed from gauging a collection of $\CD_p(G)$ theories as is classified in \cite{Kang:2021ccs}. Considering a series of superpotential deformations by relevant operators, we generate a large set of SCFTs while preserving the $a=c$ property, since, as we have discussed in Section \ref{sec:acdefs}, $a-c$ is not modified after superpotential deformation by a relevant operator. We consider all possible relevant operators to deform the theory and check whether these deformed theories flow to interacting SCFTs. For each new fixed point, we repeat this process until we exhaust all possible fixed points originating from a given seed SCFT.\footnote{One may think we can enumerate all possible relevant deformations from the ``seed'' SCFT, but some of the irrelevant operators can become relevant upon flow (called as dangerously irrelevant operators). Therefore we repeat the enumeration of such operators at each new fixed point.} This is essentially an identical procedure as in \cite{Maruyoshi:2018nod, Cho:2024civ} except that we only keep track of $a=c$ SCFTs and we often utilize known dualities \cite{Bolognesi:2015wta, Xie:2021omd,Kang:2023dsa,Maruyoshi:2023mnv,Kang:2024elv}. 
Let us explain this procedure in more detail.

\begin{enumerate}
    \item Start with a SCFT with $a=c$ constructed by gluing a collection of $\CD_{p}(G)$ theories (and/or) adjoint-valued chiral multiplets via $\CN=1$ gauging without superpotential. 
    
    \item Enumerate all relevant chiral operators. Then deform the theory by adding $\delta W = \mathcal{O}$ to the superpotential, for all relevant operators $\mathcal{O}$ carrying a non-trivial flavor charge. 

    The gauge invariant operators are built out of $u_{i,\alpha}, Q^2u_{i,\alpha}, \mu_\alpha, \phi_\nu$. 
    Some operators are subject to the chiral ring relation of $\CD_p(G)$ theories, such as \cite{Agarwal:2018zqi}
    \begin{align}
        \trace\mu^n=0\,,\quad \mu^p|_{\text{adj}}=0\,,\quad u_{i,\alpha}\mu_{\alpha}=0\,.
    \end{align}
    This removes operators of the form $\trace \mu_\alpha^2$, $\trace\mu^p\phi$, $u_{i,\alpha}\trace\mu_\alpha\phi$ from the chiral ring, for example.  
     
    \item Find the superconformal R-symmetry of the deformed theory. Suppose $\mathcal{O}$ has $R$-charge $R[\mathcal{O}]$ and a $U(1)$ flavor charge $F[\mathcal{O}]$.\footnote{This is schematic, since $\mathcal{O}$ can be charged under multiple Abelian flavor symmetries, however that does not qualitatively modify the analysis.} Then the deformed theory flows to a fixed point whose R-symmetry is shifted to:
    \begin{align}
        W = \mathcal{O}: \quad R_{\text{IR}} = R + \epsilon F \ , \quad \epsilon = \frac{2-R[\mathcal{O}]}{F[\mathcal{O}]}
    \end{align}
    If the relevant operator $\mathcal{O}$ is not charged under any flavor symmetry, we cannot find any candidate R-symmetry in the infrared. Hence a SCFT deformed by a flavor-neutral relevant operators cannot flow to a SCFT.\footnote{There is a caveat here: Sometimes we may find an accidental symmetry emerges along the RG flow and the fixed point ends up being superconformal. For example, $SU(N_c)$ SQCD with $N_f<2N_c$ flavors deformed by a quartic superpotential seemingly breaks the R-symmetry. However, once we invoke Seiberg duality, the quartic term maps into a mass term for the dual mesons. Therefore, the fixed point is described by the $SU(N_f-N_c)$ SQCD with $N_f$ flavors \cite{Strassler:2003qg}. Without evidence to the contrary, we assume that such phenomena does not happen.}
    
    \item Check if the deformed theory is ``good'', meaning that there is no gauge-invariant chiral operator $\CO$ that violates the unitarity bound $R[\CO] > 2/3$. If an operator has $R$-charge less than $2/3$, it decouples along the RG flow and becomes free. In this case, IR theory becomes a product theory of interacting part and free part \cite{Kutasov:2003iy}. When such a decoupling happens, the interacting part of the IR fixed point is no longer $a = c$ because there is an accidental symmetry that mixes with $R$.\footnote{The total central charges including the decoupled sector can still satisfy $a=c$ when the superconformal R-symmetry of the interacting part (accidentally) does not mix with the emergent symmetry. But we are mainly interested in the interacting part of the IR fixed point.}  
    
    \item Check whether there is any dual description studied in Section \ref{sec:ADdeformations}. When the constituent $\CD_p(G)$ theory is deformed by its Coulomb branch operator of lowest dimension, this gives rise to the theory of a free adjoint-valued chiral multiplet ($\phi$) coupled with superpotential $W=\trace \phi^{p+1}$. 
    But this superpotential can often (but not always) be removed upon flow since it is irrelevant. It can survive when the $R$-charge of $\phi$ is low.   
    
    \item Check whether there is any marginally irrelevant superpotential term in the superpotential $W$ that can be turned off upon flow. Such an operator $\CO$ has $R$-charge 2 with a non-trivial flavor charge at the fixed point without $\CO$ turned on, namely with the superpotential $W'=W-\CO$. Since any marginal operator that breaks the flavor symmetry is marginally irrelevant \cite{Leigh:1995ep, Gray:2013mja}, we can turn it off. In practice, we simply turn off all the marginal operators having $R$-charge 2, and keep the exactly marginal operators that do not carry any flavor charge.
    
    \item Make sure that the beta function $\beta_g$ for the gauge coupling is negative. This should be done after applying duality of deformed $\CD_p(G)$ or integrating out massive fields since that affects the sign of beta function. 
    We check this since upon deforming $\CD_p(G)$ via Coulomb branch operator, we are effectively replacing $\CD_p(G)$ by an adjoint chiral multiplet, who can potentially flip the sign of $\beta_g$. 
        
    \item Perform a refined unitarity check by computing the superconformal index \cite{Beem:2012yn,Evtikhiev:2017heo,Maruyoshi:2018nod,Kang:2022vab,Cho:2024civ} as some of the unitarity violating operators can appear outside of the chiral ring. Refer to \cite{Kang:2022vab} for the conventions and details about the computation of the index. 
    
    Ideally, we need to know the fully refined superconformal index to perform the refined check. For our specific example of gauging $\CD_2 (SU(3))$ theory, luckily we do have the full index \cite{Agarwal:2016pjo}.  
    For a generic $\CD_p(G)$ theory, we only know the Schur (or Macdonald) index \cite{Buican:2015ina, Buican:2015tda, Buican:2017uka,Cordova:2015nma, Beem:2020pry, Foda:2019guo, Watanabe:2019ssf, Song:2015wta, Song:2016yfd, Agarwal:2019crm, Song:2017oew, Xie:2019zlb}. 
    But even without knowing the full index, we can still mimic the refined check of the index from the known operator spectrum of the $\CD_p(G)$ theory. 
    Specifically, we can consider gauge-invariant operators built from the following constituents: 
    \begin{itemize}
        \item moment map operators of $\CD_{p}(G)$ theories, 
        \item adjoint chiral multiplets $\phi_\nu$, 
        \item $\psi^\dagger_\nu$ (whose R-charge is $2-r_\nu$), which is the complex conjugate of the superpartner of the adjoint chiral multiplet $\phi_\nu$.
    \end{itemize} 
    If an operator $\mathcal{O}$ built out of above constituents has $R$-charge less than 2 and contains odd number of $\psi^\dagger$, such a fermionic operator contributes to the superconformal index as $-t^{3r_\mathcal{O}}$.\footnote{We use the same convention for the fugacities as is used in \cite{Kang:2022vab}.} None of the unitary $\CN=1$ superconformal multiplets can contribute to the superconformal index by such a term. Unless there is another bosonic chiral gauge-invariant operator with the same $R$ and flavor charges to cancel out this term, unitarity is violated.  
    
    \item Repeat the process 2 to 8 for each deformed theories until none of the deformed theories contains relevant chiral operators with non-trivial flavor charges. 
\end{enumerate}

The procedure we described above exhausts all possible fixed point SCFTs with $a=c$ that can be obtained from a given ``seed $a=c$ SCFT.'' We find a large number of fixed points emanating from one gauged Argyres--Douglas theory. As we have mentioned above in step 3, we emphasize that our procedure crucially depends on knowing all the $U(1)$ symmetries that can mix with R-symmetry. Therefore we cannot rule out the possibility that there may be a non-trivial fixed point we are missing. In the following section, we will implement our general procedure for one particular seed SCFT, namely 3 copies of $\CD_2(SU(3))$ theories gauged with a single adjoint-valued chiral multiplet, as depicted in Figure \ref{fig:bigboy}. 

Before turning to the analysis of the landscape in Section \ref{sec:222landscape}, we would like to briefly comment on one particular example we find rather intriguing. 
Consider three $\CD_{p_\alpha}(G)$ theories with $(p_1,p_2,p_3)=(2,4,4)$ and one chiral multiplet, $X$, in the adjoint representation of $G=SU(2k+1)$ glued by $\CN=1$ gauging. 
This gauging is conformal (i.e., the one-loop $\beta$-function of the gauge coupling vanishes) and the gauged theory is connected to the $\widehat{E}_7(G)$ theory of \cite{Kang:2021lic} through a marginal deformation.\footnote{Actually the gauge coupling becomes free without turning on the superpotential $W=\trace X \mu_\alpha$, just as in the case of $\CN=2$ superconformal QCDs \cite{Leigh:1995ep}.} 
This theory has three flavor $U(1)$s
\begin{align}
    \mathcal{F}_1=3F_1-2F_2\,,\qquad \mathcal{F}_2=F_2-F_3\,,\qquad \mathcal{F}_3=2F_2+2F_3-3 f_1\,,
\end{align}
where $F_{1,2,3}$ are flavor symmetries from each $\CD_2(G),\, \CD_4(G),\, \CD_4(G)$ blocks respectively, and $f_1$ is the flavor symmetry rotating the adjoint chiral multiplet. There is a relevant operator $u_{1,1} u_{1,2}$ whose R-charge and flavor charges are
\begin{align}
(R,\, \mathcal{F}_1,\, \mathcal{F}_2,\, \mathcal{F}_3)[u_{1,1}u_{1,2}]=\left(\frac{11}{6},\,-4,\,-\frac{5}{2},\,-5\right).
\end{align}
Since it has a non-trivial flavor charge, the theory flows to another SCFT with $a=c$. Similarly, we can further deform the theory with a relevant operator $u_{1,1}u_{1,3}$ from the fixed point of $W=u_{1,1}u_{1,2}$. 
Finally, we can further deform the theory by $\delta W=u_{1,3}$ at the fixed point of $W=u_{1,1}u_{1,2}+u_{1,1}u_{1,3}$. Then the mixing coefficients and the R-charge of the adjoint chiral multiplet is fixed solely by the anomaly and the superpotential constraints as
\begin{align} \label{eq:def244mixing}
    \varepsilon_1=\frac{1}{3}\,,\qquad \varepsilon_2=\varepsilon_3=-\frac{7}{15}\,,\qquad R[X]=\frac{6}{5}\,.
\end{align}
These are exactly the values of the theory where, the first $\CD_2(G)$ is replaced by a free $\CN=2$ vector multiplet (the one we get after deforming via $W=Q^2 u_{1, 1}$) that decouples along the RG-flow, and the other two $\CD_4(G)$ theories are replaced by  adjoint chiral multiplets and extra superpotentials $\delta W=\trace\phi_2^5+\trace\phi_3^5$ (which we get via deformation $W=u_{1, 2} + u_{1, 3}$). Hence, the final theory is dual to $G$ gauge theory with three adjoint chiral multiplets coupled via gauge coupling and the superpotential $W=\trace \phi_2^5+\trace\phi_3^5$. 
However, this superpotential is irrelevant for the two-adjoint theory near $W=0$, so this theory may simply flow to the two-adjoint theory without superpotential which gives rise to different central charges from what we computed using the mixing coefficients in equation \eqref{eq:def244mixing} and we have emergent $SU(2)$ flavor symmetry in the IR.  
Alternatively (or radically), there may be a nontrivial UV fixed point that is far away from the $W=0$ of the two adjoint theory, that is dual to the deformed Argyres--Douglas theory we discussed so far. This is a very interesting possibility that we do not know how to either rule out or support. This structure will be discussed in more detail when we analyze the network of $a=c$ preserving superpotential deformations of the $\widehat{\Gamma}(G)$ theories in \cite{LANDSCAPEII}.

\subsection{Deformations of \texorpdfstring{$\mathcal{N}=4$}{N=4} SYM}\label{sec:defsN4}

Frequently, $\mathcal{N}=4$ super-Yang--Mills will appear in our analysis of superpotential deformations of gauged Argyres--Douglas theories. Therefore, here we briefly discuss the well-known network of relevant deformations. Consider the maximal Yang--Mills theory as an $\mathcal{N}=1$ gauge theory with three adjoint-valued chiral multiplets: $X$, $Y$, and $Z$. The six relevant operators are
\begin{equation}\label{eqn:N4rel}
    \trace X^2 \,, \qquad \trace Y^2 \,, \qquad \trace Z^2 \,, \qquad \trace XY \,, \qquad \trace XZ \,, \qquad \trace YZ \,.
\end{equation}
The mixed deformations, e.g., $\trace YZ$, integrate out two of the three chiral multiplets; the result is $G$ gauge theory coupled to a single adjoint-valued chiral multiplet, which does not flow to an SCFT in the infrared. The mass deformations, e.g., $\trace Z^2$, yields the same infrared as the $G$ gauge theory coupled to two adjoint-valued chiral multiplets. This is a 4d $\mathcal{N}=1$ SCFT with central charges
\begin{equation}
    a = c = \frac{27}{128} \dim (G) \simeq 0.2109 \, \dim(G) \,.
\end{equation}
The relevant operators of the two-adjoint theory are
\begin{equation}\label{eqn:2adjrel}
    \trace X^2 \,, \quad \trace XY \,, \quad \trace Y^2 \,, \quad \trace X^3 \,, \quad \trace X^2Y \,, \quad \trace XY^2 \,, \quad \trace Y^3 \,.
\end{equation}
Deformation by any of the quadratic operators does not lead to an SCFT in the infrared, and therefore we do not consider them further. By contrast, the cubic deformations, while they also do not trigger flows to interacting SCFTs fixed points have some interesting and heretofore unknown behavior, which we briefly highlight here.

First, we consider the deformation by the cubic superpotential in a single adjoint-valued chiral multiplet:
\begin{equation}
    W  = \trace X^3 \,.
\end{equation}
The R-charges in the infrared are fixed by anomaly cancellation, and we find
\begin{equation}
    R[X] = \frac{2}{3} \,, \qquad R[Y] = \frac{1}{3} \,.
\end{equation}
As we can see from the R-charges, the operator $\trace Y^2$ decouples along the flow into the infrared; adding the appropriate flip field, we determine that the remaining sector in the infrared is an interacting SCFT whose central charges are
\begin{align}
    a=\frac{3}{16}\dim (G)-\frac{1}{48}\,,\quad c=\frac{3}{16}\dim (G)-\frac{1}{24}\,.
\end{align}
The subtracting term in each central charge comes from the decoupled free chiral multiplet. Note that the two central charges of the remaining theory are not equal despite the two central charges of the net theory (including the decoupled free chiral) being equal to each other. Since we seek for ``interacting'' SCFTs with $a=c$, we do not count this case in our landscape.

The remaining option, up to relabeling of the chiral multiplets, is to deform the theory with the superpotential
\begin{equation}
    W = \trace XY^2 \,,
\end{equation}
which leads to the infrared R-charges being
\begin{equation}
    R[X] = 0 \,, \qquad R[Y] = 1 \,.
\end{equation}
The vanishing of the R-charge for the chiral multiplet $X$ indicates that both $\trace X^2$ and $\trace X^3$ decouple from the spectrum along the flow. Adding the necessary flip fields, we see that the central charges of the leftover sector in the infrared are
\begin{equation}
    a = \frac{3}{16} \times 2 \,, \qquad c = \frac{1}{8} \times 2 \,.
\end{equation}
This is precisely the central charges of two free $\mathcal{N}=1$ vector multiplets. We can further verify this is indeed the theory we get by computing the superconformal index of the infrared theory, which we find to be
\begin{equation}
    I=\text{PE}\left[2\times\frac{-t^3y-t^3/y+2t^6}{(1-t^3y)(1-t^3/y)}\right] \,.
\end{equation}
Unsurprisingly, this is exactly the index of two free $\mathcal{N}=1$ vector multiplets. Together with the decoupled free chiral multiplets corresponding to the $\trace X^2$ and $\trace X^3$ operators, the infrared behavior of the deformed theory appears to be that of two copies of an $\mathcal{N}=2$ vector multiplet. More generally, it turns out that the $W = XY^2$ deformation of the infrared SCFT coming from the two-adjoint $SU(N)$ gauge theory leads to $N-1$ free $\mathcal{N}=2$ vector multiplets.

\section{The landscape of \texorpdfstring{\boldmath{$(2,2,2) + \operatorname{adj}$}}{(2,2,2) + adj}}\label{sec:222landscape}

We now perform an in-depth analysis to acquire the network of superpotential deformations and interacting $a=c$ fixed points for the gauged Argyres--Douglas theory appearing in Figure \ref{fig:bigboy}. We introduce an $SU(3)$ $\mathcal{N}=1$ vector multiplet to gauge the flavor symmetry of three copies of $\mathcal{D}_2(SU(3))$ and an adjoint-valued chiral multiplet. We use $\alpha = 1, 2, 3$ to label the three Argyres--Douglas matters.

The introduced gauge coupling is asymptotically-free, and after flowing into the infrared, we write the generator of the infrared R-charge as
\begin{equation}
    R = R_0 + \sum_{\alpha=1}^3 \varepsilon_\alpha F_\alpha + \left( R[X] - \frac{2}{3} \right) f_1 \,, 
\end{equation}
where, as in Section \ref{sec:acdefs}, $\varepsilon_\alpha$, $R[X]$ are the mixing coefficients, $F_\alpha$ are the three $U(1)$ flavor symmetries coming from the $\mathcal{N}=2$ R-symmetry of the Argyres--Douglas factors, and $f_1$ is the symmetry rotating the adjoint chiral $X$. Via $a$-maximization, we determine the mixing parameters $\varepsilon_\alpha$ and the R-charge of the adjoint chiral $X$ to be
\begin{equation}
    \varepsilon_1 = \varepsilon_2 = \varepsilon_3 =  -\frac{25-6\sqrt{15}}{51}\,,\quad R[X] = \frac{3\left(7-\sqrt{15}\right)}{17} \simeq 0.5518 \,.
\end{equation}
Thus, the identical central charges are
\begin{equation}
    a=c =\frac{27\left(14+15\sqrt{15}\right)}{4624}\dim (G) \simeq 0.4210 \, \dim (G) \,.
\end{equation}
We have written $\dim (G)$ here, instead of $\dim (SU(3)) = 8$, since the result holds for any $G$ such that $\gcd(2, h_G^\vee) = 1$.\footnote{In fact, we will write $\dim (G)$ throughout when the result holds for any such $G$.}

We now enumerate the relevant operators of the resulting infrared SCFT. Note that the $\mathcal{D}_2(SU(3))$ has a rank-one Coulomb branch, and we will refer to each single Coulomb branch operator as $u_\alpha$, where $\alpha$ is the index for each Argyres--Douglas building block.\footnote{When generalizing beyond $G = SU(3)$, there are many more relevant Coulomb branch operators $u_{i>1,\alpha}$.} There are eleven relevant operators in the chiral ring:
\begin{equation}
    u_\alpha \,, \quad Q^2u_\alpha \,, \quad \trace \mu_\alpha X \,, \quad \trace X^2 \,, \quad \trace X^3 \,.
\end{equation}
This can also be verified by computing the superconformal index, following the analysis in \cite{Kang:2022vab}. Each of these relevant operators provides a relevant superpotential deformation, triggering a flow to a new infrared fixed point.

\pdfbookmark[subsection]{Deformations}{deformations}

\paragraph{\texorpdfstring{\uline{$W=u_3$ deformation}}{W = u₃}} We start by performing a relevant deformation with respect to the superprimaries of the Coulomb branch operators coming from the Argyres--Douglas factors. Without loss of generality, we consider the deformation
\begin{equation}
    W = u_3 \,.
\end{equation}
Following the Coulomb branch operator deformations explained in Section \ref{sec:ADdeformations}, after deformation we obtain the same interacting SCFT that exists in the infrared of the gauging of two $\mathcal{D}_2(SU(3))$ theories, together with two adjoint-valued chiral multiplets $X$ and $Y$:
\begin{align}\label{eqn:2D2G}
\begin{aligned}
    \begin{tikzpicture}
      \node[gaugeN1] (s0) {$SU(3)$};
      \node[d2] (c2) [left=0.6cm of s0] {$\mathcal{D}_{2}(SU(3))$};
      \node[d2] (c3) [right=0.6cm of s0] {$\mathcal{D}_{2}(SU(3))$};
      \draw (s0.west) -- (c2.east);
      \draw (s0.east) -- (c3.west);
       \draw[dashed, ->] (s0) to[out=130, in=410, looseness=4] (s0);
      \draw[dashed, ->] (s0) to[out=146, in=394, looseness=5] (s0);
    \end{tikzpicture} \,.
\end{aligned}
\end{align}
Checking the one-loop $\beta$-function, we can see that the gauge coupling is directly conformal, not asymptotically-free and the chiral superfields $X$ and $Y$ have the engineering scaling dimension $\Delta = 1$ so that $R=2/3$. We also get the superpotential $W= \trace Y^3$ from the deformation $W=u_3$, but this term is marginally irrelevant so we drop it. 

The R-charges of the operators can be determined from the following:
\begin{equation}
    R[X] = R[Y] = \frac{2}{3} \,, \qquad R[\mu_\alpha] = \frac{4}{3} \,, \qquad R[u_\alpha] = 1 \,, \qquad R[Q^2u_\alpha] = \frac{5}{3} \,.
\end{equation}
The central charges of this theory is
\begin{align}\label{eqn:mewtwo}
    a = c = \frac{3}{8} \dim (G) = 0.375 \, \dim (G) \ . 
\end{align}
In particular, we would expect that the (scalar) chiral ring operators 
\begin{equation}\label{eqn:22rel}
    \trace X^2 \,, \quad \trace XY \,, \quad \trace Y^2 \,, \quad 
    u_\alpha \,, \quad Q^2u_\alpha \,,
\end{equation}
provide relevant operators to the SCFT, whereas the marginal such operators are
\begin{equation}\label{eqn:22marg}
    u_1^2 \,, \quad u_1u_2 \,, \quad u_2^2 \,, \quad \trace X^3 \,, \quad \trace X^2Y \,, \quad \trace XY^2 \,, \quad \trace Y^3 \,, \quad \trace \mu_\alpha X \,, \quad \trace \mu_\alpha Y \,.
\end{equation}

We can verify that this provides the full spectrum of relevant and marginal operators by calculating the superconformal index. Expanding to observe the relevant and marginal operators only, the reduced superconformal index of this theory is
\begin{equation}
  \begin{aligned}
    \widehat{I}_{(2,2)}^{\,n_a = 2} &= \left(\frac{v_1^6v_2^6}{v_3^3} + v_3^3 \right) t^3 + \left( v_1^2 + v_1v_2 + v_2^2\right)t^4 -  \chi_{\bm{2}}(y)\left(\frac{v_1^2v_2^2}{v_3} + v_3 \right)t^4 \\
    &\qquad + \left(\,\frac{1}{v_3} + \frac{v_3}{v_1^2v_2^2}\,\right)t^5 -\chi_{\bm{2}}(y)(v_1 + v_2)t^5 \\
    &\qquad + \bigg( v_1^3 + v_2^3 + v_3^6 + v_1v_2^2 + v_1^2v_2 + v_1^6v_2^6 \\&\qquad\qquad\qquad + \frac{v_1^{12}v_2^{12}}{v_3^6} + \frac{v_1}{v_3^2} + \frac{v_2}{v_3^2} + \frac{v_3^2}{v_1^3v_2^4} + \frac{v_3^2}{v_1^4v_2^3} - \frac{v_1}{v_2} - \frac{v_2}{v_1} - 3 \bigg)t^6 + \cdots \,,
  \end{aligned}
\end{equation}
where $v_i$ for $i=1,2,3$ are fugacities for the three $U(1)$ flavor symmetries of the theory. A linear combination of these $U(1)$s is the Cartan of an enhanced $SU(2)$ global symmetry rotating the two adjoint-valued chiral multiplets. Thus, we see that only a subset of six out of the eleven marginal operators in equation \eqref{eqn:22marg} are exactly marginal.

We now consider the superpotential deformations by relevant operators, each in turn. We begin with the operators constructed from the adjoint-valued chiral multiplets.

\paragraph{\texorpdfstring{\uline{$W = u_3 + \trace Y^2$ deformation}}{W = u₃ + TrY²}} We first consider the mass deformation for one of the adjoint chiral multiplets, obtained by turning on the following superpotential at the $W=u_3$ fixed point:\footnote{Of course, turning on a superpotential deformation proportional to the relevant operator $\trace X^2$ triggers an RG-flow to the same fixed point discussed here.}
\begin{equation}
    \trace Y^2 \,.
\end{equation}
In the deep infrared, this theory flows to the same IR fixed point as the gauging
\begin{align}\label{eqn:2D2G1adj}
\begin{aligned}
    \begin{tikzpicture}
      \node[gaugeN1] (s0) {$G$};
      \node[d2] (c2) [left=0.6cm of s0] {$\mathcal{D}_{2}(G)$};
      \node[d2] (c3) [right=0.6cm of s0] {$\mathcal{D}_{2}(G)$};
      \draw (s0.west) -- (c2.east);
      \draw (s0.east) -- (c3.west);
       \draw[dashed, ->] (s0) to[out=130, in=410, looseness=4] (s0);
    \end{tikzpicture} \,.
\end{aligned}
\end{align}
This is a quiver with only one adjoint-valued chiral multiplet, $X$. The mixing coefficients and the effective R-charge of the chiral multiplet are given by
\begin{equation}
    \varepsilon_1 = \varepsilon_2 = \frac{\sqrt{3} - 2}{3}\,, \qquad R[X] = 1 - \frac{1}{\sqrt{3}} \simeq 0.4227 \,,
\end{equation}
and thus the identical central charges are
\begin{equation}
    a = c = \frac{3\sqrt3}{16}\dim (G) \simeq 0.3248 \, \dim (G) \,.
\end{equation}
The spectrum of relevant operators of the infrared SCFT are then
\begin{equation}\label{eqn:mudkip}
    \trace X^2 \,, \qquad \trace X^3 \,, \qquad \left(\trace X^2\right)^2 \,, \qquad \trace \mu_\alpha X \,, \qquad u_\alpha \,, \qquad Q^2 u_\alpha \,.
\end{equation}
We further study the infrared fixed points attained after superpotential deformation by each of these operators. 

\paragraph{\texorpdfstring{\uline{$W = u_3+\trace Y^2+\trace X^2$ deformation: No SCFT}}{W = u₃ + TrY² + TrX²}}
Starting from the previous theory obtained by the deformations $W = u_3+\trace Y^2$, an extra deformation by the mass term
\begin{equation}
    \trace X^2 \,,
\end{equation}
ensures that the field $X$ is integrated out below the mass scale. Thus, we simply obtain 
\begin{align}\label{eqn:2D2G0adj}
\begin{aligned}
    \begin{tikzpicture}
      \node[gaugeN1] (s0) {$G$};
      \node[d2] (c2) [left=0.6cm of s0] {$\mathcal{D}_{2}(G)$};
      \node[d2] (c3) [right=0.6cm of s0] {$\mathcal{D}_{2}(G)$};
      \draw (s0.west) -- (c2.east);
      \draw (s0.east) -- (c3.west);
    \end{tikzpicture} \,.
\end{aligned}
\end{align}
As discussed in \cite{Kang:2021ccs}, such a quiver does not flow to an interacting infrared SCFT and thus we do not consider it further here.

\paragraph{\texorpdfstring{\uline{$W = u_3+\trace Y^2+ \trace X^3$ deformation: flows to $\CN=4$ SYM}}{W = u₃ + TrY² + TrX³}} From equation \eqref{eqn:2D2G1adj} obtained by deformation $W = u_3+\trace Y^2$, we deform the theory further by 
\begin{align}\label{eqn:feebas}
     \trace X^3 \,.
\end{align}
The cancellation of the $R$-gauge-gauge anomaly fixes the infrared superconformal $R$-charge to be such that the mixing parameters are
\begin{equation}
    \varepsilon_\alpha = -\frac{1}{3} \,,
\end{equation}
and thus the R-charges of the pertinent objects are:
\begin{align}
    R[X] &= \frac{2}{3} \,, \qquad R[\mu_\alpha] = \frac{2}{3} \,, \qquad R[Q^2u_{\alpha}] = \frac{4}{3} \,.
\end{align}
After performing this particular superpotential deformation we flow to an interacting SCFT with $a = c$ in the infrared. In fact, computing the central charges directly, we find that 
\begin{equation}\label{eqn:ccsN4}
    a = c = \frac{1}{4} \dim (G) \simeq 0.25 \, \dim (G) \,.
\end{equation}
At the fixed point, there are six relevant operators 
\begin{align}\label{eqn:swiper}
    \trace  X^2 \,,\qquad Q^2u_{\alpha} \,,\qquad \trace \mu_\alpha X \,,\qquad \trace  \mu_1 \mu_2 \,.
\end{align}
The central charges together with the existence of six relevant operators indicates that the infrared theory is nothing other than an $SU(3)$ gauge theory coupled to three adjoint-valued chiral multiplets. This theory sits on the conformal manifold of $\mathcal{N}=4$ super-Yang--Mills. Since we are working with $G = SU(3)$, we can also determine the reduced superconformal index; we find
\begin{align}
    \widehat{I}=6t^4-3t^5\chi_{\bm{2}}(y)+3t^6+O(t^7) \,.
\end{align}
The coefficient of the $t^4$ term is precisely from the six operators listed in equation \eqref{eqn:swiper}. As expected, this is also the first terms in the superconformal index of $\mathcal{N}=4$ super-Yang--Mills with $G = SU(3)$. 

We can understand why we get (marginally deformed) $\CN=4$ SYM theory as the fixed point from the known duality between gauged $\CD_2(SU(2N+1))^{\otimes 3}$ theory and $\CN=4$ SYM theory with $G=SU(2N+1)$ \cite{Kang:2023dsa}. When one of the $\CD_2(SU(2N+1))$ theory is deformed by the Coulomb branch operator of lowest dimension, it flows to become an adjoint chiral multiplet with the cubic superpotential. Therefore we have two copies of $\CD_2 (SU(2N+1))$ theory gauged with an adjoint chiral multiplet $X$ with $W=\trace X^3$, which is the same setup as we discuss in this paragraph. 

The landscape of superpotential deformations of $\mathcal{N}=4$ super-Yang--Mills has already been discussed in Section \ref{sec:defsN4}.

\paragraph{\texorpdfstring{\uline{$W = u_3+\trace Y^2+(\trace X^2)^2$ deformation}}{W = u₃ + TrY² + (TrX²)²}} Next, we consider the further deformation of the SCFT associated with the gauge theory in equation \eqref{eqn:2D2G1adj} by the quartic multi-trace operator
\begin{equation}
    (\trace X^2)^2 \,.
\end{equation}
The mixing parameters and infrared R-charges of the pertinent operators are
\begin{equation}
    \varepsilon_\alpha = -\frac{1}{6} \,, \quad R[X] = \frac{1}{2} \,, \quad R[\mu_\alpha] = 1 \,, \quad R[u_{\alpha}] = R[Q^2u_{\alpha}] = \frac{3}{2} \,.
\end{equation}
Therefore, the central charges of the infrared SCFT, which is interacting and has identical central charges, are
\begin{equation}
    a = c = \frac{81}{256} \dim (G) \simeq 0.3164 \, \dim (G) \,.
\end{equation}
Intriguingly, these central charges can be written as
\begin{equation}
    a = c = \frac{27}{32} \times \frac{3}{8} \dim (G) \,,
\end{equation}
which, following \cite{Tachikawa:2009tt}, may indicate that this infrared SCFT can be obtained by starting from a 4d $\mathcal{N}=2$ SCFT with central charges 
\begin{equation}
    a = c = \frac{3}{8} \dim (G) \,,
\end{equation}
and performing a mass deformation of the adjoint-valued chiral multiplet inside of the $\mathcal{N}=2$ vector multiplet. We leave this for a future discussion. From the R-charges, we can see that the relevant operator spectrum in the infrared is as follows
\begin{equation}\label{eqn:kitties}
    \trace X^2 \,, \quad 
    \trace X^3 \,, \quad 
    \trace \mu_\alpha X \,, \quad
    u_{\alpha} \,, \quad Q^2u_{\alpha} \,.
\end{equation}
We continue by studying each of these deformations in turn.

\paragraph{\texorpdfstring{\uline{$W = u_3+\trace Y^2+(\trace X^2)^2 + \trace \mu_2 X$ deformation}}{W = u₃ + TrY² + (TrX²)² + Trμ₂X}} Let us consider the deformation of the fixed point of $W=u_3+\trace Y^2+(\trace X^2)^2$ by the relevant operator 
\begin{equation}
    \trace \mu_2 X \,,
\end{equation}
where, without loss of generality, we have taken $\alpha = 2$. The four terms in the superpotential together with anomaly cancellation uniquely fixes the infrared R-charges of the following operators to be
\begin{equation}
    R[X] = \frac{1}{2} \,, \qquad R[\mu_1] = \frac{1}{2} \,, \qquad R[\mu_2] = \frac{3}{2} \,,
\end{equation}
and the central charges to be
\begin{equation}
    a = c = \frac{135}{512} \dim (G) \simeq 0.2638 \, \dim (G) \,.
\end{equation}
The infrared fixed point is an interacting SCFT with $a = c$ and it has seven relevant operators:
\begin{equation}
     u_2\,,\quad Q^2u_{\alpha}\,,\qquad   \trace  X^2 \,, \qquad \trace  X^3 \,, \qquad 
        \trace  \mu_1 X \,, \qquad 
        \trace  \mu_1 X^2\,.
\end{equation}
We can consider deformations by each of these operators at this fixed point, however, the SCFT that we are discussing in this paragraph does not possess any continuous flavor symmetry. 
Thus, any deformation will fail to be an SCFT in the infrared since we do not have a consistent R-symmetry. However, it is still possible that there is an emergent symmetry in the infrared that can mix with R-symmetry so we end up with a SCFT. 

\paragraph{\texorpdfstring{\uline{$W = u_3+\trace Y^2+(\trace X^2)^2 + \trace X^3$ deformation: No R}}{W = u₃ + TrY² + (TrX²)² + TrX³}} There is no consistent infrared R-symmetry upon this subsequent deformation, unless there exists some emergent $U(1)$ symmetry along the flow into the infrared, and we expect this deformation does not lead to an SCFT. 

\paragraph{\texorpdfstring{\uline{$W = u_3+\trace Y^2+(\trace X^2)^2 + \trace X^2$ deformation: No R}}{W = u₃ + TrY² + (TrX²)² + TrX²}} Similarly to the $W = \trace X^3$ deformation, there does not exist a consistent infrared R-symmetry. Therefore, we expect this deformed theory does not flows to a SCFT in the infrared.

\paragraph{\texorpdfstring{\uline{$W = u_3+\trace Y^2+(\trace X^2)^2 + u_2$: Emergent R}}{W = u₃ + TrY² + (TrX²)² + u₂}} It seems there is no R-symmetry preserved upon this deformation, but we find there is an emergent symmetry in the IR. 
We can utilize the IR duality relating the $u_3$ deformation maps to an adjoint chiral multiplet $Z$. After integrating out the massive chiral multiplet $Y$, effectively the theory becomes a single $\CD_2(SU(3))$ theory gauged with 2 adjoint chirals $X,Z$ under the superpotential $W=(\trace X^2)^2 +\trace Z^3$ and we have an emergent R-symmetry at this point. 
From this superpotential, the IR R-charges are found to be
\begin{align}
    \varepsilon_1=0\,,\quad R[X]=\frac{1}{2}\,,\quad R[Z]=\frac{2}{3}\,.
\end{align}
Then we also find the central charges 
\begin{align}
    a=c=\frac{75}{256} \dim G \simeq 0.2930 \, \dim G \,.
\end{align}
Since the superpotential drained out all the flavor symmetries, one cannot find an anomaly free R-symmetry if the theory is deformed further. Therefore, if this is correct, it is one of the terminal point in the landscape of $a=c$ SCFTs.

However, we should be careful. In terms of the dual gauge theory given by $\CD_2(SU(3))$ gauged with 2 adjoint chirals have $r_X > 0.5$ so that the operator $(\trace X^2)^2$ is irrelevant at $W=0$ or $W=\trace Z^3$ fixed points. Therefore it is natural to expect that this quartic term in $X$ is turned off up on flow and we are left with only the $\trace Z^3$ term. In this case, we find the IR R-charges to be
\begin{align}
    \varepsilon_1 = \frac{1}{27} \left(2 \sqrt{79}-19\right) \,, \quad R[X] = \frac{1}{27} \left(23-\sqrt{79}\right) \simeq 0.5227 \,, \quad R[Z] = \frac{2}{3} \ , 
\end{align}
and the central charges to be
\begin{align}
    a = c = \frac{79 \sqrt{79}+442}{3888} \dim G \simeq 0.2943 \dim G \, 
\end{align}
which has slightly higher value than the previous scenario. 

We do not have a definitive argument to prefer one scenario over the other. In general, the end point of the RG flow depends on the order (or relative scalings) of the deformations. 
One argument to support the latter scenarios is that from the dual perspective,
the first scenarios requires a quartic coupling $ \lambda (\trace X^2)^2$ to have a non-trivial fixed point far away from the trivial fixed point $\lambda=0$. This theory is identical to the one that we will discuss around equation \eqref{eq:2with2adjY3} later. 

\paragraph{\texorpdfstring{\uline{$W = u_3+\trace Y^2+(\trace X^2)^2+Q^2u_{2}$ deformation: $a\neq c$}}{W = u₃ + TrY² + (TrX²)² + Q²u₂}} Additional deformation by $Q^2u_{\alpha}$, say $\alpha=2$ without loss of generality, effectively replaces one of the two $\CD_2(SU(3))$ with a free $\CN=2$ vector. The remaining interacting part of the effective theory contains one $\CD_2(SU(3))$ and one adjoint chiral $X$ that are gauged together with the superpotential
\begin{align}
 W = (\trace X^2)^2 \,.
\end{align}
The R-symmetry in the IR is immediately fixed by the anomaly-free condition as
\begin{align}
    \varepsilon_1=-\frac{2}{3}\,,\qquad R[X]=\frac{1}{2}\,.
\end{align}
The theory does not flow to an $a=c$ SCFT, since $\trace\mu_1X$ decouples.

\paragraph{\texorpdfstring{\uline{$W = u_3+\trace Y^2 + \trace \mu_2 X$ deformation: $a \neq c$}}{W = u₃ + TrY² + Trμ₂X}} This deformation is dual to the gauging of two $\mathcal{D}_2(SU(3))$ with one adjoint-valued chiral multiplet theory deformed by the following superpotential:
\begin{equation}
    W = \trace \mu_2 X \,.
\end{equation}
After deformation and application of $a$-maximization, we find that the mixing parameters for the infrared R-symmetry to be
\begin{equation}\label{eqn:torchic}
    \varepsilon_1 = \frac{\sqrt{13} - 5}{6}\,, \qquad \varepsilon_2 = \frac{\sqrt{13} - 1}{18}  \,.
\end{equation}
From this, we can see that the R-charge of the Coulomb branch operator $u_2$ falls below the unitarity bound and must be decoupled along the flow into the infrared. After decoupling, the remaining infrared theory does not have $a = c$, and thus we do not consider it further.

\paragraph{\texorpdfstring{\uline{$W = u_3+\trace Y^2 + Q^2u_2$ deformation: No SCFT}}{W = u₃ + TrY² + Q²u₂}} Following the discussion in Section \ref{sec:ADdeformations}, the $Q^2u_2$ deformation dissolves the second Argyres--Douglas factor into a free $\mathcal{N}=2$ vector multiplet, and the remaining part of the theory is the same as the infrared of one copy of $\mathcal{D}_2(SU(3))$ gauged together with a single adjoint-valued chiral multiplet. However, as noted in \cite{Kang:2021ccs}, this gauging does not flow to an interacting SCFT with $a=c$, and thus it is not pertinent to our analysis here.

\paragraph{\texorpdfstring{\uline{$W = u_3 + \trace Y^2 + u_2$ deformation}}{W = u₃ + TrY² + u₂}} This is dual to the theory appearing in equation \eqref{eqn:D2p2} (one $\CD_2(SU(3))$ with two adjoint chiral multiplets), deformed by the superpotential $W = \trace Y^3$ where $Y$ is one of the adjoint chiral multiplets. We study the network of relevant deformations later, around equation \eqref{eq:2with2adjY3}.

\paragraph{\texorpdfstring{\uline{$W = u_3 + \trace XY$ deformation: No SCFT}}{W = u₃ + TrXY}} There is one other relevant operator appearing in equation \eqref{eqn:22rel} that can be constructed out of the chiral multiplets $X$ and $Y$, and the associated superpotential deformation is
\begin{equation}
    W = \trace XY \,.
\end{equation}
Such a superpotential gives masses to both of the adjoint chirals, and leads to the same infrared fixed point as the gauging of just two $\mathcal{D}_2(SU(3))$ theories. As was explained in \cite{Kang:2021ccs}, this quiver does not flow in the infrared to an interacting SCFT, and thus we do not discuss it further here.

\paragraph{\texorpdfstring{\uline{$W= u_3+ u_2$ deformation: $a \neq c$}}{W = u₃ + u₂}} Next, we consider the sequential deformation by the Coulomb branch operators $u_3$ and then $u_2$. Gauge-anomaly cancellation and $a$-maximization leads to the following values of the mixing coefficients and the infrared R-charge of $X$:
\begin{equation}
    \varepsilon_1 = \frac{2\sqrt{79} - 17}{27} \,, \qquad \varepsilon_2 = \varepsilon_3 = -\frac{1}{3}  \,, \qquad R[X] = \frac{31-\sqrt{79}}{27} \,.
\end{equation}
As we can see, all of the chiral ring operators that we can write down satisfy the unitarity bound, however, performing a more refined check via the superconformal index demonstrates that certain operators do, in fact, decouple. The reduced index is:
\begin{align}
    \begin{gathered}
        \widehat{I}=t^{2.7412}v^6-t^{3.91373}v^2\chi_2(y)+3t^4+2t^{4.45687}v+2t^{4.91373}v^2-2t^5\chi_2(y)+t^{5.08627}v^{-2}\\-t^{5.45687}v\chi_2(y)+t^{5.48241}v^{12}-2t^{5.54313}v^{-1}-t^6+O\left(t^{6.17253}\right)\,.
    \end{gathered}
\end{align}
The term $-2t^{5.54313}v^{-1}$, which corresponds to the operators $\trace \mu_{2,3}\psi^{\dagger}$ where $\psi$ is the fermionic descendant inside of the chiral multiplet with primary $X$, violates unitarity. See \cite{Kang:2022vab} for the review of these unitarity conditions. As such, this operator must be decoupled along the flow, leading to an infrared fixed point which is not an $a=c$ SCFT. This is consistent with the deformations of the Argyres--Douglas building blocks discussed in Section \ref{sec:ADdeformations}. The deformed theory would appear to be dual to the $\mathcal{N}=1$ theory:
\vspace{-3.5em}
\begin{align}\label{eqn:3adj1DpG}
\begin{aligned}
    \begin{tikzpicture}
      \node[gaugeN1] (s0) {$G$};
      \node[d2] (c2) [left=0.6cm of s0] {$\mathcal{D}_{2}(G)$};
      \draw (s0.west) -- (c2.east);
       \draw[dashed, ->] (s0) to[out=40, in=320, looseness=4] (s0);
      \draw[dashed, ->] (s0) to[out=56, in=304, looseness=7] (s0);
      \draw[dashed, ->] (s0) to[out=63, in=297, looseness=10] (s0);
    \end{tikzpicture}
\end{aligned}
\end{align}
\vspace{-5em}

\noindent deformed by the superpotential $W=\trace Z^3$, where $Z$ is one of the adjoint-valued chiral multiplets. Since the gauge coupling is IR free in this dual theory, we would expect that the infrared merely consists of free fields and a decoupled $\mathcal{D}_2(SU(3))$. As expected, this is not an interacting infrared $a=c$ SCFT.

\paragraph{\texorpdfstring{\uline{$W=u_3+Q^2u_2$ deformation}}{W = u₃ + Q²u₂}} Finally, we can consider the theory in equation \eqref{eqn:2D2G}, together with a superpotential which is proportional to $Q^2 u_\alpha$, without loss of generality we take $\alpha = 2$.
As explained in Section \ref{sec:ADdeformations}, the second $\mathcal{D}_2(G)$ becomes a free $\mathcal{N}=2$ vector multiplet after the RG-flow. Thus, after the deformation and flow into the infrared, we are left with the same SCFT that arises in the infrared of the gauging
\vspace{-1.5em}
\begin{align}\label{eqn:D2p2}
\begin{aligned}
    \begin{tikzpicture}
      \node[gaugeN1] (s0) {$G$};
      \node[d2] (c2) [left=0.6cm of s0] {$\mathcal{D}_{2}(G)$};
      \draw (s0.west) -- (c2.east);
       \draw[dashed, ->] (s0) to[out=40, in=320, looseness=4] (s0);
      \draw[dashed, ->] (s0) to[out=56, in=304, looseness=7] (s0);
    \end{tikzpicture} 
\end{aligned}
\end{align}
The $a$-maximization yields the data of the mixing coefficient, IR R-charges, and central charges as follows:
\begin{align}
\begin{gathered}
    \varepsilon_1=-\frac{23-4\sqrt{31}}{33}\,,\quad R[X] = R[Y] =\frac{25-\sqrt{31}}{33}\,,\\
    a=c=\frac{116+31\sqrt{31}}{968}\operatorname{dim}(G) \simeq 0.2981 \, \dim (G) \,.
\end{gathered}
\end{align}
In total, there are eleven relevant operators 
\begin{align}
  \begin{gathered}
    u_1\,,\quad Q^2u_1\,,\quad \trace\mu_1 X\,,\quad \trace\mu_1 Y\,,\\ \trace X^2 \,, \quad \trace XY \,, \quad \trace Y^2 \,, \quad \trace X^3 \,, \quad \trace X^2 Y \,, \quad \trace XY^2 \,, \quad \trace Y^3 \,,
 \end{gathered}
\end{align}
by which the theory can be deformed further. We consider each of these deformations, and the further, subsequent deformations that are available, in turn. 

\paragraph{\texorpdfstring{\uline{$W=u_3+Q^2u_2 + u_1$ deformation: $\CN=4$ SYM}}{W = u₃ + Q²u₂ + u₁}} The further deformation by the Coulomb branch operator of the final remaining Argyres--Douglas matter results, following Section \ref{sec:ADdeformations}, in a theory lying on the ($\mathcal{N}=1$)-preserving conformal manifold of $\mathcal{N}=4$ super-Yang--Mills. Since the subsequent deformations of this theory have already been discussed, in Section \ref{sec:defsN4}, we do not enumerate them again here. For the sake of the analysis of upcoming sequences of deformations, we now explain the mapping of the relevant operators across this duality to $\mathcal{N}=4$ super-Yang--Mills. From $a$-maximization and gauge-anomaly cancellation, we find that
\begin{equation}
    R[X] = \frac{2}{3} \,, \qquad R[Y] = \frac{2}{3} \,, \qquad R[\mu_1] = \frac{2}{3} \,, \qquad R[Q^2u_1] = \frac{4}{3} \,.
\end{equation}
Therefore, the relevant operators belonging to the chiral ring are
\begin{equation}
    \trace X^2 \,, \qquad \trace Y^2 \,, \qquad \trace XY \,, \qquad \trace \mu_1 X \,, \qquad \trace \mu_1 Y \,, \qquad Q^2u_1 \,.
\end{equation}
Letting $X$, $Y$, and $Z$ denote the three adjoint-valued chiral multiplets of $\mathcal{N}=4$ super-Yang--Mills, it is clear that the above six relevant operators map, respectively, to
\begin{equation}\label{eqn:kabuto}
    \trace X^2 \,, \qquad \trace Y^2 \,, \qquad \trace XY \,, \qquad \trace XZ \,, \qquad \trace YZ \,, \qquad \trace Z^2 \,.
\end{equation}

\paragraph{\texorpdfstring{\uline{$W = u_3 + Q^2u_2 + Q^2u_1$ deformation}}{W = u₃ + Q²u₂ + Q²u₁}} Instead, further deforming by the relevant operator $Q^2u_1$ dissolves the Argyres--Douglas matter into a free $\mathcal{N}=2$ vector multiplet, and the remaining interacting part of the infrared is the same as that which arises from $SU(3)$ gauged theory coupled, without superpotential, to two adjoint-valued chiral multiplets. The is an endpoint in the $a=c$ landscape, as explained nearby to equation \eqref{eqn:2adjrel}.

\paragraph{\texorpdfstring{\uline{$W = u_3 + Q^2u_2 + \trace Y^2$ deformation: $a \neq c$}}{W = u₃ + Q²u₂ + TrY²}} The mass deformation removes one of the adjoint-valued chiral multiplets, which results in the same infrared physics as the gauging of a single $\mathcal{D}_2(SU(3))$ theory with a single adjoint-valued chiral multiplet. As determined in \cite{Kang:2021ccs}, such a theory does not flow in the infrared to an interacting SCFT with $a=c$.

\paragraph{\texorpdfstring{\uline{$W = u_3 + Q^2u_2 + \trace X Y$ deformation: $a \neq c$}}{W = u₃ + Q²u₂ + TrXY}} Similarly to the $\trace Y^2$ deformation, the $\trace XY$ deformation removes both chiral multiplets in the infrared. The resulting theory is therefore outside of the $a=c$ landscape, as noted in \cite{Kang:2021ccs}.

\paragraph{\texorpdfstring{\uline{$W = u_3 + Q^2u_2 + \trace \mu_1 Y$ deformation}}{W = u₃ + Q²u₂ + Trμ₁Y}} Next, we consider the deformation by $\trace \mu_1 Y$. This deformation does, in fact, lead to an interacting infrared SCFT with identical central charges. Anomaly cancellation and $a$-maximization yields the mixing coefficients and infrared R-charges of the chiral multiplets; we find
\begin{equation}
    \varepsilon_1=\frac{2\sqrt{7}-5}{9}\,,\qquad R[X]=\frac{\sqrt7-1}{3}\,,\qquad R[Y]=\frac{4\left(4-\sqrt7\right)}{9}\,,
\end{equation}
which gives us the central charges:
\begin{equation}
    a=c=\frac{5\left(7\sqrt7-10\right)}{144} \dim (G) \simeq 0.2958 \, \dim (G) \,.
\end{equation}
We can enumerate the relevant operator spectrum, and we find the following set of operators by which we can further deform:
\begin{align}
  \begin{gathered}
    u_1\,,\quad Q^2u_1\,,\quad u_1^2\,,\\ \trace X^2 \,, \quad \trace XY \,, \quad \trace Y^2 \,, \quad \trace X^3 \,, \quad \trace X^2 Y \,, \quad \trace XY^2 \,, \quad \trace Y^3 \,.
 \end{gathered}
\end{align}
It is straightforward, though again rather tedious, to discuss the subsequent deformations by each of these ten relevant operators.

\paragraph{\texorpdfstring{\uline{$W = u_3 + Q^2u_2 + \trace \mu_1 Y + u_1$ deformation: No SCFT}}{W = u₃ + Q²u₂ + Trμ₁Y + u₁}} We can see that the $\trace \mu_1 Y$ operator is a relevant operator after the $W = u_3 + Q^2u_2 + u_1$ sequence of deformations. Therefore, we would expect to get the same result by reversing the order of the final two deformations. In fact, the $W = u_3 + Q^2u_2 + u_1$ deformation leads to a point on the conformal manifold of $\mathcal{N}=4$ super-Yang--Mills, and the $\trace \mu_1 Y$ operator is mapped to the $\trace YZ$ operator; see equation \eqref{eqn:kabuto}. As discussed in Section \ref{sec:defsN4}, the $\trace YZ$ deformation of $\mathcal{N}=4$ does not lead to an infrared SCFT, and thus we do not consider this deformation any further.

\paragraph{\texorpdfstring{\uline{$W = u_3 + Q^2u_2 + \trace \mu_1 Y + Q^2u_1$ deformation: No SCFT}}{W = u₃ + Q²u₂ + Trμ₁Y + Q²u₁}} We start with the theory that exists in the infrared of $SU(3)$ gauge theory coupled to one copy of $\mathcal{D}_2(SU(3))$ and two adjoint-valued chiral multiplets, $X$ and $Y$, which is obtained after the first two deformations, as already discussed. The final two deformations would lead to an infrared R-symmetry specified by
\begin{equation}
    \varepsilon_1 = \frac{1}{3} \,, \qquad R[X] = 1 \,, \qquad R[Y] = 0 \,.
\end{equation}
Following the logic of Section \ref{sec:ADdeformations}, the $\varepsilon_1 = 1/3$ indicates that the $\mathcal{D}_2(SU(3))$ factor decouples into a free $\mathcal{N}=2$ vector multiplet, and the remaining infrared consists of the infrared of $SU(3)$ gauge theory with two adjoint-valued chiral multiplets deformed by $\trace X^2$. Since the latter is a mass deformation, we can see that no sector of the infrared is described by an $a=c$ SCFT.

There is another possibility here. Utilizing the duality of Section \ref{sec:ADdeformations}, we could expect that the infrared behavior after this section of deformations is one copy of a free $\mathcal{N}=2$ vector multiplet and the three-adjoint theory deformed by
\begin{equation}
    \trace \mu_1 Y \,.
\end{equation}
However, since this operator was removed from the spectrum when the $\mathcal{N}=2$ vector was removed, the interacting $a=c$ sector of the infrared is simply $\mathcal{N}=4$ super-Yang--Mills. Throughout this section, such an alternative possibility for the infrared behavior exists whenever there is a deformation involving operators like $\mu_i$, followed by the $Q^2u_i$ deformation, as the latter removes the former operators from the theory. While we will not mention this each time, it should be kept in mind as an alternative result of the flow.

\paragraph{\texorpdfstring{\uline{$W = u_3 + Q^2u_2 + \trace \mu_1 Y + u_1^2$ deformation}}{W = u₃ + Q²u₂ + Trμ₁Y + u₁²}} This sequence of four superpotential deformations breaks all the non-R $U(1)$ symmetries. The infrared R-charges and the central charges are
\begin{align}
\begin{gathered}
    \varepsilon_1=0\,,\qquad R[X]=\frac{1}{2}\,,\qquad R[Y]=\frac{2}{3}\,, \qquad
    a=c=\frac{75}{256} \dim (G) \simeq 0.2930 \, \dim (G) \,.
    \end{gathered}
\end{align}
There exist eight relevant operators of this $a=c$ SCFT:
\begin{align}
    u_1\,,\quad Q^2u_1\,,\quad \trace X^2Y^{n-2}\,,\quad \trace X^3\,,\quad \trace X^2Y\,,\quad\trace XY^2\,.
\end{align}
However, any deformation by one of the relevant operators immediately breaks the R-symmetry, and thus the deformed theory cannot flow to an interacting SCFT in the infrared. We reiterate that we are assuming that there are no emergent $U(1)$ symmetries in the infrared; if this assumption is violated then our conclusions about the lack of infrared SCFT can also be violated.

\paragraph{\texorpdfstring{\uline{$W = u_3 + Q^2u_2 + \trace \mu_1 Y + \trace X^2$ deformation: $a \neq c$}}{W = u₃ + Q²u₂ + Trμ₁Y + TrX²}} We find that the infrared R-charge of $Y$ vanishes: $R[Y] = 0$, and thus operators such as $\trace Y^2$ decouple along the flow into the infrared. As such, we do not obtain an IR SCFT with identical central charges.

\paragraph{\texorpdfstring{\uline{$W = u_3 + Q^2u_2 + \trace \mu_1 Y + \trace XY$ deformation: No SCFT}}{W = u₃ + Q²u₂ + Trμ₁Y + TrXY}} It is straightforward to determine that the mixing coefficient and the infrared R-charges of the chiral multiplets are
\begin{equation}
    \varepsilon_1 = -\frac{5}{3} \,, \qquad R[X] = -2 \,, \qquad R[Y] = 4 \,.
\end{equation}
It is clear that this does not provide a consistent infrared R-symmetry, and thus we do not have an IR SCFT with $a = c$.

\paragraph{\texorpdfstring{\uline{$W = u_3 + Q^2u_2 + \trace \mu_1 Y + \trace Y^2$ deformation: $a \neq c $}}{W = u₃ + Q²u₂ + Trμ₁Y + TrY²}} Solving the anomaly cancellation conditions for the infrared R-charge, we find that 
\begin{equation}
    R[X] = \frac{1}{4} \,.
\end{equation} 
Therefore, the $\trace X^2$ operator decouples when performing the $\trace Y^2$ deformation, and thus there is no $a=c$ SCFT in the infrared.

\paragraph{\texorpdfstring{\uline{$W = u_3 + Q^2u_2 + \trace \mu_1 Y + \trace X^3$ deformation: $a \neq c$}}{W = u₃ + Q²u₂ + Trμ₁Y + TrX³}} After this sequence of deformations, the naive application of $a$-maximization leads to the following mixing coefficients and infrared R-charges of the chiral multiplets:
\begin{equation}
    \varepsilon_1 = \frac{1}{9} \,, \qquad R[X] = \frac{2}{3} \,, \qquad R[Y] = \frac{4}{9} \,.
\end{equation}
As we can see, the $u_1$ operator has R-charge below the unitarity bound under this R-symmetry, and thus it decouples along the flow into the infrared. Removing this operator by adding the appropriate flip field will leave behind an infrared theory without $a=c$.

\paragraph{\texorpdfstring{\uline{$W = u_3 + Q^2u_2 + \trace \mu_1 Y + \trace X^2Y$ deformation: $a \neq c $}}{W = u₃ + Q²u₂ + Trμ₁Y + TrX²Y}} The superpotential together with gauge-anomaly cancellation fixes that the mixing coefficient and the R-charges of the chiral multiplets to be
\begin{equation}
    \varepsilon_1 = \frac{1}{3} \,, \qquad R[X] = 1 \,, \qquad R[Y] = 0 \,.
\end{equation}
The vanishing of the R-charge of the field $Y$ indicates that we do not flow to an $a = c$ SCFT in the infrared; in particular, operators like $\trace Y^2$ decouple along the flow.

\paragraph{\texorpdfstring{\uline{$W = u_3 + Q^2u_2 + \trace \mu_1 Y + \trace XY^2$ deformation}}{W = u₃ + Q²u₂ + Trμ₁Y + TrXY²}} We find that the mixing coefficients and infrared R-charges after this sequence of superpotential deformations are as follows:
\begin{equation}
    \varepsilon_1=-\frac{1}{15}\,,\qquad R[X]=\frac{2}{5}\,,\qquad R[Y]=\frac{4}{5} \,.
\end{equation}
Under this R-symmetry, there are no operators belonging to the chiral ring whose R-charge lies below the unitarity bound, and thus we expect to obtain an interacting infrared SCFT with $a = c$. It is easy to see that the central charges are
\begin{equation}\label{eqn:squirtle}
    a = c = \frac{27}{100} \dim (G) \simeq 0.27 \, \dim (G) \,.
\end{equation}
The resulting SCFT has eight relevant operators belonging to the chiral ring:
\begin{equation}
    \begin{gathered}
    u_1\,,\quad Q^2u_1\,, \quad \trace X^2 \,, \quad \trace XY \,, \quad \trace Y^2 \,, \quad \trace X^3\,,\quad \trace X^2Y\,,\quad \left(\trace X^2\right)^2 \,,
  \end{gathered}
\end{equation}
however, since the only continuous Abelian symmetry is the R-symmetry, we expect that any further deformation does not lead to superconformal symmetry in the infrared.

\paragraph{\texorpdfstring{\uline{$W = u_3 + Q^2u_2 + \trace \mu_1 Y + \trace Y^3$ deformation}}{W = u₃ + Q²u₂ + Trμ₁Y + TrY³}} As discussed already for the $u_1^2$ deformation, the further deformation by $\trace Y^3$, breaks every $U(1)$ symmetry other than the IR R-symmetry. The R-symmetry is fixed by anomaly cancellation to be
\begin{align}
    \varepsilon_1=0\,,\qquad R[X]=\frac{1}{2}\,,\qquad R[Y]=\frac{2}{3}\,,
\end{align}
which leads to the central charges
\begin{align}\label{eqn:charmander}
    a = c = \frac{75}{256} \dim (G) \simeq 0.2930 \, \dim (G) \,.
\end{align}
There are nine relevant operators
\begin{align}
  \begin{gathered}
    u_1\,,\quad Q^2u_1\,,\quad \trace \mu_1X \,, \\\trace X^2 \,, \quad \trace XY \,, \quad \trace Y^2 \,, \quad \trace X^3\,,\quad \trace X^2Y\,,\quad \trace XY^2 \,.
  \end{gathered}
\end{align}
Since the superpotential already broke every $U(1)$ flavor symmetry other than the R-symmetry, further deformation will break the R-symmetry and an interacting SCFT cannot exist at on the IR fixed point unless some $U(1)$ symmetry is emergent.

\paragraph{\texorpdfstring{\uline{$W = u_3 + Q^2u_2 + \trace Y^3$ deformation}}{W = u₃ + Q²u₂ + TrY³}} Consider now the deformation of the SCFT associated to the gauge theory in equation \eqref{eqn:D2p2} by $\trace Y^3$. This deformation does lead to an interacting infrared SCFT with identical central charges. Anomaly cancellation and $a$-maximization yields the mixing coefficients and infrared R-charges of the chiral multiplets; we find
\begin{equation}
    \varepsilon_1=\frac{2\sqrt{79}-19}{27}\,,\qquad R[X]=\frac{23 - \sqrt{79}}{27}\,,\qquad R[Y]=\frac{2}{3}\,,
\end{equation}
which gives us the central charges:
\begin{equation} \label{eq:2with2adjY3}
    a=c=\frac{442 + 79\sqrt{79}}{3888} \dim (G) \simeq 0.2943 \, \dim (G) \,.
\end{equation}
Enumerating the relevant operator spectrum we find the following set of operators by which we can further deform:
\begin{align}\label{eqn:diglett}
  \begin{gathered}
    u_1\,,\quad Q^2u_1\,,\quad \trace \mu_1 X \,, \quad \trace \mu_1 Y \,,\\ \trace X^2 \,, \quad \trace XY \,, \quad \trace Y^2 \,, \quad \trace X^3 \,, \quad \trace X^2 Y \,.
 \end{gathered}
\end{align}
We check the spectrum of relevant operators by computing the (reduced) superconformal index. It is given by
\begin{align}
\begin{split}
    \widehat{I}=&t^{3.13596}+t^{3.40787}+t^{3.56798}+t^4-t^{4.13596}\chi_{\bm{2}}(y)-t^{4.56798}\chi_{\bm{2}}(y)+t^{4.86404}\\
    &+t^{4.70394}-t^5\chi_{\bm{2}}(y)+t^{5.13596}+t^{5.29606}+t^{5.72809}-t^6+O(t^{6.13596}) \,.
    \end{split}
\end{align}
The first term $t^{3.13596}$ counts the operator $\trace X^2$. The next $t^{3.40787}$ counts the Coulomb branch operator $u_1$ of $\mathcal{D}_2(SU(3))$. The term $t^{3.56798}$ counts the operator $\trace XY$. The operator $\trace Y^2$ contributes by $t^4$. $t^{4.70394}$ counts the operator $\trace X^3$. $t^{4.86404}$ corresponds to the $Q^2u_1$ coming from the $\mathcal{D}_2(SU(3))$ theory. The next term, $t^{5.13596}$, counts the operator $\trace X^2Y$. Operators $\trace \mu X$ and $\trace \mu Y$ contribute to the index by $t^{5.29606}$ and $t^{5.72809}$. There is no marginal operator but one $U(1)$ flavor current that contributes to $t^6$ order by $-1$. So we found all the relevant operators in equation \eqref{eqn:diglett}. We now survey the theories obtained via the subsequent deformations by each of the nine relevant operators in equation \eqref{eqn:diglett}.

\paragraph{\texorpdfstring{\uline{$W = u_3 + Q^2u_2 + \trace Y^3 + u_1$ deformation: $\CN=4$ SYM}}{W = u₃ + Q²u₂ + TrY³ + u₁}} The result of this deformation is the same as that of $W = u_3 + Q^2u_2 + u_1 + \trace Y^3$; since the final $\trace Y^3$ deformation is in fact marginal, the resulting infrared SCFT lies on the same conformal manifold as the deformation $W = u_3 + Q^2u_2 + u_1$. This is nothing other than $\mathcal{N}=4$ super-Yang--Mills, whose deformations have been discussed in Section \ref{sec:defsN4}.

\paragraph{\texorpdfstring{\uline{$W = u_3 + Q^2u_2 + \trace Y^3 + Q^2u_1$ deformation: two adjoint theory}}{W = u₃ + Q²u₂ + TrY³ + Q²u₁}} The final $Q^2u_1$ deformation has the effect of dissolving the final $\mathcal{D}_2(SU(3))$ factor into a decoupled $\mathcal{N}=2$ vector multiplet, following Section \ref{sec:ADdeformations}. The remaining part of the infrared is then described by $SU(3)$ gauge theory with two adjoint-valued chiral multiplets deformed by $\trace Y^3$. As explained in Section \ref{sec:defsN4}, this is not an SCFT.

\paragraph{\texorpdfstring{\uline{$W = u_3 + Q^2u_2 + \trace Y^3 + \trace \mu_1 X$ deformation: $a \neq c$}}{W = u₃ + Q²u₂ + TrY³ + Trμ₁X}} After this sequence of deformations the operator $u_1$ has charge less than the unitarity bound under the infrared R-symmetry, and thus it should be decoupled along the flow into the infrared. Thus, the infrared is not described by an $a=c$ SCFT.

\paragraph{\texorpdfstring{\uline{$W = u_3 + Q^2u_2 + \trace Y^3 + \trace \mu_1 Y$ deformation}}{W = u₃ + Q²u₂ + TrY³ + Trμ₁Y}} This deformation leads to the same fixed point as performing the final two deformations in the opposite order. See equation \eqref{eqn:charmander} for details of the resulting $a=c$ SCFT.

\paragraph{\texorpdfstring{\uline{$W = u_3 + Q^2u_2 + \trace Y^3 + \trace X^2$ deformation: No SCFT}}{W = u₃ + Q²u₂ + TrY³ + TrX²}} We find that the infrared R-charges are
\begin{equation}
    R[X] = 1 \,, \qquad R[Y] = \frac{2}{3} \,, \qquad R[\mu_1] = -\frac{2}{3} \,.
\end{equation}
The negativity of $R[\mu_1]$ indicates that this is not a consistent superconformal R-charge, and we see that various operators, such as $\trace \mu_1 Y$, must decouple along the flow into the infrared. Thus, this deformation does not lead to an $a=c$ fixed point. This is unsurprising, since the $\trace X^2$ deformation gives a mass to one of the two chiral multiplets, and thus we arrive at the same fixed point as (a deformation of) the gauging of one $\mathcal{D}_2(SU(3))$ together with one adjoint-valued chiral multiplet. In \cite{Kang:2021ccs} it was shown that this does not flow to an SCFT.

\paragraph{\texorpdfstring{\uline{$W = u_3 + Q^2u_2 + \trace Y^3 + \trace XY$ deformation: $ a\neq c$}}{W = u₃ + Q²u₂ + TrY³ + TrXY}} This sequence of deformations fixes the R-charges of the chiral multiplets to be
\begin{equation}
    R[X] = \frac{1}{3} \,, \qquad R[Y] = \frac{2}{3} \,.
\end{equation}
Therefore, we can see that the $\trace X^2$ operator in equation \eqref{eqn:diglett} decouples along the flow triggered by the $\trace XY$ deformation. Thus, we do not obtain an interacting SCFT with $a=c$ in the infrared. 

\paragraph{\texorpdfstring{\uline{$W = u_3 + Q^2u_2 + \trace Y^3 + \trace Y^2$ deformation: No R}}{W = u₃ + Q²u₂ + TrY³ + TrY²}} We can see that this sequence of superpotential deformations does not admit a consistent assignment of infrared R-charges. As such, we do not expect the existence of superconformal symmetry in the infrared.

\paragraph{\texorpdfstring{\uline{$W = u_3 + Q^2u_2 + \trace Y^3 + \trace X^3$ deformation: $\CN=4$ SYM}}{W = u₃ + Q²u₂ + TrY³ + TrX³}} Studying the infrared R-charges and the operator spectrum of the SCFT obtained after this sequence of deformations, we can see that the resulting theory is identical to a point on the conformal manifold of $\mathcal{N}=4$ super-Yang--Mills. The subsequent network of deformations are explored in Section \ref{sec:defsN4}.

\paragraph{\texorpdfstring{\uline{$W = u_3 + Q^2u_2 + \trace Y^3 + \trace X^2Y$ deformation: $\CN=4$ SYN}}{W = u₃ + Q²u₂ + TrY³ + TrX²Y}} After deformation, the mixing coefficient and the infrared R-charges are
\begin{equation}\label{eqn:slowpoke}
    \varepsilon_1 = -\frac{1}{3} \,, \qquad R[X] = R[Y] = \frac{2}{3} \,.
\end{equation}
Therefore, as with the $\trace X^3$ deformation just discussed, this sequence of deformations again triggers a flow to a point on the conformal manifold of $\mathcal{N}=4$ super-Yang--Mills, and the further deformations are explained in Section \ref{sec:defsN4}.

\paragraph{\texorpdfstring{\uline{$W = u_3 + Q^2u_2 + \trace XY^2$ deformation}}{W = u₃ + Q²u₂ + TrXY²}} Finally, we consider the $\trace XY^2$ deformation of the SCFT associated to the gauge theory in equation \eqref{eqn:D2p2}. After this deformation, we again obtain an $a=c$ SCFT in the infrared. Anomaly cancellation and $a$-maximization gives the mixing coefficients and infrared R-charges of the chiral multiplets, as usual. In the end, we find
\begin{equation}
    \varepsilon_1=\frac{11-2\sqrt{19}}{15}\,,\qquad R[X]=\frac{2(8 - \sqrt{19})}{15}\,,\qquad R[Y]=\frac{7+\sqrt{19}}{15}\,.
\end{equation}
From these results, we determine that the identical central charges are
\begin{equation}
    a=c=\frac{28 + 19\sqrt{19}}{400} \dim (G) \simeq 0.2770 \, \dim (G) \,.
\end{equation}
The following chiral ring operators are relevant operators of the infrared SCFT:
\begin{align}
  \begin{gathered}
    u_1\,,\quad Q^2u_1\,,\quad \trace \mu_1 X \,, \quad \trace \mu_1 Y \,,\\ \trace X^2 \,, \quad \trace XY \,, \quad \trace Y^2 \,, \quad \trace X^3 \,, \quad \left(\trace X^2\right)^2 \,.
 \end{gathered}
\end{align}
For each of these nine relevant operators, we consider the further superpotential deformation and flow into the infrared.

\paragraph{\texorpdfstring{\uline{$W = u_3 + Q^2u_2 + \trace XY^2 + u_1$ deformation: $\CN=4$ SYM}}{W = u₃ + Q²u₂ + TrXY² + u₁}} Similarly to the $W = u_3 + Q^2u_2 + \trace Y^3 + u_1$ sequence of deformations, we can swap the order of the final two deformations and find that we land on the same conformal manifold as the sequence $W = u_3 + Q^2u_2 + u_1$ since the $\trace XY^2$ operator is then marginal. Thus, we obtain $\mathcal{N}=4$ super-Yang--Mills, for which the further deformations have been enumerated in Section \ref{sec:defsN4}.

\paragraph{\texorpdfstring{\uline{$W = u_3 + Q^2u_2 + \trace XY^2 + Q^2u_1$ deformation}}{W = u₃ + Q²u₂ + TrXY² + Q²u₁}} Utilizing the dualities in Section \ref{sec:ADdeformations}, the final $Q^2u_1$ deformation leads to an infrared which is a product of an $\mathcal{N}=2$ free vector multiplet and the infrared of $SU(3)$ with two adjoint-valued chiral multiplets, $X$ and $Y$, deformed by $\trace XY^2$. In Section \ref{sec:defsN4}, it was noted that the latter sector is not an SCFT.

\paragraph{\texorpdfstring{\uline{$W = u_3 + Q^2u_2 + \trace XY^2 + \trace \mu_1 X$ deformation: No SCFT}}{W = u₃ + Q²u₂ + TrXY² + Trμ₁X}} We find that the infrared R-charges of the relevant fundamental fields are
\begin{equation}
    R[\mu_1] = 2 \,, \qquad R[X] = 0 \,, \qquad R[Y] = 1 \,.
\end{equation}
The vanishing R-charge of the adjoint-valued chiral multiplet indicates the decoupling of operators, such as $\operatorname{Tr}X^2$, along the flow into the infrared. As such, this sequence of deformations does not lead to an $a=c$ SCFT.

\paragraph{\texorpdfstring{\uline{$W = u_3 + Q^2u_2 + \trace XY^2 + \trace \mu_1 Y$ deformation}}{W = u₃ + Q²u₂ + TrXY² + Trμ₁Y}} The IR physics is the same as that obtained from the deformation $W = u_3 + Q^2u_2 + \trace \mu_1 Y + \trace XY^2$, where the final two deformations are applied in the reverse order; see around equation \eqref{eqn:squirtle} for the details.

\paragraph{\texorpdfstring{\uline{$W = u_3 + Q^2u_2 + \trace XY^2 + \trace X^2$ deformation}}{W = u₃ + Q²u₂ + TrXY² + TrX²}}
After deformation, the R-charges of some of the pertinent operators are fixed to be as follows:
\begin{equation}
    R[X] = 1 \,, \qquad R[Y] = 
    \frac{1}{2} \,, \qquad R[\mu_1] = 0 \,.
\end{equation}
The operator $\trace \mu_1 Y$ belonging to the theory after the $W = u_3 + Q^2u_2 + \trace X Y^2$ decouples under the subsequent $\trace X^2$ deformation, and thus we do not obtain an interacting infrared SCFT with $a = c$. 

\paragraph{\texorpdfstring{\uline{$W = u_3 + Q^2u_2 + \trace XY^2 + \trace XY$ deformation: No SCFT}}{W = u₃ + Q²u₂ + TrXY² + TrXY}}
Consistency of the deformation leads to the following infrared R-charges for the chiral multiplets:
\begin{equation}
    R[X] = 2 \,, \qquad R[Y] = 0 \,.
\end{equation}
Since the R-charge of $Y$ vanishes, then certain operators, such as $\trace Y^2$, decouple along the flow into the infrared triggered by the $\trace XY$ deformation. As such, we do not have an interacting $a = c$ SCFT in the infrared.

\paragraph{\texorpdfstring{\uline{$W = u_3 + Q^2u_2 + \trace XY^2 + \trace Y^2$ deformation: No SCFT}}{W = u₃ + Q²u₂ + TrXY² + TrY²}}
Anomaly cancellation yields the following infrared R-charges for the chiral multiplets:
\begin{equation}
    R[X] = 0 \,, \qquad R[Y] = 1 \,.
\end{equation}
As we can see, the R-charge of the adjoint-valued chiral multiplet $X$ vanishes after the $\trace Y^2$ deformation, leading to operators decoupling from the theory along the flow. Therefore we do not obtain an SCFT with $a = c$ in the infrared.

\paragraph{\texorpdfstring{\uline{$W = u_3 + Q^2u_2 + \trace XY^2 + \trace X^3$ deformation: $\CN=4$ SYM}}{W = u₃ + Q²u₂ + TrXY² + TrX³}} We see that we can swap the order of the final two deformations and thus the infrared SCFT is the same as that obtained via the deformation
\begin{equation}
    W = u_3 + Q^2u_2 + \trace Y^3 + \trace X^2Y \,.
\end{equation}
As explained around equation \eqref{eqn:slowpoke}, this is nothing other than a point on the conformal manifold of $\mathcal{N}=4$ SYM, for which the further deformations are given in Section \ref{sec:defsN4}.

\paragraph{\texorpdfstring{\uline{$W = u_3 + Q^2u_2 + \trace XY^2 + (\trace X^2)^2$ deformation}}{W = u₃ + Q²u₂ + TrXY² + (TrX²)²}} The R-charges of the chiral multiplets and the mixing coefficient are fixed by gauge-anomaly cancellation; we find
\begin{equation}
    \varepsilon_1 = - \frac{1}{6} \,, \qquad R[X] = \frac{1}{2} \,, \qquad R[Y] = \frac{3}{4} \,.
\end{equation}
None of the chiral ring operators cross the unitarity bound during this flow, and thus we obtain an infrared SCFT with $a = c$. The central charges are
\begin{equation}
    a = c = \frac{567}{2048} \operatorname{dim}(G) \simeq 0.2769 \, \dim (G) \,.
\end{equation}
Since this theory only has a single $U(1)$ global symmetry, the superconformal R-symmetry, any further deformation is expected, modulo the possibility of emergent Abelian symmetries, to not give rise to an SCFT. 

\paragraph{\texorpdfstring{\uline{$W = \trace \mu_3 X$ deformation}}{W = Trμ₃X}} After this time, where we have exhaustively enumerated the subsequent superpotential deformations of the $W = u_3$ deformation of the infrared SCFT arising from the gauge theory in Figure \ref{fig:bigboy}, we now turn to the 
\begin{equation}
    W = \trace \mu_3 X \,,
\end{equation}
deformation of Figure \ref{fig:bigboy}, and all subsequent deformations. As usual, $a$-maximization and anomaly cancellation fixes the mixing coefficients and the infrared R-charge of the adjoint-valued chiral. We find
\begin{equation}
    \varepsilon_1 = \varepsilon_2 = -\frac{12-\sqrt{109}}{21} \,, \qquad \varepsilon_3 = \frac{2\sqrt{109} - 17}{63} \,, \qquad R[X] = \frac{4(19 - \sqrt{109})}{63} \,.
\end{equation}
Thus, the central charges of the deformed SCFT are
\begin{equation}
    a = c = \frac{323 + 109\sqrt{109}}{3528} \dim(G) \simeq 0.4141 \, \dim (G) \,.
\end{equation}
There are twelve relevant operators belonging to the theory:
\begin{align}
    u_{\alpha}\,,\quad Q^2u_{\alpha}\,,\quad \trace X^2\,,\quad \trace X^3\,,\quad \trace \mu_{1,2} X\,,\quad u_3\trace X^2\,,\quad u_3^2\,. 
\end{align}
We analyze the theory deformed by each of the relevant operators.

\paragraph{\texorpdfstring{\uline{$W = \trace \mu_3 X + u_2$ deformation}}{W = Trμ₃X + u₂}} Swapping the order of the deformations, we see that this deformation leads to the same SCFT as that obtained by the conformal gauge theory written in equation \eqref{eqn:2D2G}, together with the deformations
\begin{equation}
    \trace Y^3 + \trace \mu_2 X \,.
\end{equation}
As explained in equation \eqref{eqn:22marg}, these operators are both marginal, and thus we land on a point on the conformal manifold of the $W = u_3$ deformation of the original gauge theory in Figure \ref{fig:bigboy}. The subsequent deformations have been discussed subsequent to equation \eqref{eqn:2D2G}.

\paragraph{\texorpdfstring{\uline{$W = \trace \mu_3 X + u_3$ deformation: $a \neq c$}}{W = Trμ₃X + u₃}} The infrared R-charge is specified via the following parameters:
\begin{equation}
    \varepsilon_1 = \varepsilon_2 = -\frac{2}{3} \,, \qquad \varepsilon_3 = -\frac{1}{3} \,, \qquad R[X] = \frac{4}{3} \,.
\end{equation}
As we can see, $R[\mu_1] = R[\mu_2] = 0$, and thus operators such as $\trace \mu_1 \mu_2$ decouple along the flow into the infrared triggered by the $u_3$ deformation. Thus, we obtain no $a=c$ SCFT.

\paragraph{\texorpdfstring{\uline{$W = \trace \mu_3 X + Q^2u_2$ deformation: $a \neq c$}}{W = Trμ₃X + Q²u₂}} Performing the deformations in the opposite order, we first see by the dualities in Section \ref{sec:ADdeformations} that the $W=Q^2u_2$ deformation leads to an infrared which is a product of the gauging of two $\mathcal{D}_3(SU(3))$s with one adjoint-valued chiral multiplet and a free $\mathcal{N}=2$ vector multiplet. The $\trace \mu_3 X$ operator remains as a relevant operator of the  interacting $a=c$ part of the infrared, as we can see from equation \eqref{eqn:mudkip}. The result of the $\trace \mu_3 X$ deformation of this theory has been discussed around equation \eqref{eqn:torchic}, where it was observed that the infrared is not described by an $a=c$ SCFT.

\paragraph{\texorpdfstring{\uline{$W = \trace \mu_3 X + Q^2u_3$ deformation: IR free}}{W = Trμ₃X + Q²u₃}} 
Solving the gauge-anomaly cancellation conditions for the mixing coefficients and the infrared R-charge, we find that the R-symmetry is fixed via
\begin{equation}\label{eqn:banette}
    \varepsilon_1 = \varepsilon_2 = \varepsilon_3 = \frac{1}{3} \,, \qquad R[X] = 0 \,.
\end{equation}
We can see that a slew of operators decouple along the flow into the infrared; for example, we can see that
\begin{equation}
    R[u_\alpha] = 0 \,.
\end{equation}
We can flip each of these operators, along with $\trace X^2$ and $\trace X^3$ which also decouple along the flow. 
As discussed in Section \ref{sec:ADdeformations}, the $\varepsilon = 1/3$ indicates that the associated $\mathcal{D}_2(SU(3))$ factor decomposes into an $\mathcal{N}=2$ vector multiplet. In fact, the R-charges in equation \eqref{eqn:banette} are the same as those obtained from the sequence of deformations:
\begin{equation}
    W = Q^2u_3 + Q^2u_2 + Q^2u_1 \,.
\end{equation}
By the logic of Section \ref{sec:ADdeformations}, the infrared is three $\mathcal{N}=2$ vector multiplets, together with the infrared of $SU(3)$ gauge theory with one adjoint-valued chiral multiplet. The latter is known to flow to two $\mathcal{N}=2$ vector multiplets, and thus we expect the infrared to be five free $\mathcal{N}=2$ vector multiplets.

\paragraph{\texorpdfstring{\uline{$W = \trace \mu_3 X + \trace X^2$ deformation}}{W = Trμ₃X + TrX²}} The mixing parameters and the R-charge of the chiral multiplet are fixed to be
\begin{equation}
    \varepsilon_1 = \varepsilon_2 = -\frac{5}{12} \,, \qquad \varepsilon_3 = -\frac{1}{6} \,, \qquad R[X] = 1 \,.
\end{equation}
Accordingly, the central charges of the infrared SCFT after deformation are
\begin{equation}
    a = c = \frac{243}{1024} \operatorname{dim}(G) \simeq 0.2373 \, \dim (G) \,.
\end{equation}
This $a=c$ SCFT has seven relevant operators, which can be written as
\begin{equation}
    u_3 \,, \quad Q^2u_\alpha \,, \quad \trace \mu_1\mu_2 \,, \quad \trace \mu_1\mu_3 \,, \quad\trace \mu_2\mu_3 \,.
\end{equation}
We study each of these further deformations in turn.

\paragraph{\texorpdfstring{\uline{$W = \trace \mu_3 X + \trace X^2 + u_3$ deformation: No R}}{W = Trμ₃X + TrX² + u₃}} This deformation does not admit a consistent superconformal R-symmetry, and thus we do not expect a flow to an SCFT in the infrared.

\paragraph{\texorpdfstring{\uline{$W = \trace \mu_3 X + \trace X^2 + Q^2u_2$ deformation: $a \neq c$}}{W = Trμ₃X + TrX² + Q²u₂}} Since the $\trace \mu_3 X + \trace X^2$ and $Q^2u_2$ building blocks do not involve any of the same fields, we can perform the deformations in the opposite order and take advantage of the dualities in Section \ref{sec:ADdeformations}. The $Q^2u_2$ deformation leads to an infrared which is a product of the SCFT in equation \eqref{eqn:2D2G1adj} and a free $\mathcal{N}=2$ vector. The operators $\trace \mu_3 X$ and $\trace X^2$ are relevant operators of the interacting $a=c$ sector of the infrared, however, as already discussed, each deformation individually leads to decoupling, and thus does not lead to an $a=c$ SCFT in the infrared.

\paragraph{\texorpdfstring{\uline{$W = \trace \mu_3 X + \trace X^2 + Q^2u_3$ deformation: No R}}{W = Trμ₃X + TrX² + Q²u₃}} Attempting to solve the gauge-anomaly cancellation conditions as well as the R-charge constraints coming from the superpotential terms, we see that there is no consistent assignment of R-charges. Thus, we do not expect an interacting SCFT with $a = c$ after this sequence of deformations.

\paragraph{\texorpdfstring{\uline{$W = \trace \mu_3 X + \trace X^2 + \trace \mu_1 \mu_2$ deformation: No R}}{W = Trμ₃X + TrX² + Trμ₂μ₃}} This deformation also does not admit a consistent superconformal R-symmetry built out of the UV Abelian symmetries. As such, we do not expect an infrared SCFT.

\paragraph{\texorpdfstring{\uline{$W = \trace \mu_3 X + \trace X^2 + \trace \mu_2 \mu_3$ deformation: two adjoint theory}}{W = Trμ₃X + TrX² + Trμ₂μ₃}} The infrared R-symmetry is fixed by the following mixing coefficients and R-charge of the chiral multiplet:
\begin{equation}\label{eqn:dusklops}
    \varepsilon_1 = - \frac{2}{3} \,, \qquad \varepsilon_2 = \varepsilon_3 = - \frac{1}{6} \,, \qquad R[X] = 1 \,.
\end{equation}
We can see that none of the operators belonging to the chiral ring decouple under these charges, and thus we obtain an $a=c$ SCFT in the infrared with 
\begin{equation}
    a = c = \frac{27}{128} \operatorname{dim}(G) \simeq 0.2109 \, \dim (G) \,.
\end{equation}
This is the same central charges as the $SU(3)$ gauge theory with two adjoint-valued chiral multiplets. While it is not a priori obvious that this sequence of deformations should lead to the two-adjoint theory, we can provide a cross-check by computing the superconformal index; we find that the index of the deformation agrees with that of the two-adjoint theory at least up to order nine. Therefore, we conclude that this deformation leads to the two-adjoint theory, which, as discussed in Section \ref{sec:defsN4}, is an endpoint of the $a=c$ landscape.

\paragraph{\texorpdfstring{\uline{$W = \trace \mu_3 X + \trace X^3$ deformation}}{W = Trμ₃X + TrX³}} After this sequence of deformations, we find that 
\begin{equation}
    \varepsilon_1 = \varepsilon_2 = -\frac{1}{6} \,, \qquad \varepsilon_3 = 0 \,, \qquad R[X] = \frac{2}{3} \,.
\end{equation}
After verifying that no operators decouple along the flow into the infrared, we see that the resulting SCFT has identical central charges:
\begin{equation}\label{eqn:pineapple}
    a = c = \frac{51}{128} \operatorname{dim}(G) \simeq 0.3984 \, \dim (G) \,.
\end{equation}
The SCFT has the following relevant operators:
\begin{equation}
    u_\alpha \,, \qquad Q^2u_\alpha \,, \qquad \trace X^2 \,, \qquad \trace \mu_1 X \,, \qquad \trace \mu_2 X\,.
\end{equation}
We consider the subsequent deformations by each of these relevant operators in turn. 

\paragraph{\texorpdfstring{\uline{$W = \trace \mu_3 X + \trace X^3 + u_2$ deformation}}{W = Trμ₃X + TrX³ + u₂}} Gauge-anomaly cancellation together with the constraints from the superpotentials reveals that the mixing coefficients and the R-charge of the chiral multiplet are
\begin{equation}
    \varepsilon_1 = \varepsilon_3 = 0 \,, \qquad \varepsilon_2 = -\frac{1}{3} \,, \qquad R[X] = \frac{2}{3} \,.
\end{equation}
Since there is no decoupling that we can observe, the infrared $a=c$ SCFT has central charges:
\begin{equation}
    a = c = \frac{3}{8} \operatorname{dim}(G) \simeq 0.3750 \, \dim (G) \,.
\end{equation}
We note that this is the same central charge as appeared in equation \eqref{eqn:mewtwo} for the $W = u_3$ deformation of the original theory in Figure \ref{fig:bigboy}. This is expected from the dualities in Section \ref{sec:ADdeformations}; swapping the order of the deformations, the $u_2$ deformation first leads to the theory in equation \eqref{eqn:2D2G}, which is further deformed by
\begin{equation}
    \trace Y^3 + \trace \mu_3 X + \trace X^3 \,.
\end{equation}
Since each of these operators are marginal (see equation \eqref{eqn:22marg}), then we land on the same conformal manifold as the $W = u_3$ deformation. This superpotential deformations of this fixed point have been enumerated subsequent to equation \eqref{eqn:2D2G}.

\paragraph{\texorpdfstring{\uline{$W = \trace \mu_3 X + \trace X^3 + u_3$ deformation}}{W = Trμ₃X + TrX³ + u₃}} There is no consistent infrared R-symmetry built out of the Abelian symmetries evident in the ultraviolet. Thus, we do not expect to obtain an $a=c$ SCFT from this deformation.

\paragraph{\texorpdfstring{\uline{$W = \trace \mu_3 X + \trace X^3 + Q^2u_2$ deformation}}{W = Trμ₃X + TrX³ + Q²u₂}} Performing the $Q^2u_2$ deformation first, we find that one of the $\mathcal{D}_2(SU(3))$ factors decouples into an $\mathcal{N}=2$ vector multiplet. The superpotential deformations of the $a=c$ factor of the infrared orthogonal to the $\mathcal{N}=2$ vector multiplet have been discussed under the
\begin{equation}
    W = Q^2u_3 + \trace Y^2 \,,
\end{equation}
deformation. As we can see following equation \eqref{eqn:2D2G1adj}, the subsequent deformations by either $\trace \mu_2 X$ or $\trace X^3 + \trace \mu_2 X$ do not lead to an interacting SCFT fixed point. Thus, the sequence of deformations analyzed in this paragraph also does not lead to an interacting $a=c$ SCFT.

\paragraph{\texorpdfstring{\uline{$W = \trace \mu_3 X + \trace X^3 + Q^2u_3$ deformation: No R}}{W = Trμ₃X + TrX³ + Q²u₃}} These superpotential deformations do not admit a consistent superconformal R-symmetry, unless there exists some emergent Abelian symmetry along the flow into the infrared. Therefore, we do not expect a flow to an interacting infrared SCFT with $a = c$.

\paragraph{\texorpdfstring{\uline{$W = \trace \mu_3 X + \trace X^3 + \trace X^2$ deformation: No R}}{W = Trμ₃X + TrX³ + TrX²}} Without decoupling, a consistent infrared R-charge is not obtained from this sequence of deformations. Thus, this deformation does not lead to an infrared SCFT with $a = c$.

\paragraph{\texorpdfstring{\uline{$W = \trace \mu_3 X + \trace X^3 + \trace \mu_2 X$ deformation}}{W = Trμ₃X + TrX³ + Trμ₂X}} We find that
\begin{equation}
    \varepsilon_1 = -\frac{1}{3} \,, \qquad \varepsilon_2 = \varepsilon_3 = 0 \,, \qquad R[X] = \frac{2}{3} \,.
\end{equation}
The central charges are identical:
\begin{equation}\label{eqn:raichu}
    a = c = \frac{3}{8} \operatorname{dim}(G) \simeq 0.3750 \, \dim (G)\,.
\end{equation}
In fact, we can see that the infrared SCFT is nothing other than 
\begin{equation}
    \begin{aligned}
    \begin{tikzpicture}
      \node[gaugeN1] (s0) {$SU(3)$};
      \node[d2] (c2) [left=0.6cm of s0] {$\mathcal{D}_{2}(SU(3))$};
      \node[d2] (c3) [right=0.6cm of s0] {$\mathcal{D}_{2}(SU(3))$};
      \draw (s0.west) -- (c2.east);
      \draw (s0.east) -- (c3.west);
       \draw[dashed, ->] (s0) to[out=130, in=410, looseness=4] (s0);
      \draw[dashed, ->] (s0) to[out=146, in=394, looseness=5] (s0);
    \end{tikzpicture} \,,
\end{aligned}
\end{equation}
which is obtained by the deformation of the original theory in Figure \ref{fig:bigboy}:
\begin{equation}
    W = u_3 \,,
\end{equation}
and whose further superpotential deformations have been discussed around equation \eqref{eqn:2D2G}. This alternative description can be explained with recourse to the dualities in Section \ref{sec:ADdeformations}. Since $\varepsilon_1$ is $-1/3$, the infrared is the same as if we removed the first $\mathcal{D}_2(SU(3))$ factor and replaced it by an adjoint-valued chiral multiplet with superpotential $\trace Y^3$. Thus, we get equation \eqref{eqn:2D2G} deformed by
\begin{equation}
    \trace Y^3 + \trace X^3 + \trace \mu_1 X + \trace \mu_2 X \,,
\end{equation}
however, each of these four operators is marginal. Thus, as expected, we land on the same conformal manifold as the $W = u_3$ deformation.

\paragraph{\texorpdfstring{\uline{$W = \trace \mu_3 X + \trace \mu_2X$ deformation}}{W = Trμ₃X + Trμ₂X}} Performing $a$-maximization determines the mixing coefficients and the R-charge of the chiral multiplet after deformation. We find
\begin{equation}
    \varepsilon_1 = - \frac{7 - 2\sqrt{7}}{9} \,, \qquad \varepsilon_2 = \varepsilon_3 = \frac{\sqrt{7}-2}{9} \,, \qquad R[X] = \frac{2(5 - \sqrt{7})}{9} \,.
\end{equation}
This flow leads to a theory with identical central charges, which are
\begin{equation}
    a = c = \frac{10 + 7\sqrt{7}}{72} \operatorname{dim}(G) \simeq 0.3961 \, \dim (G) \,.
\end{equation}
The infrared SCFT possesses a variety of relevant operators with which we can further deform the theory:
\begin{equation}
  \begin{gathered}
    u_\alpha \,, \quad Q^2u_\alpha \,, \quad u_2^2 \,, \quad u_2u_3 \,, \quad u_3^2 \,, \quad \trace \mu_1 X \,, \\ \trace X^2 \,, \quad \trace X^3 \,, \quad u_2 \trace X^2 \,, \quad u_3 \trace X^2 \,.
  \end{gathered}
\end{equation}
We now consider each of these deformations in turn. Since, by symmetry considerations, it is clear that we can swap the order of the $W = \trace \mu_3 X + \trace \mu_2X$ deformations, we will not separately consider further deformations that are symmetric under the interchange of the second and third $\mathcal{D}_2(SU(3))$ factors.

\paragraph{\texorpdfstring{\uline{$W = \trace \mu_3 X + \trace \mu_2X + u_1$ deformation}}{W = Trμ₃X + Trμ₂X + u₁}} After this sequence of deformations we find that the R-charge of $X$ and mixing coefficients are
\begin{equation}
    \varepsilon_1 = - \frac{1}{3} \,, \qquad \varepsilon_2 = \varepsilon_3 = 0 \,, \qquad R[X] = \frac{2}{3} \,,
\end{equation}
and the central charges of the infrared SCFT are thus
\begin{equation}
    a = c = \frac{3}{8} \operatorname{dim}(G) \simeq 0.3750 \, \dim (G) \,.
\end{equation}
In fact, the infrared is nothing other than a point on the conformal manifold of the gauging
\begin{equation}
    \begin{aligned}
    \begin{tikzpicture}
      \node[gaugeN1] (s0) {$SU(3)$};
      \node[d2] (c2) [left=0.6cm of s0] {$\mathcal{D}_{2}(SU(3))$};
      \node[d2] (c3) [right=0.6cm of s0] {$\mathcal{D}_{2}(SU(3))$};
      \draw (s0.west) -- (c2.east);
      \draw (s0.east) -- (c3.west);
       \draw[dashed, ->] (s0) to[out=130, in=410, looseness=4] (s0);
      \draw[dashed, ->] (s0) to[out=146, in=394, looseness=5] (s0);
    \end{tikzpicture} \,.
\end{aligned}
\end{equation}
with trivial superpotential. The network of subsequent deformations of this IR SCFT is given following equation \eqref{eqn:2D2G}. This is not surprising as we can swap the order of the deformations. Taking the $u_1$ deformation first, we transform the first $\mathcal{D}_2(SU(3))$ factor into an adjoint-valued chiral multiplet, $Y$, and thus end up with equation \eqref{eqn:2D2G} deformed by the superpotential $\trace Y^3$. As discussed around equation \eqref{eqn:2D2G}, $\trace Y^3$, as well as $\trace \mu_2 X$ and $\trace \mu_3 X$, are marginal, and thus we land on the same conformal manifold as equation \eqref{eqn:2D2G}.

\paragraph{\texorpdfstring{\uline{$W = \trace \mu_3 X + \trace \mu_2X + u_3$ deformation: No SCFT}}{W = Trμ₃X + Trμ₂X + u₃}}
This sequence of deformations leads to the anomaly-free R-symmetry being fixed by the parameters
\begin{equation}
    \varepsilon_1 = -1 \,, \quad \varepsilon_2 = \varepsilon_3 = -\frac{1}{3} \,, \quad R[X] = \frac{4}{3} \,.
\end{equation}
Naively, the central charges arising from such a superconformal R-symmetry vanish, signaling that this final $u_3$ deformation does not trigger a flow to sensible SCFT. Again, this is not surprising, utilizing the dualities described in Section \ref{sec:ADdeformations}. Performing the $u_2$ deformation first, we turn the second $\mathcal{D}_2(SU(3))$ factor into an adjoint-valued chiral multiplet, $Y$, deformed by $\trace Y^3$. Under this duality the $\trace \mu_2 X$ operator maps to $\trace XY$. Since $\trace Y^3$ and $\trace \mu_1 X$ are marginal (as explained around equation \eqref{eqn:22marg}), the IR is the same as the theory in equation \eqref{eqn:2D2G} deformed by $\trace XY$, which we already established does not flow to an infrared SCFT.

\paragraph{\texorpdfstring{\uline{$W = \trace \mu_3 X + \trace \mu_2X + Q^2u_1$ deformation: $a \neq c$}}{W = Trμ₃X + Trμ₂X + Q²u₁}} As usual, we use the duality in Section \ref{sec:ADdeformations} to argue that the infrared consists of a free $\mathcal{N}=2$ vector together with the 
\begin{equation}
    \trace \mu_2 X + \trace \mu_1X \,,
\end{equation}
deformation of the SCFT associated to the gauge theory in equation \eqref{eqn:2D2G1adj}. As already discussed, such deformations do not lead to an interacting $a=c$ SCFT in the infrared.

\paragraph{\texorpdfstring{\uline{$W = \trace \mu_3 X + \trace \mu_2X + Q^2u_3$ deformation: No SCFT}}{W = Trμ₃X + Trμ₂X + Q²u₃}} After this deformation, the superpotential terms together with gauge-anomaly cancellation yields the following infrared R-symmetry:
\begin{equation}
    \varepsilon_1 = \varepsilon_2 = \varepsilon_3 = \frac{1}{3} \,, \qquad R[X] = 0 \,. 
\end{equation}
This is the same R-symmetry obtained via the deformation
\begin{equation}
    Q^2u_3 + Q^2u_2 + Q^2u_1 \,,
\end{equation}
as discussed around equation \eqref{eqn:banette}. Similarly to what was discussed there, we do not expect an SCFT in the infrared.

\paragraph{\texorpdfstring{\uline{$W = \trace \mu_3 X + \trace \mu_2X + u_2u_3$ deformation}}{W = Trμ₃X + Trμ₂X + u₂u₃}} We find that the superconformal R-symmetry is fixed via
\begin{equation}\label{eqn:trubbish}
    \varepsilon_1 = -\frac{1}{3} \,, \qquad \varepsilon_2 = \varepsilon_3 = 0 \,, \qquad R[X] = \frac{2}{3} \,.
\end{equation}
The central charges are identical:
\begin{equation}
    a = c = \frac{3}{8} \operatorname{dim}(G)  \simeq 0.3750 \, \dim (G)\,.
\end{equation}
This is the same central charges as those of the conformal gauge theory around equation \eqref{eqn:mewtwo}. Following from the dualities described in Section \ref{sec:ADdeformations}, we know that when we have a $\mathcal{D}_2(SU(3))$ factor where the mixing coefficient for the $U(1)$ flavor inside of the $\mathcal{N}=2$ R-symmetry is $\varepsilon = -1/3$, after gauging and flow into the infrared, then the infrared is the same as replacing the $\mathcal{D}_2(SU(3))$ factor with an adjoint valued chiral multiplet $Y$ together with the $\trace Y^3$ superpotential. Therefore, the infrared is the same as the theory in equation \eqref{eqn:2D2G} deformed by
\begin{equation}
    \trace \mu_3 X + \trace \mu_2 X + u_2u_3 + \trace Y^3 \,.
\end{equation}
Each of these operators is, in fact, marginal, as we can see from equation \eqref{eqn:22marg}. Thus, the infrared of the 
\begin{equation}
    W = \trace \mu_3 X + \trace \mu_2X + u_2u_3 \,,
\end{equation}
deformation belongs to the same conformal manifold as the infrared of the $W = u_3$
deformation. We have studied the further network of deformations around equation \eqref{eqn:2D2G}.

\paragraph{\texorpdfstring{\uline{$W = \trace \mu_3 X + \trace \mu_2X + u_3^2$ deformation}}{W = Trμ₃X + Trμ₂X + u₃²}} We find the exact same infrared R-symmetry as in the $W = \trace \mu_3 X + \trace \mu_2X + u_2u_3$ deformation just discussed; the mixing coefficients and R-charge of $X$ are as in equation \eqref{eqn:trubbish}. By similar arguments to those appearing in the previous paragraph, the infrared behavior of this deformed SCFT is the same as that of the $W = u_3$ deformation, up to moving on the conformal manifold. As such, we refer the reader to equation \eqref{eqn:2D2G} and what follows for the details of the further deformation landscape.

\paragraph{\texorpdfstring{\uline{$W = \trace \mu_3 X + \trace \mu_2X + \trace \mu_1 X$ deformation: IR free}}{W = Trμ₃X + Trμ₂X + Trμ₁X}} Solving for the mixing coefficients and the infrared R-charge of the chiral multiplets, we find:
\begin{equation}
    \varepsilon_1 = \varepsilon_2 = \varepsilon_3 = \frac{1}{3} \,, \qquad R[X] = 0 \,.
\end{equation}
It is clear that several operators, such as $\trace X^2$ and $\trace X^3$ decouple along the flow into the infrared triggered by the final deformation: $\trace \mu_1 X$. Thus we do not obtain an interacting fixed point with $a = c$. In fact, we can see that this theory flows to a theory of five free $\mathcal{N}=2$ vector multiplets in the infrared.

\paragraph{\texorpdfstring{\uline{$W = \trace \mu_3 X + \trace \mu_2X + \trace X^2$ deformation}}{W = Trμ₃X + Trμ₂X + TrX²}} The infrared R-symmetry is specified by the following:
\begin{equation}
    \varepsilon_1 = - \frac{2}{3} \,, \qquad \varepsilon_2 = \varepsilon_3 = - \frac{1}{6} \,, \qquad R[X] = 1 \,.
\end{equation}
These is the same R-symmetry that appeared in the deformation
\begin{equation}
    W = \trace \mu_3 X + \trace X^2 + \trace \mu_2\mu_3 \,.
\end{equation}
This is expected since we can use the mass deformation to remove the chiral multiplet $X$. See around equation \eqref{eqn:dusklops} for a description of the infrared behavior.

\paragraph{\texorpdfstring{\uline{$W = \trace \mu_3 X + \trace \mu_2X + \trace X^3$ deformation}}{W = Trμ₃X + Trμ₂X + TrX³}} This sequence of deformations leads to the same infrared fixed point as that of
\begin{equation}
    W = \trace \mu_3 X + \trace X^3 + \trace \mu_2X \,.
\end{equation}
This theory, and its further possible deformations, was discussed around equation \eqref{eqn:raichu}.

\paragraph{\texorpdfstring{\uline{$W = \trace \mu_3 X + \trace \mu_2X + u_3 \trace X^2$ deformation}}{W = Trμ₃X + Trμ₂X + u₃TrX²}} After the deformations, the infrared R-symmetry is specified via
\begin{equation}
    \varepsilon_1 = - \frac{5}{21} \,, \qquad \varepsilon_3 = \varepsilon_3 = \frac{1}{21} \,, \qquad R[X] = \frac{4}{7} \,.
\end{equation}
We can see that no operator decouples along the flow, so we obtain an infrared SCFT with identical central charges:
\begin{equation}\label{eqn:weedle}
    a = c = \frac{135}{343} \operatorname{dim}(G)  \simeq 0.2637 \, \dim (G)\,.
\end{equation}
There are a number of relevant operators:
\begin{equation}
    \begin{gathered}
        \trace X^2 \,, \quad \trace X^3 \,, \quad \trace \mu_\alpha X \,, \quad u_\alpha \,, \quad Q^2u_\alpha \,, \quad 
        u_2^2 \,, \quad u_2u_3 \,, \quad u_3^2 \,,
    \end{gathered}
\end{equation}
however, since there is only a single $U(1)$ global symmetry of the SCFT, any further deformation is expected to not lead to an interacting SCFT, modulo emergent symmetries.

\paragraph{\texorpdfstring{\uline{$W = \trace \mu_3 X + u_3\trace X^2$ deformation}}{W = Trμ₃X + u₃TrX²}} We find that the infrared R-symmetry is specified by
\begin{equation}
    \varepsilon_1 = \varepsilon_2 = -\frac{2}{21} \,, \qquad \varepsilon_3 = \frac{1}{21} \,, \qquad R[X] = \frac{4}{7} \,.
\end{equation}
We can see that no chiral ring operators decouple along the flow, and thus we believe the infrared is described by an SCFT with central charges
\begin{equation}
    a = c = \frac{81}{196} \operatorname{dim}(G)  \simeq 0.4133 \, \dim (G) \,.
\end{equation}
The SCFT has eleven relevant chiral ring operators:
\begin{equation}
    u_\alpha \,, \quad Q^2u_\alpha \,, \quad \trace X^2 \,, \quad \trace X^3 \,, \quad \trace \mu_1 X \,, \quad \trace \mu_2 X \,, \quad u_3^2 \,.
\end{equation}
We study each of these deformations in turn. 

\paragraph{\texorpdfstring{\uline{$W = \trace \mu_3 X + u_3\trace X^2 + u_2$ deformation: $a \neq c$}}{W = Trμ₃X + u₃TrX² + u₂}} After this sequence of deformations, the infrared R-symmetry is fixed via
\begin{equation}
    \varepsilon_1 = \frac{1}{21} \,, \qquad \varepsilon_2 = -\frac{1}{3} \,, \qquad \varepsilon_3 = \frac{1}{7} \,, \qquad R[X] = \frac{4}{7} \,.
\end{equation}
Under this R-symmetry, the R-charge of the $u_1$ operator falls below the unitarity bound, and thus it is expected that it decouples along the flow triggered by the $u_2$ deformation. As such, we do not get an interacting SCFT with $a=c$.

\paragraph{\texorpdfstring{\uline{$W = \trace \mu_3 X + u_3\trace X^2 + u_3$ deformation: No R}}{W = Trμ₃X + u₃TrX² + u₃}} There does not exist a consistent infrared R-symmetry, unless there exists some emergent Abelian symmetry along the RG flow into the infrared. As such, we do not expect this sequence of deformations to flow to an interacting $a = c$ SCFT.

\paragraph{\texorpdfstring{\uline{$W = \trace \mu_3 X + u_3\trace X^2 + Q^2u_2$ deformation}}{W = Trμ₃X + u₃TrX² + Q²u₂}} As already discussed around equation \eqref{eqn:torchic}, the deformation $W = \trace \mu_3 X + Q^2u_2$ does not lead to a infrared containing an $a=c$ SCFT sector. The further deformation by $u_3\trace X^2$, thus, is also not an $a=c$ SCFT.

Alternatively, the statement of the duality from Section \ref{sec:ADdeformations} indicates that the infrared R-charge should be formally the same as the gauging of two copies of $\mathcal{D}_2(SU(3))$ together with one adjoint-valued chiral multiplet and deformed by
\begin{equation}
    \trace \mu_3 X + u_3\trace X^2 \,.
\end{equation}
While $u_3\trace X^2$ is seemingly irrelevant from the perspective of the this description at $W=0$ fixed point, if we assume that there is a fixed point when turning both couplings (because it might be dangerously irrelevant), we can determine the R-symmetry fixed by this deformation. 
Then we find that it is specified by
\begin{equation}
    \varepsilon_1 = - 
    \frac{11}{21} \,,  \qquad \varepsilon_2 = \frac{1}{21} \,, \qquad  R[X] = \frac{4}{7} \,.
\end{equation}
Surprisingly, no operators that we can detect decouple under this R-symmetry, and thus we would predict a fixed point with central charges
\begin{equation}
    a = c = \frac{81}{343}\operatorname{dim}(G) \,. 
\end{equation}
Strikingly, this hints us of a rather unexpected phenomena that 
we could conjecture the existence of a non-trivial UV fixed point for the gauging of two $\mathcal{D}_2(SU(3))$ with an adjoint; the existence of the duality under $Q^2u_2$ deformation may be taken as some evidence in this direction. However, we do not have sufficient information to determine which is the correct scenario.

\paragraph{\texorpdfstring{\uline{$W = \trace \mu_3 X + u_3\trace X^2 + Q^2u_3$ deformation: No R}}{W = Trμ₃X + u₃TrX² + Q²u₃}} This sequence of superpotential deformations does not admit a consistent superconformal R-symmetry. Therefore, we do not expect a flow to an interacting infrared SCFT with $a = c$.

\paragraph{\texorpdfstring{\uline{$W = \trace \mu_3 X + u_3\trace X^2 + \trace X^2$ deformation: No R}}{W = Trμ₃X + u₃TrX² + TrX²}} The conditions imposed by anomaly cancellation together with this sequence of three superpotential deformations cannot be solved for the $\varepsilon_3$ and $R[X]$; thus, we do not obtain an $a=c$ SCFT in the infrared.

\paragraph{\texorpdfstring{\uline{$W = \trace \mu_3 X + u_3\trace X^2 + \trace X^3$ deformation: No R}}{W = Trμ₃X + u₃TrX² + TrX³}} Similarly to the previous case, there is no assignment of R-charges, built out of the UV Abelian symmetries, such that this superpotential deformation has a consistent superconformal R-symmetry. Thus, we do not flow to an interacting fixed point with $a = c$.

\paragraph{\texorpdfstring{\uline{$W = \trace \mu_3 X + u_3\trace X^2 + \trace \mu_2 X$ deformation}}{W = Trμ₃X + u₃TrX² + Trμ₂X}} Swapping the order of the deformations, this leads to the same $a=c$ fixed point as $W = \trace \mu_3 X + \trace \mu_2 X + u_3\trace X^2$, which was discussed around equation \eqref{eqn:weedle}.

\paragraph{\texorpdfstring{\uline{$W = \trace \mu_3 X + u_3\trace X^2 + u_3^2$ deformation: No R}}{W = Trμ₃X + u₃TrX² + u₃²}} Such a sequence of deformations does not admit a consistent superconformal R-symmetry formed out of the $U(1)$ symmetries present in the UV. As such, this deformation does not trigger a flow to an infrared SCFT with $a = c$.

\paragraph{\texorpdfstring{\uline{$W = \trace \mu_3 X + u_3^2$ deformation}}{W = Trμ₃X + u₃²}} The infrared superconformal R-symmetry is fixed by the following:
\begin{equation}
    \varepsilon_1 = \varepsilon_2 = -\frac{1}{6} \,, \qquad \varepsilon_3 = 0 \,, \qquad R[X] = \frac{2}{3} \,,
\end{equation}
which leads to the identical central charges of the infrared SCFT being
\begin{equation}
    a = c = \frac{51}{128} \operatorname{dim}(G)  \simeq 0.3984 \, \dim (G) \,.
\end{equation}
This leads to the same infrared R-symmetry and central charges as the deformation 
\begin{equation}
    W = \trace \mu_3 X + \trace X^3 \,,
\end{equation}
discussed around equation \eqref{eqn:pineapple}. Indeed, one can check that the relevant operator spectrum is the same, and that the non-R global symmetries also agree; thus, we conclude that these two deformations lead to the same infrared fixed point. The network of subsequent deformations has been discussed around equation \eqref{eqn:pineapple}.

\paragraph{\texorpdfstring{\uline{$W = \trace X^3$ deformation}}{W = TrX³}} We now consider the cubic deformation with respect to the adjoint-valued chiral multiplet of the starting point SCFT whose gauge theory description was given in Figure \ref{fig:bigboy}:
\begin{equation}
    \trace  X^3 \,.
\end{equation}
After deformation, $a$-maximization and anomaly cancellation fix the mixing coefficients and the R-charge of the chiral multiplet to be:
\begin{equation}
    \varepsilon_1 = \varepsilon_2 = \varepsilon_3 = - \frac{1}{9} \,, \qquad R[X] = \frac{2}{3} \,,
\end{equation}
and thus we find that the central charges of the infrared SCFT are
\begin{equation}
    a = c = \frac{11}{27} \dim (G)  \simeq 0.4074 \, \dim (G) \,.
\end{equation}
We can enumerate the spectrum of relevant chiral ring operators of this deformed SCFT, and we find that they are the following
\begin{equation}
    u_\alpha \,, \qquad Q^2 u_\alpha \,, \qquad \trace  \mu_\alpha X \,, \qquad \trace X^2 \,.
\end{equation}
We consider these deformations, and their subsequent deformations, in turn. Without loss of generality, we will take $\alpha = 3$.

\paragraph{\texorpdfstring{\uline{$W = \trace X^3 + u_3$ deformation}}{W = TrX³ + u₃}} Next, we consider the sequential deformations 
\begin{equation}
    W = \trace X^3 + u_3 \,.
\end{equation}
As explained in Section \ref{sec:ADdeformations}, the $u_3$ deformation has the effect of dissolving the associated Argyres--Douglas factor into a new adjoint-valued chiral multiplet, together with a cubic superpotential term. Thus, this deformation leads us to a point on the conformal manifold of the infrared of the gauging of two copies of $\mathcal{D}_2(SU(3))$ together with two adjoint-valued chiral multiplets:
\begin{equation}
    \begin{aligned}
    \begin{tikzpicture}
      \node[gaugeN1] (s0) {$SU(3)$};
      \node[d2] (c2) [left=0.6cm of s0] {$\mathcal{D}_{2}(SU(3))$};
      \node[d2] (c3) [right=0.6cm of s0] {$\mathcal{D}_{2}(SU(3))$};
      \draw (s0.west) -- (c2.east);
      \draw (s0.east) -- (c3.west);
       \draw[dashed, ->] (s0) to[out=130, in=410, looseness=4] (s0);
      \draw[dashed, ->] (s0) to[out=146, in=394, looseness=5] (s0);
    \end{tikzpicture} \,.
\end{aligned}
\end{equation}
The deformations of this SCFT have already been discussed from equation \eqref{eqn:2D2G} onwards.

\paragraph{\texorpdfstring{\uline{$W = \trace X^3 + Q^2u_3$ deformation: $\CN=4$ SYM}}{W = TrX³ + Q²u₃}} Next, we consider further enacting the $Q^2u_3$ deformation. As explained in detail in Section \ref{sec:ADdeformations}, this deformation leads to an infrared which is a product of an interacting $a=c$ SCFT and a free $\mathcal{N}=2$ vector multiplet. The interacting part of the infrared is the same as the $W = \trace X^3$ deformation of the infrared SCFT of the following gauge theory:
\begin{equation}
    \begin{aligned}
    \begin{tikzpicture}
      \node[gaugeN1] (s0) {$SU(3)$};
      \node[d2] (c2) [left=0.6cm of s0] {$\mathcal{D}_{2}(SU(3))$};
      \node[d2] (c3) [right=0.6cm of s0] {$\mathcal{D}_{2}(SU(3))$};
      \draw (s0.west) -- (c2.east);
      \draw (s0.east) -- (c3.west);
       \draw[dashed, ->] (s0) to[out=130, in=410, looseness=4] (s0);
    \end{tikzpicture} \,.
\end{aligned}
\end{equation}
As explained around equation \eqref{eqn:feebas}, this is nothing other than $\mathcal{N}=4$ super-Yang--Mills; the deformations of this theory have been discussed in Section \ref{sec:defsN4}.

\paragraph{\texorpdfstring{\uline{$W = \trace X^3 + \trace \mu_3 X$ deformation}}{W = TrX³ + Q²u₃ + Trμ₃X}} Continuing, we consider the further deformation by the $\trace \mu_3 X$ operator. The resulting theory is the same as that obtained by performing the deformations in the opposite order, that is, 
\begin{equation}
    W = \trace \mu_3 X + \trace X^3 \,.
\end{equation}
The network of subsequent deformations has appeared around equation \eqref{eqn:pineapple}.

\paragraph{\texorpdfstring{\uline{$W = \trace X^3 + \trace X^2$ deformation: No R}}{W = TrX³ + TrX²}} 
Finally, we must consider the deformation by the $\trace X^2$ operator. After such a deformation, it is not possible to find a consistent superconformal R-symmetry, assuming that there are no emergent $U(1)$ symmetries along the flow, and thus we assume that the endpoint of the flow does not have superconformal symmetry. 

\paragraph{\texorpdfstring{\uline{$W = Q^2u_3$ deformation}}{W = Q²u₃}} Next, we look at relevant deformations of the SCFT with UV description as given in Figure \ref{fig:bigboy} via the operators $Q^2u_\alpha$, where, without loss of generality, we take $\alpha = 3$:
\begin{equation}
    W = Q^2 u_3 \,.
\end{equation}
As described in Section \ref{sec:ADdeformations}, such a deformation dissolves the third $\mathcal{D}_2(SU(3))$ factor into a single free $\mathcal{N}=2$ vector multiplet, and the interacting $a=c$ sector of the infrared is given just by the infrared of the $SU(3)$ gauging of two $\mathcal{D}_2(SU(3))$ theories, together with one adjoint-valued chiral multiplet and trivial superpotential:
\begin{equation}
    \begin{aligned}
    \begin{tikzpicture}
      \node[gaugeN1] (s0) {$SU(3)$};
      \node[d2] (c2) [left=0.6cm of s0] {$\mathcal{D}_{2}(SU(3))$};
      \node[d2] (c3) [right=0.6cm of s0] {$\mathcal{D}_{2}(SU(3))$};
      \draw (s0.west) -- (c2.east);
      \draw (s0.east) -- (c3.west);
       \draw[dashed, ->] (s0) to[out=130, in=410, looseness=4] (s0);
    \end{tikzpicture} \,.
\end{aligned}
\end{equation}
We have discussed the network of superpotential deformations of this theory already; starting from equation \eqref{eqn:2D2G1adj}.

\paragraph{\texorpdfstring{\uline{$W = \trace X^2$ deformation: $\CN=4$ SYM}}{W = TrX²}} Finally, we can consider the mass deformation of the infrared SCFT associated with the asymptotically-free gauge theory depicted in Figure \ref{fig:bigboy}, which integrates out $X$ in the infrared:
\begin{equation}
    W = \trace X^2 \,.
\end{equation}
Such a deformation leads to the same infrared SCFT as that of the asymptotically-free $SU(3)$ gauge theory obtained via gauging three copies of $\mathcal{D}_2(SU(3))$. As was shown in \cite{Kang:2023dsa}, this is nothing other than (a point on the ($\mathcal{N}=1$)-preserving conformal manifold of) $\mathcal{N}=4$ super-Yang--Mills with gauge group $SU(3)$; the network of superpotential deformations arising from this infrared fixed point have already been discussed. See Section \ref{sec:defsN4} for details.

\section{Discussion}\label{sec:discussion}

In this paper, we have started from the asymptotically-free $\mathcal{N}=1$ gauge theory in Figure \ref{fig:bigboy}, and we have determined all sequences of relevant deformations of the infrared SCFT which give rise to SCFTs with identical central charges. We find twenty-one different interacting $a=c$ fixed points in this landscape, which we have summarized in Table \ref{tbl:summary} and depicted as a graph in Figure \ref{fig:deformations}. This is one example of the landscape of $a=c$ SCFTs obtained from superpotential deformations of gauged Argyres--Douglas theories; the landscape of generic gauged Argyres--Douglas theories will be discussed in future work \cite{LANDSCAPEII,LANDSCAPEIII}.

Interestingly, most of our SCFTs in the landscape are connected either to $\CN=4$ super-Yang--Mils or to the $\CN=1$ theory with two adjoint chiral multiplets with the latter being the one with the lowest central charge. It would be desirable to have a supergravity interpretation of our RG flows via finding domain wall solutions interpolating the fixed points. We leave this supergravity connection for future work.

\begin{table}[p]
    \centering
    \begin{threeparttable}
    \begin{minipage}{0.465\textwidth}
    \centering
    \renewcommand{\arraystretch}{1.1}
    \resizebox{\textwidth}{!}{%
    \begin{tabular}{cc}
        \toprule
        $W$ & $\begin{gathered}\dfrac{a = c}{\operatorname{dim}(G)}\end{gathered}$  \\\midrule
        $0$ & $\dfrac{27(14 + 15\sqrt{15})}{4624}$\\\midrule 
        $\trace \mu_3 X$ & $\dfrac{323 + 109\sqrt{109}}{3528}$ \\\midrule
        $\trace \mu_3 X + u_3\trace X^2$ & $\dfrac{81}{196} $ \\\midrule
        $\trace X^3$ & $\dfrac{11}{27}$ \\\midrule
        $\trace \mu_3 X + \trace X^3$ & \multirow{3}{*}{$\dfrac{51}{128} $} \\
        $\trace \mu_3 X + u_3^2$ & \\
        $\trace X^3 + \trace \mu_3 X$ & \\\midrule
        $\trace \mu_3 X + \trace \mu_2 X$ & $\dfrac{10 + 7\sqrt{7}}{72}$ \\\midrule
        $\trace \mu_3 X + \trace \mu_2 X + u_3\trace X^2$ & \multirow{2}{*}{$\dfrac{135}{343}$} \\
        $\trace \mu_3 X + u_3\trace X^2 + \trace \mu_2 X$ & \\\midrule
        $u_3$ & \multirow{9}{*}{$\dfrac{3}{8}$} \\
        $\trace \mu_3X + u_2$ & \\
        $\trace \mu_3X + \trace X^3 + u_2$ & \\
        $\trace \mu_3X + \trace X^3 + \trace \mu_2 X$ & \\
        $\trace \mu_3X + \trace \mu_2 X + \trace X^3$ & \\
        $\trace \mu_3X + \trace \mu_2 X + u_1$ & \\
        $\trace \mu_3X + \trace \mu_2 X + u_2u_3$ & \\
        $\trace \mu_3X + \trace \mu_2 X + u_3^2$ & \\
        $\trace X^3 + u_3$ & \\\midrule 
        $u_3 + \trace Y^2$ & \multirow{2}{*}{$\dfrac{3\sqrt{3}}{16}$} \\
        $Q^2u_3$ & \\\midrule 
        $u_3 + \trace Y^2 + (\trace X^2)^2$ & $\dfrac{81}{256}$ \\\midrule 
        $u_3 + Q^2u_2$ & $\dfrac{116+31\sqrt{31}}{968}$ \\\bottomrule 
    \end{tabular}} \vspace{21.5pt}
    \end{minipage}\qquad 
    \begin{minipage}{0.465\textwidth}
    \centering
    \renewcommand{\arraystretch}{1.08}
    \resizebox{\textwidth}{!}{%
    \begin{tabular}{cc}
        \toprule
        $W$ & $\begin{gathered}\dfrac{a = c}{\operatorname{dim}(G)}\end{gathered}$  \\\midrule
        $u_3 + Q^2u_2 + \trace \mu_1 Y$ & $\dfrac{5(7\sqrt{7}-10)}{144}$ \\\midrule 
        $u_3 + Q^2u_2 + \trace Y^3$ & \multirow{2}{*}{$\dfrac{442+79\sqrt{79}}{3888}$} \\
        $u_3 + \trace Y^2 + u_2$ & \\\midrule
        $u_3 + Q^2u_2 + \trace \mu_1 Y + \trace Y^3$ & \multirow{4}{*}{$\dfrac{75}{256}$} \\
        $u_3 + Q^2u_2 + \trace Y^3 + \trace \mu_1 Y$ & \\
        $u_3 + Q^2u_2 + \trace \mu_1 Y + u_1^2$ & \\
        $u_3+\trace Y^2+(\trace X^2)^2 + u_2$ & \\\midrule
        $u_3 + Q^2u_2 + \trace XY^2$ & $\dfrac{28+19\sqrt{19}}{400}$ \\\midrule
        $u_3 + Q^2u_2 + \trace XY^2 + (\trace X^2)^2$ & $\dfrac{567}{2048}$ \\\midrule
        $u_3 + Q^2u_2 + \trace \mu_1 Y + \trace XY^2$ & \multirow{2}{*}{$\dfrac{27}{100}$} \\
        $u_3 + Q^2u_2 + \trace XY^2 + \trace \mu_1 Y$ & \\\midrule 
        $u_3 + \trace Y^2 + (\trace X^2)^2 + \trace \mu_2 X$ & $\dfrac{135}{512}$ \\\midrule 
        $u_3 + \trace Y^2 + \trace X^3$ & \multirow{9}{*}{$\dfrac{1}{4}$} \\
        $u_3 + Q^2u_2 + u_1$ & \\
        $u_3 + Q^2u_2 + \trace Y^3 + \trace X^3$ & \\
        $u_3 + Q^2u_2 + \trace Y^3 + \trace X^2Y$ & \\
        $u_3 + Q^2u_2 + \trace Y^3 + u_1$ & \\
        $u_3 + Q^2u_2 + \trace XY^2 + u_1$ & \\
        $u_3 + Q^2u_2 + \trace XY^2 + \trace X^3$ & \\
        $\trace X^3 + Q^2u_3$ & \\
        $\trace X^2$ & \\\midrule 
        $\trace \mu_3 X + \trace X^2$ & $\dfrac{243}{1024} $ \\\midrule
        $\trace \mu_3 X + u_3\trace X^2 + Q^2u_2$ & $\dfrac{81}{343} $ \\\midrule
        $u_3 + Q^2u_2 + Q^2u_1$ & \multirow{3}{*}{$\dfrac{27}{128}$} \\
        $\trace \mu_3 X + \trace X^2 + \trace \mu_2\mu_3$ & \\
        $\trace \mu_3 X + \trace \mu_2 X + \trace X^2 $ & \\\bottomrule 
    \end{tabular}} 
    \end{minipage}
    \end{threeparttable}
    \caption{The $a=c$ fixed points belonging to the landscape obtained by superpotential deformations of the gauged Argyres--Douglas theory in Figure \ref{fig:bigboy}. The entries are ordered by decreasing central charge. Note that we have not included entries for deformations that first land on $\mathcal{N}=4$ super-Yang--Mills and then are further deformed to the two-adjoint theory. For sequences of deformations where it was unclear what fixed point is landed on, we have written only one here; see Section \ref{sec:222landscape} for the full details.}
    \label{tbl:summary}
\end{table}

\subsection{Novel dualities}

In the process of deriving this landscape of $a=c$ theories, we discovered a variety of interesting dualities. For dualities amongst the $a=c$ fixed point, these are straightforwardly read off from Table \ref{tbl:summary}. Some examples would be the following duality:
\begin{equation}
    \begin{gathered}
    \scalebox{0.8}{
      \begin{tikzpicture}
      \node[gaugeN1] (s0) {$SU(3)$};
      \node[d2] (c1) [left=0.6cm of s0] {$\mathcal{D}_{2}(SU(3))$};
      \node[d2] (c2) [right=0.6cm of s0] {$\mathcal{D}_{2}(SU(3))$};
      \node[d2] (c3) [below=0.4cm of s0] {$\mathcal{D}_{2}(SU(3))$};
      \node (tt) [below=1.2cm of s0] {$W = \trace \mu_3X + \trace \mu_2X + \trace X^3$};
      \draw (s0.east) -- (c2.west);
      \draw (s0.west) -- (c1.east);
      \draw (s0.south) -- (c3.north);
      \draw[dashed, ->] (s0) to[out=130, in=410, looseness=4] (s0);
    \end{tikzpicture}}
    \end{gathered} \quad \simeq \quad 
    \begin{gathered}
      \scalebox{0.8}{\begin{tikzpicture}
      \node[gaugeN1] (s0) {$SU(3)$};
      \node[d2] (c2) [left=0.6cm of s0] {$\mathcal{D}_{2}(SU(3))$};
      \node[d2] (c3) [right=0.6cm of s0] {$\mathcal{D}_{2}(SU(3))$};
      \draw (s0.west) -- (c2.east);
      \draw (s0.east) -- (c3.west);
       \draw[dashed, ->] (s0) to[out=130, in=410, looseness=4] (s0);
      \draw[dashed, ->] (s0) to[out=146, in=394, looseness=5] (s0);
    \end{tikzpicture}}
    \end{gathered} \quad \,,
\end{equation}
or
\begin{equation}
    \begin{gathered}
    \scalebox{0.8}{
      \begin{tikzpicture}
      \node[gaugeN1] (s0) {$SU(3)$};
      \node[d2] (c1) [left=0.6cm of s0] {$\mathcal{D}_{2}(SU(3))$};
      \node[d2] (c2) [right=0.6cm of s0] {$\mathcal{D}_{2}(SU(3))$};
      \node (tt) [below=0.3cm of s0] {$W = \trace X^3$};
      \draw (s0.east) -- (c2.west);
      \draw (s0.west) -- (c1.east);
      \draw[dashed, ->] (s0) to[out=130, in=410, looseness=4] (s0);
    \end{tikzpicture}}
    \end{gathered} \quad \simeq \quad 
    \begin{gathered}
        \mathcal{N}=4 \text{ super-Yang--Mills } \\
        G = SU(3) 
    \end{gathered} \quad \,.
\end{equation}
Many of these dualities can be explained via recourse to Section \ref{sec:ADdeformations}, where it is explained that, under the right conditions, a copy of $\mathcal{D}_2(SU(3))$ can be exchanged with a chiral multiplet deformed via a cubic superpotential, and vice versa. 

We are also led to dualities that do not involve $a=c$ SCFTs in the infrared. For example, we found that a superpotential deformation of the two-adjoint theory leads to a collection of free $\mathcal{N}=2$ vector multiplets:
\begin{equation}
    \begin{gathered}
    \scalebox{0.8}{
      \begin{tikzpicture}
      \node[gaugeN1] (s0) {$SU(N)$};
      \node (tt) [below=0.3cm of s0] {$W = \trace XY^2$};
      \draw[dashed, ->] (s0) to[out=130, in=410, looseness=4] (s0);
      \draw[dashed, ->] (s0) to[out=146, in=394, looseness=5] (s0);
    \end{tikzpicture}}
    \end{gathered} \quad \simeq \qquad 
    \begin{gathered}
        (N-1) \text{ free } \\
        \mathcal{N}=2 \text{ vector multiplets }
    \end{gathered} \quad \,.
\end{equation}
Another example where the infrared behavior is simply a collection of free vector multiplets is the following deformed gauged Argyres--Douglas theory:
\begin{equation}
    \begin{gathered}
    \scalebox{0.8}{
      \begin{tikzpicture}
      \node[gaugeN1] (s0) {$SU(3)$};
      \node[d2] (c1) [left=0.6cm of s0] {$\mathcal{D}_{2}(SU(3))$};
      \node[d2] (c2) [right=0.6cm of s0] {$\mathcal{D}_{2}(SU(3))$};
      \node[d2] (c3) [below=0.4cm of s0] {$\mathcal{D}_{2}(SU(3))$};
      \node (tt) [below=1.2cm of s0] {$W = \trace \mu_3X + Q^2u_3$};
      \draw (s0.east) -- (c2.west);
      \draw (s0.west) -- (c1.east);
      \draw (s0.south) -- (c3.north);
      \draw[dashed, ->] (s0) to[out=130, in=410, looseness=4] (s0);
    \end{tikzpicture}}
    \end{gathered} \quad \simeq \qquad 
    \begin{gathered}
        5 \text{ free } \\
        \mathcal{N}=2 \text{ vector multiplets }
    \end{gathered} \quad \,.
\end{equation}
As expected, studies of the landscape of superpotential deformations of supersymmetric quantum field theories novel structures and relations between such theories, including non-trivial dualities. Indeed, it is particularly interesting that, in the landscape explored in this paper, most deformations pass through a locus involving enhanced $\mathcal{N}=4$ supersymmetry; this feature is replicated in the study of the landscape of other gauged Argyres--Douglas SCFTs \cite{LANDSCAPEII,LANDSCAPEIII}. It is tempting to suggest that this behavior is related to the $a=c$ property.

\subsection{Infrared fate}

The main focus in this paper has been to analyze when the infrared of a deformed $a=c$ SCFT is another $a=c$ SCFT. However, during this analysis, we discovered many deformations of $a=c$ SCFTs which do not appear to lead to an infrared SCFT as the ABJ anomaly breaks the only remaining classical $U(1)$ global symmetry; thus, modulo emergent symmetries, there does not exist a $U(1)_R$ R-symmetry in the infrared, and therefore the infrared does not realize superconformal symmetry. A natural question then is: what is the infrared behavior after such a sequence of deformations? One way to constrain the infrared behavior is to study the anomalies in the ultraviolet and use 't Hooft anomaly matching \cite{tHooft:1979rat} to learn about the anomalies of the infrared theory. Such constraints also hold for the anomalies of generalized global symmetries \cite{Gaiotto:2014kfa}.

Several examples of generalized symmetries of 4d gauge theories constraining the infrared behavior are well-known. Consider first pure $\mathcal{N}=1$ super-Yang--Mills with gauge group $SU(N)$: the theory has a one-form symmetry $\mathbb{Z}_N$, a zero-form symmetry $\mathbb{Z}_{2N}$, and a mixed anomaly between them. One infrared behavior satisfying the anomaly is spontaneous symmetry breaking of the $\mathbb{Z}_{2N}$ to $\mathbb{Z}_2$, where there are $N$ distinct confining vacua; this behavior is confirmed via gaugino condensation \cite{Witten:1982df,Veneziano:1982ah}. In another example \cite{Gaiotto:2017yup}, 4d pure Yang--Mills theory with gauge group $SU(2N)$ with $\theta$-angle $\pi$ has a mixed anomaly between the $\mathbb{Z}_{2N}$ one-form symmetry and time-reversal symmetry; such a anomaly enforces, at least, that the infrared is not trivially gapped.

More examples arising from deformations of 4d $\mathcal{N}=1$ Lagrangian gauge theories were discovered in \cite{Kang:2024elv}. As an indicative example, it was shown that $Spin(12)$ gauge theory with a single chiral multiplet, $Q$, in the $\bm{77}$ representation deformed in the infrared by a superpotential $W = Q^4$ leads to an infrared which is neither an SCFT nor trivially gapped due to the presence of a mixed anomaly between the $\mathbb{Z}_2 \oplus \mathbb{Z}_2$ one-form symmetry and the $\mathbb{Z}_4$ zero-form symmetry. One possibility, consistent with the anomaly, is that the infrared involves spontaneous symmetry breaking where $\mathbb{Z}_4 \rightarrow \mathbb{Z}_2$ with two confining vacua. 

When considering the gauged Argyres--Douglas theories that have been reviewed in Section \ref{sec:acdefs}, the GCD-condition in equation \eqref{eqn:GCDCOND} generally leads to the resulting SCFT after gauging the global symmetry $G$ having a one-form symmetry which is the center of $G$: $Z(G)$.\footnote{More precisely, the intermediate defect group of the gauged theory is $\mathbb{D} = Z(G) \oplus Z(G)$ and the one-form symmetry is fixed after picking a choice of Lagrangian subgroup of $\mathbb{D}$. Here, we assumed that we introduced a simply-connected gauge group $G$, which is equivalent to a choice of Lagrangian subgroup which makes the one-form symmetry $Z(G)$. For further discussion on this subtlety, see \cite{Gukov:2020btk,Lawrie:2023tdz,Kang:2024elv}.} Furthermore, there is a non-trivial discrete zero-form symmetry, $\mathbb{Z}_L$, which survives from the $U(1)$ broken by the ABJ anomaly. For simplicity, we assume that $G = SU(N)$. Denoting the background two-form gauge field for the $\mathbb{Z}_N$ one-form symmetry as $B$, we can see that, under a $\mathbb{Z}_L$ zero-form symmetry transformation, the partition function transforms as follows:
\begin{equation}\label{eqn:frenchtoast}
    Z[B] \rightarrow \exp\left( 2 \pi i \ell \frac{N-1}{2N} \int_X \mathcal{P}(B) \right) Z[B] \,.
\end{equation}
Here $\mathcal{P}(B)$ is the Pontryagin square, $\ell$ is the element of $\mathbb{Z}_L$ parametrizing the zero-form transformation, and $X$ is an arbitrary spacetime manifold over which the theory is defined. Since the phase is generally non-trivial, the SCFT has a mixed anomaly between the zero-form and one-form symmetries. Turning on relevant deformations typically breaks the zero-form symmetry to a subgroup $\mathbb{Z}_L \rightarrow \mathbb{Z}_{L'}$; if the generator of $\mathbb{Z}_{L'}$ is such that the phase in equation \eqref{eqn:frenchtoast} remains non-trivial then the mixed anomaly survives the deformation.

It is straightforward to apply this analysis to study the mixed anomalies of the R-symmetry-breaking deformations of the theory in Figure \ref{fig:bigboy}, which have been enumerated throughout this paper. We leave an investigation of such constraints on the infrared physics for the future.

\subsection*{Acknowledgements}

M.J.K.,~K.H.L.~and C.L.~thank KAIST for hospitality during the initiatory and the final phases of this work; M.J.K.~is grateful to the Aspen Center for Physics where the work is performed in part, which is supported by a grant from the Simons Foundation (1161654, Troyer). M.J.K., C.L., and J.S.~thank the Simons Summer Workshop 2024 for hospitality, especially Martin Roček for 35 Beach Road, 
during the middle stage of this work. 
M.J.K.~is supported by the U.S.~Department of Energy, Office of Science, Office of High Energy Physics, under Award Numbers DE-SC0013528 and QuantISED Award DE-SC0020360.
C.L.~acknowledges support from DESY (Hamburg, Germany), a member of the Helmholtz Association HGF; C.L.~also acknowledges the Deutsche Forschungsgemeinschaft under Germany's Excellence Strategy - EXC 2121 ``Quantum Universe'' - 390833306 and the Collaborative Research Center - SFB 1624 ``Higher Structures, Moduli Spaces, and Integrability'' - 506632645. 
K.H.L.~and J.S.~are supported by the National Research Foundation of Korea (NRF) grant RS-2023-00208602, and also by POSCO Science Fellowship of POSCO TJ Park Foundation. K.H.L.~is also supported by the Israel Science Foundation under grant no.~759/23. J.S is also supported by the National Research Foundation of Korea (NRF) grant RS-2024-00405629.

\bibliography{references}{}
\bibliographystyle{sortedbutpretty}
\end{document}